# Superconductivity in Iron Compounds


G. R. Stewart

Department of Physics, University of Florida, Gainesville, FL 32611-8440



Abstract:  Kamihara and coworkers' report of superconductivity at $T_c$ = 26 K in fluorine-doped LaFeAsO inspired a worldwide effort to understand the nature of the superconductivity in this new class of compounds.  These iron pnictide and chalcogenide (FePn/Ch) superconductors have Fe electrons at the Fermi surface, plus an unusual Fermiology that can change rapidly with doping, which lead to normal and superconducting state properties very different from those in standard electron-phonon coupled 'conventional' superconductors.    Clearly superconductivity and magnetism/magnetic fluctuations are intimately related in the FePn/Ch - and even coexist in some.  Open questions, including the superconducting nodal structure in a number of compounds, abound and are often dependent on improved sample quality for their solution.  With $T_c$ values up to 56 K, the six distinct Fe-containing superconducting structures exhibit complex but often comparable behaviors.  The search for correlations and explanations in this fascinating field of research would benefit from an organization of the large, seemingly disparate data set.  This review attempts to provide an overview, using numerous references, with a focus on the materials and their superconductivity.




## Contents













## I. Introduction

The report of superconductivity at 26 K in LaFeAsO doped with F on the oxygen site in 2008 (Kamihara et al., 2008) was not the first discovery of an iron-containing superconductor, nor even the first reported superconducting iron pnictide (LaFePO, $T_c \approx 5$ K, Kamihara et al., 2006). Although iron has been considered deleterious to superconductivity due to its strong local magnetic moment, a number of superconducting compounds containing iron in which the iron is non-magnetic have long been known. $Th_7Fe_3$ ($T_c$=1.8 K, Matthias, Compton and Corenzwit 1961), $U_6Fe$ ($T_c$=3.9 K Chandrasekhar and Hulm, 1958), $Lu_2Fe_3Si_5$ ($T_c$=6.1 K, Braun 1980), and β''-(bedt-ttf)$_4$[(H$_2$0)Fe(C$_2$0$_4$)$_3$]-PhCN ($T_c$=8.5 K, Graham, Kurmoo and Day 1995) are all examples of Fe-containing superconductors. In fact, Fe itself under pressure is a superconductor, with $T_c \sim 1.8$ K at 20 GPa (Shimizu et al., 2001).

However, the discovery of Kamihara et al. is ground breaking for a number of reasons. One is that – just like the discovery of superconductivity at 35 K in Ba-doped $La_2CuO_4$ (Bednorz and Müller 1986) – it led to the almost immediate further discovery of even higher $T_c$ materials, with the current record ~ 56 K observed in $Gd_{0.8}Th_{0.2}FeAsO$ (C. Wang et al., 2008), $Sr_{0.5}Sm_{0.5}FeAsF$ (G. Wu et al., 2009) and in $Ca_{0.4}Nd_{0.6}FeAsF$ (Cheng et al., 2009). The path to this higher transition temperature was also similar to that in the high $T_c$ cuprates, where pressure experiments (Chu et al., 1987) first increased the $T_c$ in Ba-doped $La_2CuO_4$ from 35 to 53 K. This was followed by 'chemical pressure' experiments where $T_c$ was raised to 93 K (Wu et al., 1987) by replacing La with the smaller Y to make a multi-phase sample containing $YBa_2Cu_3O_{7-\delta}$. In the case of F-doped LaFeAsO, Takahashi et al. (2008a) found that 4 GPa pressure increased the $T_c$ from 26 K



to 43 K.  This result then inspired researchers to use chemical pressure (replacing the La

with the smaller rare earths Gd, Sm, Nd, Pr, Ce), first reaching $T_c$=43 K in

$SmFeAsO_{0.85}F_{0.15}$ (X. H. Chen et al., 2008) and then less than a month later $T_c$=55 K in

the oxygen deficient $SmFeAsO_{0.85}$ prepared by high pressure synthesis (Ren et al.,

2008a).

A second reason why the work of Kamihara et al. is so seminal is that it has led to

a new *class* of high temperature superconductors, the so-called iron pnictides ('FePn',

where Pn is As or P), which have already been extended to include iron chalcogenides

('FeCh', where Ch includes S, Se and Te).  The list of these compounds has expanded

rapidly from the original LaFeAsO '1111' structure (of which there are over 150 Rare

Earth/Transition Metal/Pnictide/O examples, see Pöttgen and Johrendt, 2008 for a

review) first explored by Kamihara et al. and successors for superconductivity.  The next

iron-containing superconductor structure includes members of the $MFe_2As_2$ ('122')

family (of which there are over 450 distinct compounds, Villars and Calvert, 1985),

where Rotter, Tegel and Johrendt (2008) discovered $T_c$=38 K in K-doped $BaFe_2As_2$,

$Ba_{0.6}K_{0.4}Fe_2As_2$.  The MFeAs ('111') family (X. C. Wang et al., 2008, $T_c$=18 K), the iron

chalcogenide FeSe ('11') family (Hsu et al., 2008, $T_c$=8 K), the $Sr_2MO_3FePn$,

M=Sc,V,Cr  ('21311') family (M=Sc and Pn=P, Ogino et al., 2009, $T_c$=17 K; M=V and

Pn=As, Zhu et al., 2009b, $T_c$=37 K) and the defect structure $A_{0.8}Fe_{1.6}Se_2$ ($T_c$≈32 K, A=K,

Rb, Cs, Tl) related to the 122 structure and called '122*' herein round out the established

list.   The last four families all exhibit superconductivity *without* doping an additional

atom type and as yet have only a few members known, although this is changing.  For

example, Ogino et al. (2010c) reported an alteration of the 21311 structure and found



$Ca_2(Mg_{0.25}Ti_{0.75})_{1.5}O_{-4}FeAs$ to have $T_c^{onset} = 47$ K. As will be discussed, it is not just the 1111 structure whose initial $T_c$ has been greatly enhanced by further work. For example, $T_c$ of FeSe under 7 GPa pressure increases more than fourfold to 37 K (Margadonna et al., 2009b) as discussed below in section IID and with Te doping ($Fe(Se_{1-x}Te_x)$) increases almost twofold to ~ 15 K (Yeh et al., 2008), section IIB3 .

A third, and perhaps the most interesting, aspect of these new iron-containing superconductors (the subject of over 2000 publications in just 3 years) from a basic physics point of view is that the superconducting pairing mechanism may be related to the coexistent magnetism in the phase diagram. Current thinking is that the pairing is not primarily phonon mediated, although due to the coupling of the magnitude of the Fe moments to the Fe-Pn/Ch bond length and the presence of an isotope effect (discussed in Section IVA), the magnetoelastic coupling is thought to be important for superconductivity. See, e. g., Cano et al., 2010, for a discussion of the magnetoelastic coupling. Theoretical alternatives to phonon coupling include various *electronic* excitations that could mediate the superconducting pairing, e. g. spin fluctuations (as is suggested by inelastic neutron scattering data) or inter-orbital pair hopping. If this is indeed the case, such a pairing mechanism may promise even higher temperature superconductivity since the transition temperature, $T_c$, would be proportional to a characteristic energy scale potentially significantly larger than the BCS scale dependence on the average phonon frequency, $T_c^{BCS} \propto <\omega>$.

Fourth, as will be clear in this review, the properties of the FePn/Ch superconductors are fundamentally different both from those of a conventional electron-phonon coupled superconductor and also from those of the cuprates.



In a clean conventional superconductor, the electronic excitations are (exponentially) suppressed in the superconducting state by the gap, while in unconventional superconductors like the FePn/Ch there are many examples of compounds with nodal (gap zero) points or lines leading to finite electronic excitations remaining as $T\rightarrow0$. Although the pairing symmetry in the superconducting state is still under debate, it is apparently not conventional s-wave in many of the FePn/Ch since neutron scattering measurements provide convincing (but see Onari, Kontani and Sato, 2010) evidence for a sign change in the superconducting energy gap $\Delta$ on different parts of the Fermi surface in a number of compounds. In certain samples, neutron scattering data imply a direct coupling between the superconductivity and the magnetism, as seen in, for example, the unconventional heavy Fermion superconductor $UPt_3$. As a more mundane (but perhaps fundamentally interesting) comparison with conventional - e. g. elemental or A-15 - superconductors, the discontinuity in the specific heat at $T_c$, $\Delta C$, scales differently in the FePn/Ch superconductors: $\Delta C \propto T_c^3$ vs $T_c^2$ for conventional superconductors.

In comparing to the cuprates, it seems clear that – although the FePn/Ch are unconventional superconductors - they are different in many respects from the cuprates. The cuprates have strong electron correlations, while the FePn/Ch show in general relatively weak correlations, see, e. g., Yang et al. (2009b) who find in representative 1111 and 122 FePn/Ch that the onsite Coulomb repulsion $U \leq 2$ eV vs a bandwidth for the Fe conduction band states of $\sim 4$ eV. Using thermoelectric power (TEP) measurements, Wang, Lei and Petrovic (2011a) argue for relatively weak electronic correlations in 122* $K_xFe_{2-y}Se_2$ while Pourret et al.'s (2011) TEP data are interpreted as



showing that 11 $FeTe_{0.6}Se_{0.4}$ – uniquely among the FePn/Ch and in agreement with DMFT calculations (Korshunov, Hirschfeld, and Mazin, 2011) - has electronic correlations comparable in strength to the cuprates. The cuprates are much more anisotropic and have d-wave gap symmetry vs primarily s-wave for the FePn/Ch. The cuprates have a much different Fermiology that remains relatively constant (at least for hole-doping) with doping vs the Fermiology in the FePn/Ch (whose Fermiology is believed key for the superconducting pairing - see section IV). The cuprates have - barring some spin glass behavior (perhaps disorder induced, Andersen et al., 2007) – no coexistent long range magnetic order and superconductivity as do at least the 122, 11 $FeSe_{1-x}Te_x$, the 122* and perhaps (Sefat et al., 2010) the 21311. The cuprates exhibit a rapid decrease in $T_c$ upon doping in the CuO planes vs the relative insensitivity of the FePn/Ch layer superconductivity to doping. Thus, doping and its effect on $T_c$, $T_S$ and $T_{SDW}$ is an important tool for understanding the pairing mechanism in the FePn/Ch. A comparison between the cuprates and the FePn/Ch that is highlighted by the recent discovery of superconductivity in the defect-driven 122* structure $A_{0.8}Fe_{1.6}Se_2$ compounds is that, with the exception of the 122*'s, the FePn/Ch do not appear to have an insulating phase anywhere nearby in the phase diagram to the superconducting compositions, while the cuprates do. Lastly, it is well to remember that the FePn/Ch superconductors mechanically are metals, without the brittleness of the ceramic cuprates, making applications more tractable. The cuprates are in daily application (e. g. the SuperLink® filters on cell phone towers) and researchers are actively investigating application (see section Vb) of the FePn/Ch materials. For reviews of the high $T_c$ cuprates, see M. A. Kastner et al. (1998), Basov and Timusk (2005), Lee, Nagaosa and



Wen (2006), Barzykin and Pines (2009), and Armitage, Fournier, and Greene (2010); for an early comparison of the cuprates with the FePn/Ch, see Sawatsky et al. (2009) and Mazin and Johannes (2009).

An important guiding organizational principle throughout this review is that, despite a great diversity of behavior, the new iron superconductors have a number of properties in common. These common properties presumably hold the clue to understanding the relatively high temperature of the superconductivity. It is naturally hoped that achieving this understanding will help lead to discovery of even higher $T_c$'s. A representative list of these common properties (together with the exceptions) would include:

1. All six families of iron-containing superconductors have 2 dimensional planes of FePn/Ch tetrahedra, and the angle of the bonds in the tetrahedra as well as the height of the Pn/Ch above the Fe are indicators of $T_c$.

2. The Fe 3d electrons are – in contrast to the earlier superconductors containing Fe – at the Fermi energy, and clearly taking part in the superconductivity.

3. In most FePh/Ch, the Fe 3d electrons are magnetic in some part of the phase diagram either close to or even coexistent with superconductivity. Although there are examples of FePn/Ch superconductors without magnetism in their phase diagrams, e. g. LiFeAs, FeSe, and – based on the limited data to date - the 21311's (but see the calculation of the susceptibility of $Sr_2VO_3FeAs$ by Mazin, 2010 and data from Sefat et al., 2010), it is arguably the case that the



superconducting properties of this new class of superconductor are fundamentally influenced by the Fe and its magnetic fluctuations.

4. Both hole and electron doping of the non-superconducting 1111 and 122 parent compounds cause superconductivity, with electron-doping causing in general the higher $T_c$'s in the 1111's while hole-doping causes higher $T_c$'s in the 122's.

5. For the undoped 1111 and the 122 compounds, there are both a spin density wave transition and a structural phase transition, $T_S$, (tetragonal to orthorhombic upon cooling). There is neither an SDW nor a structural transition in the Li 111 material but both occur in the Na 111, while superconducting FeSe displays a structural transition (tetragonal – orthorhombic) at 90 K (McQueen et al., 2009b) but no magnetic transition.  $Fe_{1+y}Se_xTe_{1-x}$, which is superconducting for $x \geq 0.05$, has both a structural – tetragonal to monoclinic - and a coincident magnetic transition (at 72 K for x=0) (Fruchart et al., 1975, Martinelli et al., 2010.)  The spin density wave (antiferromagnetic) transition in the 1111 and the 122 has a two sublattice structure with parallel "stripes" of parallel moments running along the orthorhombic b-axis, vs a double stripe arrangement in FeTe.  These parallel moments are aligned perpendicularly to the stripes with each successive stripe's moments opposite to those in the previous one, giving an antiferromagnetic moment in the a-axis direction perpendicular to the stripes (Kitagawa et al., 2008).  In the 122* there is a defect ordering temperature which changes the structure from one tetragonal symmetry to another a few tens of Kelvin above the antiferromagnetic transition which, unlike the other FePn/Ch structures, has the moment along the c-axis.



6. The two transitions are at different temperatures in the undoped 1111's (e. g. $T_S$=155 K vs $T_{SDW}$=140 K in CeFeAsO - Zhao et al., 2008a - although this difference is shrinking with better sample quality – Jesche et al., 2010), but coincide in temperature in the undoped 122's (see section II and Table 1). $T_S/T_{SDW}$ values for the $MFe_2As_2$ are similar to those in the 1111's and range from 140 to 205 K.  This coincidence of the structural and magnetic transitions in the 122's disappears with doping on the Fe and As sites, although the case of isoelectronic Ru doping of the Fe in $BaFe_2As_2$ is under debate (Thaler et al., 2010 and Rullier-Albenque et al., 2010).

7. Inelastic neutron scattering (INS) has found (similar to results in the cuprates) a spin fluctuation resonance in the 1111, 122, and 11 structure superconductors below $T_c$.  These experiments may provide evidence (that is still undergoing refinement) for a causal link between the spin fluctuations (which are directionally in the Fermi surface pocket nesting direction) and the pairing that opens the superconducting gap.

8. Measurement of angular resolved photoemission spectroscopy (ARPES) of the FePn/Ch finds a Fermiology consisting typically of five separate pockets, with varying degrees of interpocket nesting ranging from very strong in the undoped 122 parent compounds to totally absent in overdoped (but still superconducting) $BaFe_{2-x}Co_xAs_2$ and LiFeAs.  The importance of the five Fe 3d bands at the Fermi energy in these materials is well established, with good agreement between measurement and calculation.



These common factors (with the exception of the five-fold Fermiology) have their analogs in the well-studied high $T_c$ cuprates. All the cuprate derivative structures have Cu-O planes in common, the Cu electrons are involved in the superconductivity, there is magnetism in the undoped, non-superconducting compound phase diagrams, both hole and electron doping cause superconductivity with hole doping being more effective in raising $T_c$, and pressure is known, as already mentioned, to have a large effect on $T_c$.

There are however important differences between the new iron superconductors and the cuprates, as have already been discussed as one of the main points of interest for studying the FePn/Ch. In the final analysis, although analogy with the huge body of knowledge collected on the cuprates can be of help in choosing which investigations might yield essential insights, the FePn/Ch appear to be – in much of their fundamental behavior – categorically *different* from the cuprates.

A strict effort has been made to make this review an organized whole, to provide easy navigation to topics of interest for the non-specialist reader interested in understanding FePn/Ch superconductivity. Each of the succeeding main topics sections II-V begins with an introduction *and* summary, as do most of the major subsections. The organization at the level of the presentation of detailed results is based on the six FePn/Ch structures, generally in the order of discovery (1111 . . . 122*) presented above. There are numerous references to specialized reviews for further in-depth reading on selected topics. Several compendia of papers on the field of FePn/Ch superconductors exist, including Superconductor Science and Technology **23**, May 2010 (focus on electromagnetic properties), Physica C **469**, 313-674 (2009), Physica C **470** Supplement

1, S263-S520 (2010), New Journal of Physics **11**, February 2009, and J. Phys. Soc. Japan **77**, Supplement C, 1-159 (2008)**.**  The Journal of the Physical Society of Japan has a banner "Iron-Pnictide and Related Superconductors" on their home web page that links to an detailed index with links to 32 separate subject areas organizing all of the articles in the journal on this subject.  Early reviews by Norman (2008) and Ishida, Nakai and Hosono (2009) give a good overview of the beginning work and understanding thereof in this field.  More recent reviews include those by Lumsden and Christianson, 2010 (magnetic properties), Mizuguchi and Takano, 2010 (the iron chalcogenides), Mandrus et al., 2010 ($BaFe_2As_2$ and dopings thereof), Paglione and Greene, 2010 (overview), Johnston, 2010 (comprehensive overview, emphasis on normal state properties) and Korshunov, Hirschfeld and Mazin, 2011 (theory).  Lastly, in the modern multi-media age there is a video of a slide presentation on this subject at the March, 2010 APS meeting by Norman viewable at:  http://physics.aps.org/videos/2010-norman-iron-age_superconductors.



## II. Structural and Electronic Properties, Part One – $T_c$ and its Dependencies/Correlations

As discussed in the Introduction, all of the iron pnictide and chalcogenide superconductors have structural and physical properties in common. The present section focuses on the superconductivity, its connection with the structural and magnetic phase transitions (phase diagrams), the important question of coexistence of magnetism and superconductivity, and the influences of pressure and magnetic field on $T_c$. First, the structure (section A) of these materials is presented. The structure is crucial in any attempt to understand the superconductivity, particularly since there are aspects of the structure in the FePn/Ch which influence $T_c$ where similarities and correlations have been found. Then, the large body of data about the phase diagrams of these compounds (section B) is presented, with graphs of $T_c$, the structural phase transition temperature $T_S$, and $T_{SDW}$ as a function of doping. There appear to be two distinct kind of phase diagrams vis-à-vis whether the magnetism is suppressed by doping before superconductivity is induced. Further, in the 'coexistent' kind of phase diagram, there are again two distinct types. These are distinguished by whether the magnetic transition temperature, $T_{SDW}$, ever sinks down to $T_c$ at a given composition or whether $T_{SDW}$ remains larger than $T_c$.

Section C considers the important topic of microscopic vs phase-separated coexistence of the magnetism and the superconductivity after the experimental evidence for coexistence in section B is established. Coexistence is obviously of interest for understanding the pairing mechanism. Finally, sections D and E discuss the pressure and field dependence of $T_c$, and the insights therefrom for understanding the superconductivity.



## A. Structure/T_c vs lattice spacing

The original discovery of superconductivity at 26 K by Kamihara, et al. (2008) was in $LaFeO_{1-x}F_x$, which has the tetragonal, tP8 ('t' means tetragonal, 'P' means 'primitive' or no atoms in either the body or face centers, 8 atoms per unit cell) ZrCuSiAs (=prototypical compound) structure with 2D layers of FeAs shown in Fig. 1.

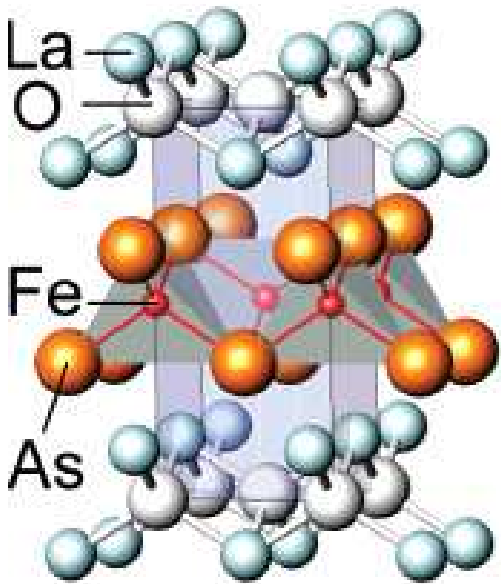
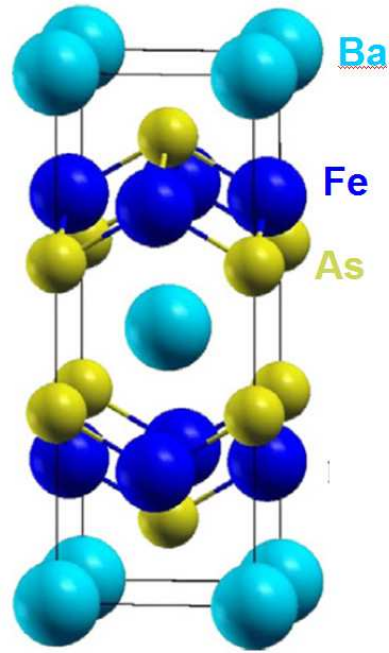

Fig. 1 (color online)   From Kamihara et al. (2008), the lattice structure of 1111 LaFeAsO.

Fig. 2 (color online) Shein and Ivanovskii (2009a), lattice structure of 122 $BaFe_2As_2$.

The second FePn/Ch structure discovered to be superconducting, also tetragonal with 2D FeAs planes, was K-doped $BaFe_2As_2$, with the tetragonal tI10 ('I' means there is an atom at the center of the 10 atom unit cell, see Fig. 2) $ThCr_2Si_2$ structure (Fig. 2) and $T_c$=38 K (Rotter, Tegel and Johrendt, 2008). This is a well known and well studied structure in materials superconductivity and is the same structure as the first discovered heavy Fermion superconductor, $CeCu_2Si_2$ (Steglich et al., 1979.) The third and fourth FePn/Ch



superconducting structures to be discovered, Figs. 3 and 4, also both with 2D planes (FeAs and FeSe respectively), were the MFeAs, '111', (X. C. Wang et al., 2008, M=Li, $T_c$=18 K) with the tetragonal tP6 $Cu_2Sb$ structure and the iron chalcogenide FeSe ('11') family (Hsu et al., 2008, $T_c$=8 K) with the tetragonal tP4 PbO structure. The fifth structure with FePn planes to join this superconducting set of materials is the so-called 21311 (sometimes called the 42622) structure. The first member found, $Sr_2ScO_3FeP$ (Ogino et al., 2009 in (pictured in Fig. 5) had a 17 K $T_c$. Replacement of Sc

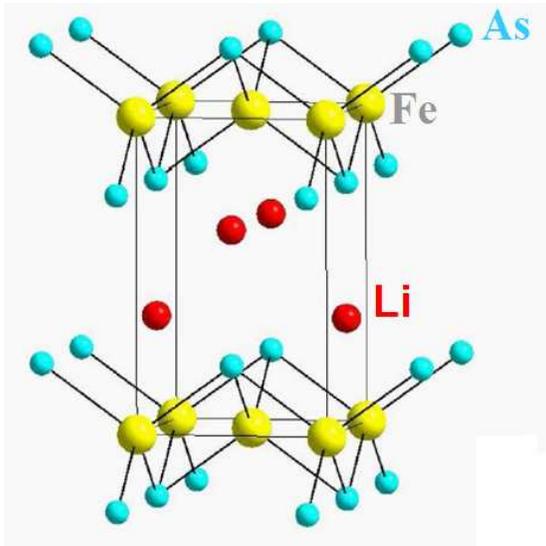

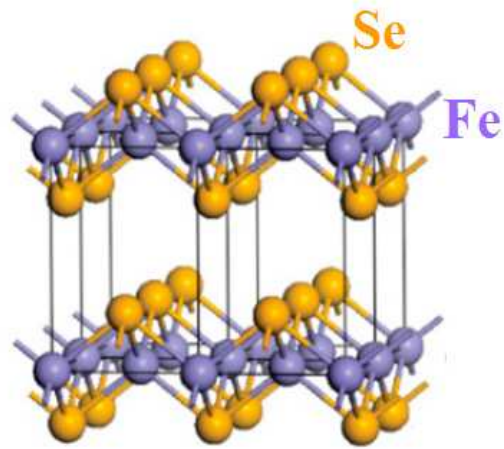

Fig. 3 (color online) Deng et al. (2009), structure of 111 LiFeAs.

Fig. 4 (color online) Hsu et al. (2008), structure of FeSe.



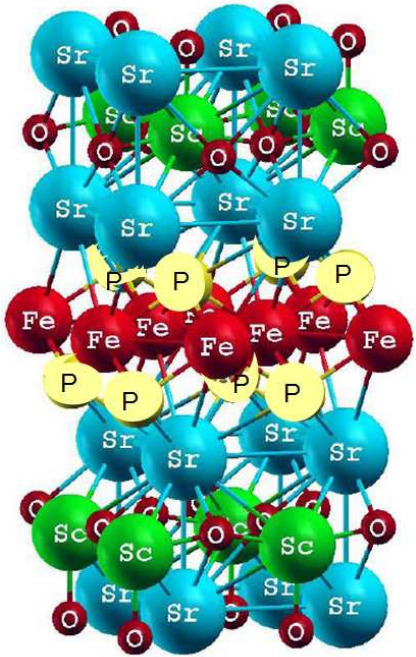

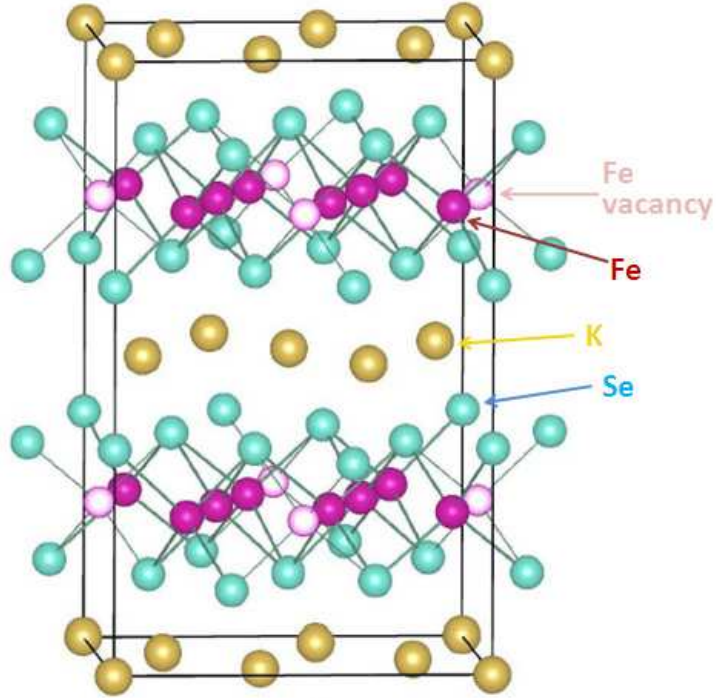

Fig.5(color online) Shein and Ivanovskii(2009b),$Sr_2ScO_3FeP$

Fig. 6 (color online) $K_{0.8}Fe_{1.6}Se_2$ Bao et al. (2011b). The solid maroon Fe atoms, on the Fe2 site (16/unit cell), are all interior to the unit cell (marked with black lines) while the pink open circle Fe vacancies are on the Fe1 site (4/unit cell) and are all on faces, i. e. shared with neighboring unit cells. Note the enlarged unit cell with respect to the 122 structure in Fig. 2.

with Cr or V, and P with As, has increased the $T_c$ up to 37 K in $Sr_2VO_3FeAs$ (Zhu et al., 2009b), while $Sr_2Mg_{0.2}Ti_{0.8}O_3FeAs$ has $T_c$=39 K (Sato et al., 2010). The structure in Fig. 5 can be visualized as layers of 122 $SrFe_2P_2$ alternating with perovskite $Sr_3Sc_2O_6$ layers. Intercalation of further layers of atoms between the FeAs layers to try to increase $T_c$ by expanding the c-axis has so far (Ogino et al., 2010c, discussed in section IIB3a) resulted in $T_c$'s up to 47 K. The most recent FePn/Ch structure discovered (Fig. 6) with superconductivity ($T_c \approx$32 K) is an ordered-defect alteration of the 122 $BaFe_2As_2$ structure (called the '122*' structure herein), written $A_{0.8}Fe_{1.6}Se_2$ or sometimes $A_xFe_{2-y}Se_2$ (A=K, Rb, Cs, Tl), where the ordered arrangement of Fe vacancies below $T_S$ on the inequivalent



Fe sites (in the ideal case Fe2 sites are fully occupied, Fe1 sites are fully unoccupied) has important influence (Bao et al., 2011b; Ye et al., 2011) on the measured properties, including superconductivity. Zavalij et al. (2011) state that below the ordering temperature $T_S$ the Fe1 site may have nonzero (3.2-7.8 %) occupation, although they speculate that this could be due to small, fully Fe1 occupied domains. Another way to interpret this structure is as FeSe intercalated with K, Rb, Cs, Tl or combinations thereof. The unit cell for the tetragonal 122* ordered defect structure is larger than that for the tetragonal 122 by $\sqrt{5}$ x $\sqrt{5}$ x 1 in the a, b, and c-axis directions respectively, see Bao et al. (2011a, 2011b) for further diagrams.

Thus, all of the discovered FePn/Ch superconductors are tetragonal with planes of tetrahedra of Fe and either As or P (pnictogens) or S, Se or Te (chalcogenides). The rather short (2.67 Å in the 11, 2.77 Å in the 122* - Guo et al., 2010 - up to 2.84 Å in the 21311 and 2.85 Å in the 1111, Ikida, Nakai and Hosono, 2009; Ogino et al., 2009) Fe-Fe spacings insure that the 3d Fe electrons take part in band formation. Various calculations of the electronic structure result in the consensus that these Fe d-bands dominate the rather large density of states near the Fermi energy (see Raghu et al., 2008, for a discussion of the basic features of a band model). Together with nesting on the Fermi surface, these Fe bands can lead to magnetic ordering (Cao, Hirschfeld, and Cheng, 2008; Dong et al., 2008a; Singh, 2009) as discussed below in Section IIB. Four of the six structures have the same space group, P4/nmm, space group number 129. The exceptions are the 122, $MFe_2As_2$ structure - which has I4/mmm (space group number 139) due to the body centered M atom shown in Fig. 2 – and the ordered defect 122* structure, $A_{0.8}Fe_{1.6}Se_2$. The 122* structure has the reduced I4/m symmetry (space group 87) below



the defect ordering transition $T_S$ (vs I4/mmm of the 122 structure at higher temperature) since as may seen from Fig. 6 the ordered defect 122* structure loses the mirror plane symmetries in the x- and y-directions of the 122 structure in Fig. 2 when the Fe1 sites are empty. In this symmetry notation, 'P' and 'I' mean primitive and body centered respectively, just as in the structure notation, '4' means that the structure is identical under fourfold rotation (by 90 °) around the c-axis, 'mmm' means that the structure is identical when mirrored in planes perpendicular to all three of the orthogonal tetragonal axes, and 'nmm' means symmetric about mirror planes perpendicular to the two equal tetragonal axes (a and b) and that for the third, unequal tetragonal axis (c-axis) the symmetry operations that bring the crystal back to itself are called glide plane symmetry, where the n-glide involves reflecting about a mirror plane parallel to the c-axis followed by a translation along 1/2 of the face diagonal. These symmetry operations can be followed in Figs. 1-6. The space groups, numbered from 1-230, are all unique and describe all possible crystal symmetries.

The influence of lattice structure on $T_c$ has been the focus of various authors and is clearly an important issue. The FeAs$_4$ (FeSe$_4$) building blocks common to all the structures form tetrahedra (see Figs. 1-6), that are 'regular' (meaning the four faces are equilateral triangles) if the As-Fe-As bond angle, $\alpha$, is 109.47°. Lee et al. (2008) pointed out that $T_c$ plotted vs $\alpha$ for a wide range of doped 1111 and 122 FePn superconducting samples shows a sharp peak at the regular tetrahedron bond angle, indicating that local symmetry around the Fe and As is decisive for the superconductivity. Putting this dependence of superconductivity on the lattice structure on a theoretical basis, Kuroki et al. (2009) discussed how nesting among pieces of the Fermi surface (see section IVA2



below for a discussion of the experimental determination of the Fermiology), which are determined by the lattice structure, determine not only the size of $T_c$ but also the symmetry of the gap function. Thus, Kuroki et al. point out that the nature of the gap symmetry, nodal vs fully gapped (see section IV for a discussion of the theory and experiments), is controlled by the height of the arsenic (or more generally the pnictogen or chalcogen) above the iron plane. Small pnictogen height favors nodal behavior (LaFePO), vs large pnictogen height which favors more fully gapped behavior (LaFeAsO$_{1-x}$F$_x$).

The correlation in the high $T_c$ cuprates that $T_c$ scales with the CuO interplanar spacing was at least part of the motivation for investigating the 21311 materials, e. g. Sr$_2$ScO$_3$FeP, but the resultant c-axis spacing (15.543 Å vs 8.73 Å for LaFeO$_{1-x}$F$_x$), with the concomitant much larger Fe-Fe interlayer spacing, and relatively low (17 K) $T_c$ indicates that other factors are also playing a role. For a discussion of the lattice parameters for the first four FePn/Ch structures, see the review by Ishida, Nakai, and Hosono (2009); for the 21311, see Ogino et al., 2010b; for the defect 122* structure see Zavalij et al. (2011) and Bao et al. (2011b).

Within a given structure, various correlations between lattice spacing and $T_c$ have been noted. Shirage et al. (2008) noted in electron doped, oxygen deficient LnFeAsO$_{1-x}$ and La$_{1-y}$Y$_y$FeAsO$_{1-x}$ that $T_c$ scales with the a-axis spacing (see Fig. 7). In terms of hole doping of the 1111's, this is somewhat of an open question as there have been conflicting reports since annealing of hole doped samples to optimize the superconductivity can also lead to oxygen deficiency (equivalent to electron doping). Specifically, Wen et al. (2008) measure $T_c$ as a function of doping in hole doped La$_{1-x}$Sr$_x$FeAsO and find that $T_c$

remains unusually constant (within 10%) at ~ 25 K as a function of x between 0.1 and 0.2.

G. Wu et al. (2008b) argue that Sr-doping of LaFeAsO does not cause bulk

superconductivity, that only annealing which then produces an oxygen deficiency results

in bulk behavior.

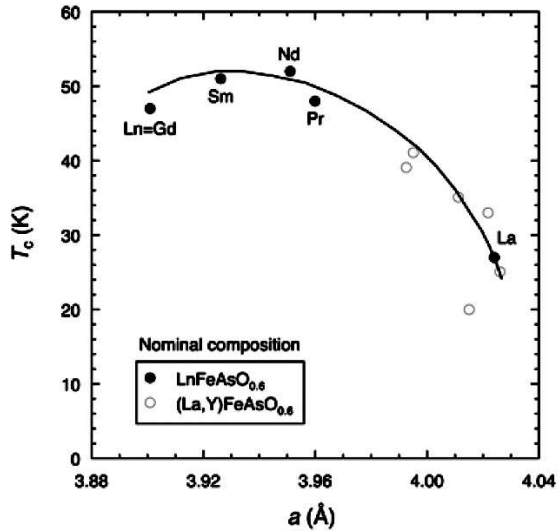

Fig. 7 Eisaki et al. (2008), $T_c$ vs a-axis spacing in LnFeAsO$_{0.6}$ and (La,Y)FeAsO$_{0.6}$. See also Miyazawa et al. (2009) for a follow up work that includes Dy and Tb, a=3.86 and 3.875 Å respectively. Note the open circles corresponding to Y replacing La.

In the 122's, with decreasing transition metal electron doping (Canfield and

Bud'ko, 2010) on the Fe site in BaFe$_2$As$_2$ (e. g. Co in Ba(Fe$_{1-x}$Co$_x$)$_2$As$_2$, 0≤x≤0.11) , the

c-axis increases (just as it does in electron doped LaFeAsO) while the ratio a/c falls

(corresponding to an almost constant a-axis spacing), both monotonically as $T_c$ falls with

decreasing concentration (see Fig. 12 below in section B for $T_c$ vs x in Ba(Fe$_{1-x}$Co$_x$)$_2$As$_2$.)

With increasing doping (Rotter, Tegel and Johrendt, 2008) of BaFe$_2$As$_2$ with K on the Ba

site, the a-axis shrinks while the c-axis expands. Thus, the behavior of the c-axis

(proportional to the interlayer spacing) as doping concentration is varied from large

electron (Co, x≈0.11) doping through x=0.00 and further to increasing hole (K) doping is



monotonically increasing, while $T_c$ is approximately 'V-shaped', i. e. has a minimum at x=0.00 and rises approximately linearly with either electron or hole doping.

## B.  $T_c$, $T_S$ and $T_{SDW}$ vs doping/Phase diagrams

**Introduction:**  After the 2D layers common to the FePn/Ch superconductors, the search for other commonalities to explain the superconductivity focused on the magnetic behavior of the various systems as they were discovered.  Using neutron scattering, de la Cruz et al. (2008) report for the undoped parent compound LaFeAsO spin density wave (SDW) antiferromagnetism at $T_{SDW}$ = 137 K with a low temperature moment of 0.36 $\mu_B$/Fe atom, preceded by a structural distortion from tetragonal to orthorhombic (originally indexed as monoclinic, but corrected by Nomura et al., 2008) at $T_S$ = 155 K.  Both these transitions are suppressed in the discovery compound of Kamihara et al. (2008), LaFeAsO$_{0.92}$F$_{0.08}$, $T_c$=26 K.  $T_{SDW}$ and $T_S$ are depressed by intermediate doping at approximately the same rate, so that $T_S$ remains greater than $T_{SDW}$, discussed below.   LaFePO, which is superconducting at 5-6 K in the undoped state, is not magnetic – Carlo et al., 2009.  As a note of historical interest, the discovery of Kamihara et al. (2008) of superconductivity at 26 K in F-doped LaFeAsO was foreshadowed by the discovery of superconductivity around 5 K in LaFePO (Kamihara et al. 2006) by more than just LaFePO having the same 1111 structure.  The work in 2006, although this is little commented upon, reported that $T_c$ increased up to ≈10 K with 6% F-doping on the O-site in LaFePO.

Undoped BaFe$_2$As$_2$ was reported (Rotter et al., 2008b) to have an SDW transition at 140 K, as well as a tetragonal-orthorhombic structural distortion at the same temperature.  Later neutron scattering work (Huang et al, 2008) determined the low



temperature moment to be 0.87 $\mu_B$/Fe atom.  Both this measured local moment and that

for LaFeAsO (0.36 $\mu_B$/Fe atom) are significantly smaller than those calculated by density

functional theory (DFT) band structure calculations (Mazin and Johannes, 2009).   Since

DFT calculations do not properly include electronic correlations (see Yin, Haule and

Kotliar, 2011 for a comparison of DFT with DFT+DMFT Fermi surface calculations),

this difference in the determined magnetic moment implies that such correlations may be

important in the FePh/Ch.  In the discovery work, upon doping with K, $Ba_{0.6}K_{0.4}Fe_2As_2$

became superconducting at 38 K with no structural transition down to at least 20 K

(Rotter, Tegel and Johrendt 2008).  Later work, discussed below in the subsection (IIB2)

on the 122 structure, delineated the decrease in $T_{SDW}$ and the structural transition

temperature, $T_S$, with doping on all three of the sites in $MFe_2As_2$. This later work found a

clear consensus that there is a separation, with $T_S > T_{SDW}$, upon doping either the Fe (with

the possible exception of Ru-doping) or the As sites, but with some disagreement

regarding doping on the M site.  Thus, upon doping the 122's on either the Fe or the

Pn/Ch site, they are clearly comparable to the 1111 compounds in the separation of $T_S$

and $T_{SDW}$, while there is only limited evidence in the 122's for the splitting of $T_S$ and

$T_{SDW}$ for doping on the M site.

The next 2D layered FePn superconductor discovered, LiFeAs, shows bulk

superconductivity at $T_c$=18 K but has neither a magnetic nor a structural transition,

although there are very strong magnetic fluctuations (Jeglic et al., 2010).  The other

known superconducting 111 material, $Na_{1-\delta}FeAs$, shows a broad ($\Delta T_c$ up to 15 K)

resistive transition at $T_c$=23 K, and shows two transitions above $T_c$ (G. F. Chen et al.,

2009).  The lower temperature transition had been earlier identified as a magnetic



transition (~ 40 K, μSR data from Parker et al., 2009), with an estimate of the local moment of 0.1-0.2 $\mu_B$. A follow up work determined a local Fe moment of $0.09 \pm 0.04$ $\mu_B$ (elastic neutron scattering data from S. Li et al., 2009a) and a tetragonal to orthorhombic structural transition (at ~ 50 K, S. Li et al., 2009a). This low value of the local ordered moment is the lowest in the magnetically ordered parent FePn/Ch compounds. Whether $Na_{1-\delta}FeAs$ is a bulk superconductor and the role of Na defects will be discussed below in section IIB3. As will be discussed in several sections, LiFeAs is different from the other FePn/Ch superconductors in numerous ways, not just in its lack of structural or magnetic transition in comparison to $Na_{1-\delta}FeAs$. The small Li ionic radius compared to that of Na (1.55 vs 1.90 Å) is presumably part of the reason – LiFeAs is already "pre-compressed" (see Section IID on $T_c$ as a function of pressure). The LiFeAs tetrahedral As-Fe-As bond angle, $\alpha$, is 113.7$^o$ (Pitcher et al., 2008), far from the regular tetrahedron value of 109.47$^o$ where Lee et al. (2008) pointed to a maximum in the $T_c$'s of the 1111's.

The '11' structure $FeSe_{1-x}$, $T_c=8$ K, shows a structural transition (just like the 1111 and 122 structures, tetragonal to orthorhombic) at 90 K (McQueen et al., 2009b) with no magnetic transition (confirmed in McQueen et al., 2009a who prefer '$Fe_{1+\delta}Se$') while $FeSe_xTe_{1-x}$, $T_c=15$ K, has both a structural (tetragonal to monoclinic) and magnetic transition (both at 72 K for x=0) (Fruchart et al., 1975, see also R. Viennois et al., 2010.) The low temperature magnetic moment of non-superconducting $Fe_{1.068}Te$ is 2.25 $\mu_B$/Fe atom (S. Li et al., 2009b). The physical properties of $Fe_{1+x}Te$ depend on the amount of excess Fe, with the low temperature structure becoming orthorhombic rather than



monoclinic below $T_S \sim 63$ K and the magnetic ordering becoming incommensurate for x=0.141 (Bao et al., 2009).

The 21311 structure, represented by $Sr_2VO_3FeAs$, $T_c = 37$ K, apparently does not have a structural transition but does show a transition (that is preparation dependent) consistent with magnetism at $\sim 155$ K with a moment less than $\sim 0.1$ $\mu_B$ (Sefat et al., 2010; Cao et al., 2010; Tegel et al., 2010).

The ordered defect 122* $K_{0.8}Fe_{1.6}Se_2$ structure, $T_c \approx 32$ K, has (Bao et al., 2011b) an Fe-sublattice order-disorder transition at $T_S \approx 578$ K, followed by antiferromagnetic order at $T_N \approx 559$ K with a low temperature ordered local moment of 3.31 $\mu_B$ per Fe atom. Both the high magnetic ordering temperature and the size of the local moment are records for the FePn/Ch superconductors.  Liu et al. (2011), using resistivity, $\rho$, and magnetic susceptibility, $\chi$, report $T_S$ and $T_N$ for all of the superconducting $A_{0.8}Fe_{2-y}Se_2$, A=K, Cs, Rb, (Tl,K), and (Tl,Rb), and found $T_N$ values between 540 K (A=K) and 496 K (A=(Tl,K)). As a comparison, in insulating $TlFe_{1.6}Se_2$, Sales et al. (2011), using inelastic neutron scattering, found $T_N$=430 K with the Fe sublattices slightly disordered (90% of the Fe2 sublattice and 30% of the Fe1 sublattice were occupied) below $T_S \approx T_N$.  Sales et al. found that the ordered moment in the insulating compound peaks at 2.1 $\mu_B$ – significantly smaller than Bao et al.'s (2011b) result of 3.31 $\mu_B$ for the superconducting ordered 122* structure - at 140 K but then decreases to 1.3 $\mu_B$ at low temperatures after two (still under investigation) phase transitions at 140 and 100 K.

Unlike the 1111, the 122, and the 11 structures, the low temperature crystal structure of the superconducting ordered defect 122* structure $A_{0.8}Fe_{1.6}Se_2$ remains



tetragonal, although with a lower symmetry (see Fig. 6) than the high temperature structure (I4/m vs I4/mmm respectively) due to the Fe sublattice ordering.

It is interesting to note that, although both calculations (Subedi et al., 2008) and ARPES measurements (Xia et al. 2009) of the Fermi surfaces of the undoped 11 compounds indicate nesting similar to that of the undoped 1111 and 122 materials (see section IVB2 below for a discussion of the ARPES data), the ordered wavevector in the 11's is different as shown in Fig. 8. (In the 122*, the ordered moment is – instead of being in the ab plane – along the c-axis, Bao et al., 2011a.)

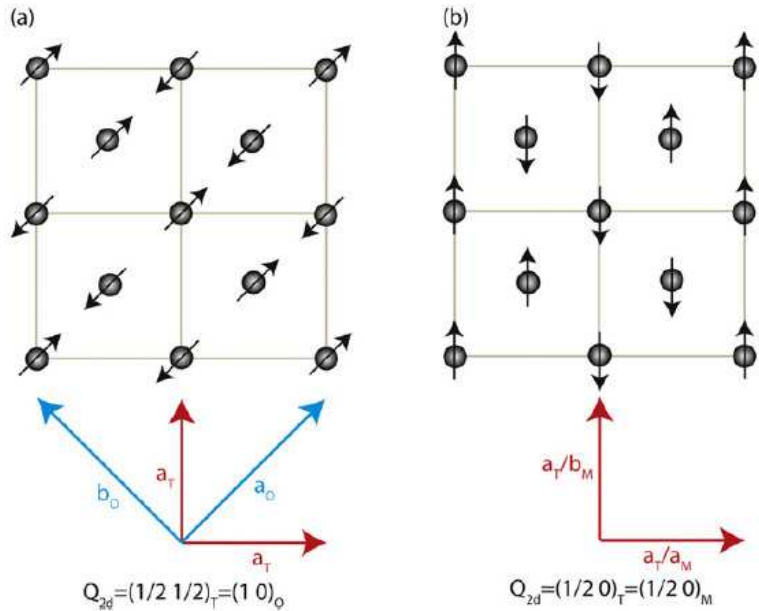

Fig. 8 (color online). In plane magnetic spin arrangement for undoped 1111 and 122 materials, part (a) and for 11 materials, part (b). The colored vectors denote the tetragonal ('T'), orthorhombic ('O') and monoclinc ('M') structures. From Lumsden and Christianson, 2010. Note that some authors use tetragonal notation for the ordering wave vector ((½, ½) while others use orthorhombic (1,0).

Johannes and Mazin, 2009, using LAPW calculations, calculate the stabilization energies for various magnetic configurations in the undoped 11 and 122 structures and find that the observed $(1/2,1/2)_T$ wavevector in the 122's is energetically favored while it



is energetically approximately the same as the $(1/2,0)_T$ wavevector observed in the 11's. Thus, the authors argue that, based on the calculated *and* observed difference in ordered wavevectors for the 11's vs the 122 compounds despite the similar nesting, the magnetic ordering is not driven by the nesting in the 122's (and, by extension, in the 1111's). However, this logic can be inverted – since according to ARPES measurements there is no Fermi surface nesting in LiFeAs (Borisenko et al., 2010) which is non-magnetic, *ergo* one could argue that nesting is important for the magnetic ordering. Hsieh et al. (2008), based on ARPES measurements in $SrFe_2As_2$, also argue that nesting is important for the magnetic order. Johannes and Mazin (2009) conclude that instead of superexchange between neighboring spins, the magnetic wavevector is due to a combination of local moments and long range itinerant interactions.

Based on the above short discussion of local vs itinerant for the magnetic order in the FePn/Ch, it is apparent that - as discussed more thoroughly in the review of magnetism in Fe-based superconductors by Lumsden and Christianson (2010) – this is still a topic of "considerable debate". There are a number of experimental and theoretical works on both sides of this question. For the experimental side, one of the main experimental probes is of course neutron scattering. See, e. g., neutron studies on $CaFe_2As_2$ by McQueeney et al. (2008) and Zhao et al. (2009) for conflicting points of view on the itinerancy of the magnetism, as well as the discussion in the review by Lumsden and Christianson (2010)). However, there are also results from other measurement techniques, see, e. g., angular resolved photoemission spectroscopy (ARPES) work in $(Ba,Sr)Fe_2As_2$ of Yi et al. (2009) and optical spectroscopy work on 122 parent compounds by Hu et al. (2008). For discussion of the theory on both sides of this



question see, e. g., in addition to Johannes and Mazin (2009) discussions by Goswami et al. (2010), M. J. Han et al. (2009), and Knolle et al. (2010).

There is an interesting theoretical argument by Fernandes and Schmalian, based on the reentrant (magnetic→paramagnetic) behavior in the phase diagram of Ba(Fe$_{1-x}$Co$_x$)$_2$As$_2$ discussed below in section IIB2b (see also Fig. 12), that at least in some systems the magnetic order must be partially itinerant. There are also theoretical (Lee, Yin and Ku, 2009; Lv, Wu and Phillips, 2009; C.-C. Chen et al., 2010; Kontani, Saito and Onari, 2011) and experimental (Akrap et al., 2009; Shimojima et al., 2010; Dusza et al., 2010) works which propose that the observed magnetic ordering *and* the structural phase transition are related to the orbital structure of the FePn/Ch (see also the discussions below in Sections IIB2b and IIIA).

Moon et al. (2010), in a combined optical spectroscopy and density functional calculation work, as well as Lumsden and Christianson (2010), argue in agreement with Johannes and Mazin for the best description being a combination of localized and itinerant magnetism. This is certainly in agreement with the thermodynamically determined entropy of ordering, $\Delta S$, at $T_{SDW}$ which, in the systems where high temperature specific heat data exist, is relatively small compared to that expected for full local moment ordering (5.76 J/moleK or Rln2 of entropy for a spin 1/2 local moment.) On the other hand, for a fully itinerant magnetic moment, there would be essentially no entropy of ordering at the transition temperature as is observed, e. g., in the itinerant ferromagnet ZrZn$_2$, where $\Delta S \sim 0.02$ J/moleK (Yelland et al., 2005). Values for $\Delta S$ at $T_{SDW}$ for BaFe$_2$As$_2$ (Ba$_{0.8}$K$_{0.2}$Fe$_2$As$_2$), SrFe$_2$As$_2$, EuFe$_2$As$_2$ and Fe$_{1.1}$Te (obtained by analyzing the published specific heat heat data) are respectively 0.85 (0.18) J/moleK



(Kant et al., 2010), $\approx$1 J/moleK (Krellner et al., 2008), 1.5 J/moleK (Jeevan et al., 2008a) and 2.4 J/moleK (Westrum, Chou and Gronvold, 1959). Further, this measured entropy of the magnetic moment ordering is intertwined with the entropy of structural ordering at the coincident $T_S$ and thus is even smaller. It is interesting to note, however, that the neutron-scattering-determined local moments for these compounds (see Lumsden and Christianson, 2010) approximately scale with $\Delta S$, since the moments for $BaFe_2As_2$ and $SrFe_2As_2$ are $\approx$1 $\mu_B$ while measured values for $Fe_{1.1}Te$ range between 1.96 and 2.25 $\mu_B$.

Leaving now the discussion of local vs itinerant magnetic order, some aspects of the magnetic ordering and the spin excitations in the FePn/Ch, particularly in the 122's where larger single crystal arrays are available (see the discussion below in section VC), have in contrast been decided. The magnetic interactions determined by inelastic neutron scattering (INS), in contrast to the 2D interactions in the cuprates (Kastner et al., 1998), are 3D in nature, with some anisotropy. For example, the ratio of the spin wave velocity perpendicular to the plane ($v_\perp$) to that in the plane ($v_\parallel$) is (McQueeney et al. 2008) at least half in $CaFe_2As_2$, with similar values in $BaFe_2As_2$ ($v_\perp / v_\parallel \sim 0.2$, Matan et al., 2009), $SrFe_2As_2$ ($v_\perp / v_\parallel \sim 0.5$, Zhao et al., 2008d), and underdoped (before the ordering is suppressed) $BaFe_{1.92}Co_{0.08}As_2$ ($v_\perp / v_\parallel \sim 0.2$, Christianson et al., 2009), vs $v_\perp / v_\parallel = 1$ for isotropic 3D and $v_\perp / v_\parallel = 0$ for purely 2D excitations. After the long range magnetic order in $BaFe_2As_2$ is suppressed with sufficient Co-doping (optimally and overdoped samples), there is a significant decrease in c-axis spin correlations, moving toward more 2D behavior (Lumsden et al., 2009). Whether the more 2D nature of the fluctuations at the highest $T_c$ (optimally doped) part of the phase diagram is a significant consideration for understanding the superconductivity is at this point speculative. In $CaFe_2As_2$,



measurements of INS to shorter wavelengths out to the zone boundary (Zhao et al., 2009) have been able to determine the signs of the exchange coupling constants $J_{1a}$ and $J_{1b}$ in the plane, with the result that the former is antiferromagnetic and the latter is ferromagnetic. For a discussion of these data and the question of local vs itinerant magnetism and of the question of magnetic frustration, see Schmidt, Siahatgar, and Thalmeier (2010). Several theoretical works (Ma, Lu and Xiang, 2008, Si and Abrahams, 2008, Yildirim, 2008) in the 1111 materials argue for the importance of frustration.

After this Introduction, we now discuss the composition dependence of $T_c$ - and $T_S$ and $T_{SDW}$ where they exist - for the FePn/Ch superconductors structure by structure (as each section in this review is organized) where doping has been used to vary the superconductivity. The response of $T_c$, $T_S$ and $T_{SDW}$ to doping has been the subject of intense study in the search for understanding the basic mechanism of the superconductivity, and thus there is a mass of data to summarize below – much of it still waiting for unifying insight. For an example where this effort has made notable progress, see e. g. the discussion of $Ba(Fe_{1-x}Co_x)_2As_2$ in Section IIB2b.

## 1.) 1111 Structure

The samples discussed in this section were all prepared in polycrystalline form unless otherwise stated. With the exception of $SmFeAsO_{1-x}F_x$, which is still under debate as discussed below, both $T_S$ and $T_{SDW}$ are suppressed by doping in 1111's before superconductivity appears. There are only a few examples of hole-doping-caused superconductivity in the 1111's, primarily in $Ln_{1-x}Sr_xFeAsO$, with G. Wu et al. (2008b) arguing for oxygen deficiency and thus effective electron doping in the Ln=La case. There is one example of "isoelectronic-doped," $CeFeAs_{1-x}P_xO$, where $T_c$ remains zero



(Luo et al., 2010; de la Cruz et al., 2010) for $0 \leq x \leq 1$ unlike P-doping on the As site in

$BaFe_2As_2$ discussed in Section IIB2 below. Otherwise, the doping in 1111's has been

electron doping, with $T_c$'s found above 50 K.

This section on the $T_c$ vs doping (subsection a.) and on the correlations between

$T_c$, $T_S$, and $T_{SDW}$ (subsection b.) in the 1111's attempts to present a thorough review of all

the data so that the reader can gain an overview. Table 1 and Figs. 9 and 10 below are

aids in this goal. Unfortunately, due to difficulty of preparation and sample quality

questions, the 1111's present a much less cohesive picture than the 122's in section IIB2

following.

**a.) $T_c$ vs doping:** Electron doping LnFeAsO (Ln=La, Dy, Tb, Gd, Sm, Nd, Pr,

Ce), via either the discovery method (F partially replacing O) of Kamihara et al. (2008)

where superconductivity starts at 4% F doping or via oxygen deficiency achieved with

high pressure synthesis, was the first focus of study in 1111 FePn superconductivity.

The choice of smaller Lanthanide elements (see Fig. 7) to increase $T_c$, as discussed above

in the Introduction, was inspired by the increase in $T_c$ of $LaFeAsO_{1-x}F_x$, x=0.11, from 26

to 43 K under pressure observed by Takahashi et al. (2008a). Eisaki et al. (2008) showed

early (Fig. 7) that $T_c$ in $LnFeAsO_{1-y}$ was not actually a function of the electronic nature of

the lanthanide element, but rather of the a-axis lattice spacing since they could achieve

the same $T_c$ progression by simply doping the smaller Y for La in $LaFeAsO_{1-y}$. Peak $T_c$'s

found for oxygen deficiency were in $NdFeAsO_{0.85}$, $T_c$=53.5 K and in $SmFeAsO_{0.85}$,

$T_c$=55 K (Ren et al., 2008a, using high pressure synthesis) and for the fluorine doped

$SmFeAsO_{0.9}F_{0.1}$, $T_c$=55 K (Ren et al., 2008b).



Interestingly, Zhu et al. (2009a) found $T_c^{onset} \sim 32$ K in $Sr_{0.6}La_{0.4}FeAsF$ (La provides electron doping of SrFeAsF, which has a positive Hall coefficient, Han et al., 2008). Further, G. Wu et al. (2009) found $T_c$ in $Sr_{0.5}Sm_{0.5}FeAsF$ at ~56 K, and Cheng et al. (2009) find the same 56 K $T_c$ in $Ca_{0.4}Nd_{0.6}FeAsF$, i. e. all three systems have no oxygen at all.

Next, electron and hole doped $Ln_{1-x}M_xFeAsO$ was studied. Substitution of 4-valent Th for 3-valent Gd (i. e. electron doping) in $Gd_{0.8}Th_{0.2}FeAsO$ leads to $T_c=56$ K (C. Wang et al., 2008). Hole doping has been primarily studied in $Ln_{1-x}Sr_xFeAsO$, with Ln=La ($T_c=25$ K, Wen et al., 2008), Pr ($T_c=15$ K, Mu et al., 2009b; Ju et al., 2009), and Nd ($T_c=13.5$ K, Kasperkiewicz et al., 2009). Thus, at least from these few measurements, hole doping in 1111 structure FePn superconductors is much less effective at raising $T_c$ than electron doping. G. Wu et al. (2008b) argue that $La_{1-x}Sr_xFeAsO$, in which $T_c$ is reported (Wen et al., 2008) to be unusually constant with doping, is only superconducting with oxygen deficiency.

In electron doped $LnFe_{1-x}Co_xAsO$, Sefat et al. (2008a) was the first to discover that – unlike the high $T_c$ cuprate CuO planes – the superconducting FeAs planes can tolerate significant disorder (this is also the case, discussed below, for the 122 structure). This is a key point (and thus doping on the Fe site is thoroughly discussed here) in understanding the superconductivity in the FePn/Ch and will be further discussed below. For Ln=La and a Co concentration of x~0.05, $T_{SDW}$ is suppressed and $T_c$ starts at ~11 K, rising up to 14 K at x=0.11 before falling back to $T_c=6$ K at x=0.15. Single crystal $LaFe_{0.92}Co_{0.08}AsO$ had $T_c=9$ K (Yan et al., 2009). See also Cao, et al. (2009) who, besides LaFeAsO doped with Co, also studied $SmFe_{1-x}Co_xAsO$, with $T_c(x=0.1) = 17$ K.



Single crystal electron-doped NdFe$_{0.95}$Co$_{0.05}$AsO has T$_c$=25 K (S. K. Kim et al., 2010). Y. Qi et al. (2009b) substituted Ir for Fe in LaFeAsO and found a maximum T$_c$~12 K for 7.5% Ir. Co-doping of SrFeAsF creates a maximum T$_c$ of 4 K (Matsuishi et al., 2008a) while Co-doping of the related CaFeAsF gives the much higher T$_c$ of 22 K for 10% replacement of Fe by Co (Matsuishi et al., 2008b). The higher T$_c$ in Co-doped CaFeAsF vs SrFeAsF is argued by Nomura et al. (2009) to be due to Co-doping causing the FeAs$_4$ tetrahedra to become more regular (angle approaches 109.47$^o$) in CaFe$_{1-x}$Co$_x$AsF but more distorted in SrFe$_{1-x}$Co$_x$AsF.

Finally, 'isoelectronic' doping (where Ru has the same valency as Fe) was studied (McGuire et al., 2009) in polycrystalline PrFe$_{1-x}$Ru$_x$AsO, with total suppression of the structural/magnetic transitions by x=0.67. Possible distortion of the Fe-As tetrahedral by the larger Ru atom was suggested as an explanation for the lack of superconductivity down to 2 K. As will be seen in Section IIB2a and in Table 2 below, Ru substitution does cause superconductivity when substituted for Fe in the 122's.

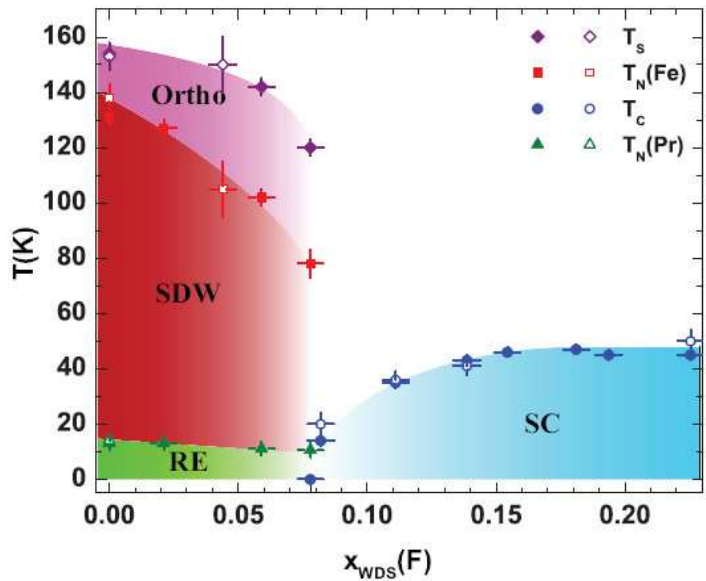



Fig. 9 (color online) The structural, magnetic and superconducting phase diagram of electron doped PrFeAsO$_{1-x}$F$_x$, $0 \le x \le 0.225$ as determined from synchrotron X-ray powder diffraction, magnetization and resistivity measurements (Rotundu et al., 2009). T$_{SDW}$ for x=0 determined from $\rho$ data is 140 K, while from ac susceptibility data is 130 K. Note that T$_c$ is not a sensitive function of doping level for x≥0.14, i. e. the superconducting "dome" is relatively flat. This insensitivity of T$_c$ to composition over a broad range is typical of the 1111's. 'RE' in the diagram is the rare earth Pr antiferromagnetic ordering.

**b.) Correlation between T$_c$, T$_S$ and T$_{SDW}$:** The progression of T$_c$, T$_S$ and T$_{SDW}$ with fluorine doping in LnFeAsO$_{1-x}$F$_x$, Ln=Pr, La, Ce, and Sm varies in two distinct fashions, depending on the Lanthanide atom. For Ln=Nd, there have not been complete phase diagram studies as a function of fluorine doping as yet. Both van der Beek et al. (2010), for NdFeAsO$_{0.9}$F$_{0.1}$, T$_c$~36 K, and Qiu et al. (2008), for NdFeAsO$_{0.8}$F$_{0.2}$, T$_c$=50 K, report no coexistence of magnetism and superconductivity at the superconducting compositions studied. For a list of the undoped 1111 T$_S$/T$_{SDW}$ values, see Table 1.

For Pr/La (Rotundu et al., 2009/Luetkens et al., 2009) the two slightly different ordering temperatures - T$_S$ (154/158 K for x=0) for the tetragonal to orthorhombic lattice distortion and T$_{SDW}$ (~135/134 K for x=0) for the ordering of the Fe ions - decrease gradually while T$_c$ remains zero up to x~0.07/0.04, and then T$_S$ and T$_{SDW}$ vanish to lowest temperature abruptly with further fluorine doping, x=0.08/0.05, while at these compositions superconductivity appears at ~ 20 K and rises in a rather flat "dome" shape to over 40 K, as shown in Fig. 9 for Ln=Pr. Note that for Ln=Pr, there is antiferromagnetic ordering of the Pr ions at low temperature, T$_N$~13 K for x=0, that is absent for the non-magnetic Ln=La. Otherwise, the two phase diagrams are comparable. In PrFeAsO the Fe local moment in the ordered SDW state is 0.48 $\mu_B$ and the Pr local ordered moment at 5 K is 0.84 $\mu_B$ (Zhao et al., 2008b.)

For Ce (Fig. 10)/Sm, T$_S$ and T$_{SDW}$ vary more gradually with fluorine doping in LnFeAsO$_{1-x}$F$_x$, falling continuously to T=0; for Ce (Zhao et al., 2008a), T$_c$ becomes finite



only after $T_S$ and $T_{SDW} \rightarrow 0$. For SmFeAsO$_{1-x}$F$_x$, the question of whether the magnetic

order disappears before superconductivity appears with increasing electron doping is not

yet entirely resolved. Drew et al. (2009) used a microscopic probe, μSR, to determine

that magnetism existed in at least 90 % of their x=0.12 and 0.13 samples ($T_{SDW}$ ~ 40 and

30 K respectively), with clear superconducting resistive transitions where $\rho \rightarrow 0$ at

approximately 9 and 13 K respectively. However, the diamagnetic indications of

superconductivity in these two samples were weak, leading Drew et al. to leave open the

possibility of phase separation between superconducting and magnetic regions.

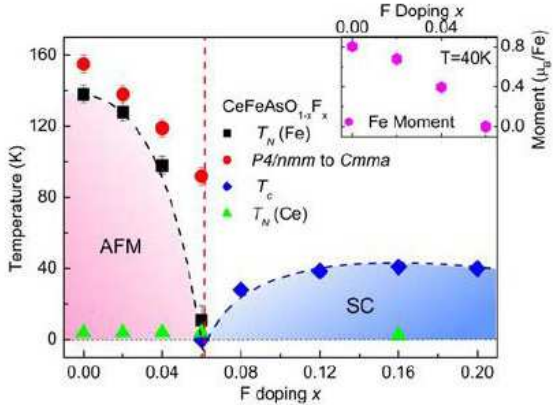

Fig. 10 (color online) Phase Diagram for
CeFeAsO$_{1-x}$F$_x$ Zhao et al. (2008a)

Kamihara et al. (2010) present resistivity data on SmFeAsO$_{1-x}$F$_x$ which show apparent

coexistence of superconductivity and magnetism in only a very narrow composition

range, with $T_{SDW}$~120 K and $T_c$ slightly below 1.8 K (lowest temperature of

measurement) for x=0.037 and no indications of magnetism from the resistivity for

x=0.045, where $\rho \rightarrow 0$ at ~ 22 K. Kamihara et al. present Mössbauer data, which is a

better measure of magnetic order, which show clear lack of magnetic behavior to their



lowest temperature of measurement (4.2 K) for x=0.069, but do not report Mössbauer

data for any lower x (e. g. 0.045) values except for x=0.  Kamihara et al. describe their

data around x=0.04 in $SmFeAsO_{1-x}F_x$ as evidence for disorder and conclude that there is

no coexistence of magnetism and superconductivity in $LnFeAsO_{1-x}F_x$, Ln=Sm.   Ignoring

the compositional disagreement between the two works as simply due to sample variation

issues, what is certain is that $SmFeAsO_{1-x}F_x$ is not a definite example of coexistent

magnetism and superconductivity unlike all of the 122 materials to be discussed next.

Since the other $LnFeAsO_{1-x}F_x$ discussed here, Ln=Nd, Pr (Fig. 9), La, and Ce (Fig. 10) do

not exhibit coexistent magnetism and superconductivity, it may be concluded that the

$LnFeAsO_{1-x}F_x$ 1111 superconducting system does not offer clear coexistence evidence.

In addition to these rather complete fluorine doping results there are data for

electron doping via introducing oxygen deficiency in $LnFeAsO_{1-y}$, Ln=La, Nd, where $T_c$

becomes finite at about y=0.08 (Ishida et al., 2010), a concentration (considering the

respective valencies) not inconsistent with the fluorine doping results.  The authors argue

for coexistence of magnetism (based on structure around 140 K in $\rho$) and

superconductivity for y=0.08 and 0.10.  However, the structure in $\rho$ is unusually constant

in temperature vs the supposed monotonic increase in y, nor is there any investigation of

possible microscopic phase separation.

Therefore, it may be that the 1111 materials, with respect to coexistence of

superconductivity and magnetism, are fundamentally different from the 122's.   See

section IIC below for a summary discussion of coexistence in the FePn/Ch.

CeFeAsO, SmFeAsO and NdFeAsO (phase diagram not shown, see Table 1)

show antiferromagnetic ordering of the rare earth ion moments below 4, 5 and 6 K



respectively.  Below $T^* = 15$ K, Tian et al., 2010, report for the Nd compound – similar to results for Pr (Kimber et al., 2008) and Ce (Zhao et al., 2008a) but with more precise determination of $T^*$ - that the c-axis Fe ordering below $T_{SDW}$=141 K changes from antiferromagnetic to ferromagnetic, indicating an interaction with the rare earth magnetic fluctuations and a delicate balance of the Fe c-axis exchange couplings.

## Table 1:  Structural and Magnetic Transition Temperatures for Undoped 1111, 122, 111, 11, and 122* Parent Compounds

| Material | $T_S$ (K) | $T_{SDW}$ (K) | ref. |
|---|---|---|---|
| LaFeAsO | 158 | 134 | Luetkens et al., 2009 |
| PrFeAsO | 154 | 135 | Rotundu et al., 2009 |
| CeFeAsO | 155 | 140 | Zhao et al., 2008[a] |
|  | 151 | 145 | Jesche et al., 2010 |
| NdFeAsO | 150 | 141 | Qiu et al., 2008/Y. Chen et al., 2008 |
|  | 143 | 137 | Tian et al., 2010 |
| SmFeAsO | 130* | 135* | Margadonna et al., 2009a/Drew et al., 2009 |
| GdFeAsO | 135 |  | C. Wang et al., 2008 |
| SrFeAsF | 180 | 133 | Xiao et al., 2010 |
| CaFeAsF | 134 | 114 | Xiao et al., 2009b |
| BaFe$_2$As$_2$ | 142 | = | Huang et al., 2008 |
| SrFe$_2$As$_2$ | 205 | = | Krellner et al., 2008 |
| CaFe$_2$As$_2$ | 171 | = | Ronning et al., 2008 |
| EuFe$_2$As$_2$ | 190 | = | Tegel et al., 2008b |
| Na$_{1-\delta}$FeAs | 50 | 40 | S. Li et al., 2009a/Parker et al., 2009 |
| FeTe | 72 | = | Fruchart et al., 1975 |
| K$_{0.8}$Fe$_{2-y}$Se$_2$ | 578/551 | 559/540 | Bao et al., 2011b/Liu et al., 2011 |
| Rb$_{0.8}$Fe$_{2-y}$Se$_2$ | 540 | 534 | Liu et al., 2011 |
| Cs$_{0.8}$Fe$_{2-y}$Se$_2$ | 525 | 504 | Liu et al., 2011 |

**\***reversal of $T_S > T_N$, see discussion in text

In the case of Sm, the determination of $T_{SDW}$ (Drew et al., 2009) and $T_S$ (Margadonna et al., 2009a) in separate works results in $T_{SDW} = 135$ K for undoped SmFeAsO and $T_S$=130 K, i. e. reversed from the behavior seen in all the other 1111's (Table 1).  If this is born out by further measurements on the *same* high quality sample,



this reverse ordering of $T_S$ and $T_{SDW}$ would profoundly contradict our theoretical understanding of the link between the structural and magnetic transitions in the FePn/Ch.

Since the work of Zhao et al. (2008a) on polycrystalline $CeFeAsO_{1-x}F_x$ shown in Fig. 10, higher quality samples of the undoped starting compound CeFeAsO in single crystal form have been prepared (Jesche et al., 2010). The separation between $T_S$ and $T_{SDW}$ observed in the polycrystalline material (155 and 140 K respectively) has shrunk by more than half, with values of 151 and 145 K respectively. Thus, the question was posed (Jesche et al., 2010) as to how much the separation of $T_S$ and $T_{SDW}$ in *all* the undoped 1111's is intrinsic, and how much is due to defects. Recently, high quality single crystals of NdFeAsO have been prepared (Yan et al., 2009), with $T_S=142$ K and $T_{SDW}=137$ K (Tian et al., 2010) vs previous values on polycrystalline material of $T_S=150$ K (Qiu et al., 2008) and $T_{SDW}=141$ K (Y. Chen et al., 2008) – see Table 1. Thus, the shrinkage of the difference in $T_S$ and $T_{SDW}$ with increasing sample quality in the 1111's suggested by Jesche et al. (2010) is borne out in NdFeAsO. It would be interesting to see if single crystals of SrFeAsF, where as shown in Table 1 the difference in polycrystalline material between $T_S$ and $T_{SDW}$ is 47 K (Xiao et al., 2010) – the largest separation of any 1111, would also see a decrease in the difference $T_S$ - $T_{SDW}$ with improved sample quality.

In their work on single crystal CeFeAsO, Jesche et al. (2010) analyze the structural transition to be second order, and the magnetic transition to be possibly a broadened first order phase transition. Tian et al. (2010) identify the magnetic transition in their single crystal sample of NdFeAsO as being second order. These two 1111 compounds display different behavior than will be discussed below for the undoped 122's,

where the question of the thermodynamic order of the two coincident-in-temperature transitions has been more of a focus.

## 2.) 122 Structure

Due to the ease by which the 122's can be prepared in single crystal form (see section V), a much larger variety of transition metal dopings – see Table 2 - on the Fe sites have been studied. In the properties discussed in this section, the 122's are often unlike the 1111's: 1.) $T_S$ and $T_{SDW}$ in general are the same in the undoped 122 $M(TM)_2(Pn)_2$ compounds (as listed in Table 1), but then do split upon doping upon the transition metal and the pnictide site, with some disagreement about splitting upon doping on the M-site. 2.) While a number of 1111's have magnetic ordering of the lanthanide site rare earth ion (Pr, Ce, Nd, Sm) in addition to the ordering of the Fe as discussed above, in the 122 undoped parent compounds there is only $EuFe_2As_2$ where in addition to the Fe ordering at 190 K, the Eu orders antiferromagnetically below 19 K (Xiao et al., 2009a). As an additional contrast, in $EuFe_2(As_{1-x}P_x)_2$, for x≥0.22, the Eu ordering becomes ferromagnetic (Jeevan et al., 2011). 3.) The structural transition in the undoped $MFe_2As_2$ compounds appears, based on hysteresis in the specific heat transition and on the jump in unit cell volume determined by neutron scattering or x-ray diffraction, to be first order in the following cases: M=Ba, $T_S$=142 K, (see early work by Huang et al., 2008 and recent data on an annealed single crystal by Rotundu et al., 2010); M=Sr, $T_S$ =205, (Krellner et al., 2008; Zhao et al., 2008c); M=Ca, $T_S$=171 K (Ronning et al., 2008; Goldman et al., 2008; Kumar et al., 2009a). This is consistent with Landau theory, which states that two simultaneous phase transitions that interact with each other (i. e. are not simultaneous due to coincidence) and break different symmetries result in a first



order transition. (See Sections IIB2b and IIIA below for a discussion of the possible connection between the magnetic and structural phase transitions.) However, Wilson et al. (2009), in their neutron scattering experiments on a high quality single crystal of $BaFe_2As_2$, find that both the structural and magnetic transitions at 136 K are second order, with a possible weak first order transition within their error bar. Tegel et al. (2008b) argue from their measurements of the lattice order parameter (P=(a-b)/(a+b), where a and b are the orthorhombic axes' lengths) in M=Sr ($T_S$=203 K) and Eu ($T_S$=190 K) that - despite their measured cell volume discontinuity at $T_S$ in $SrFe_2As_2$ – *all* of the $MFe_2As_2$ starting compounds undergo in fact second order structural phase transitions. Tegel et al. find that P in their data scales with $[(T_S-T)/T]^\beta$ where $\beta$, although small, remains finite – i. e., implying that the transition, despite its abruptness, remains second order. If this is the case, and in light of the prediction of Landau theory, then either the simultaneity of $T_S$ and $T_{SDW}$ are coincidental (see discussion in IIB2b and IIIA) or there should be some higher temperature precursor of one of the transitions that breaks that transition's symmetry at a higher temperature. Yi et al. (2011), in an ARPES study of Co-doped $BaFe_2As_2$ single crystals under uniaxial stress (which of course intrinsically provides symmetry breaking) to detwin the orthorhombic state, find electronic anisotropy well above the structural phase transition. In any case, the structural transitions in the samples that have been measured to date in the 122's definitely show a more rapid variation of the lattice structure with temperature at $T_S$ than those in the 1111's. 4.) Unlike all the $LnFeAsO_{1-x}F_x$ except possibly for Ln=Sm, magnetism and superconductivity coexist quite generally in the lower ('underdoped') portion of the superconducting dome for the 122's. The question of whether this coexistence is at the microscopic or phase separated



level will be discussed.  5.)  Finally, hole doping raises $T_c^{max}$ in the 122's to a significantly higher value than electron doping, 38 K vs 25 K.

   **a.) $T_c$ vs doping:**  The discovery of superconductivity in the 122 structure was via K-doping (hole doping) of $BaFe_2As_2$ (Rotter, Tegel and Johrendt, 2008).  Three other non-superconducting $MFe_2As_2$ (M=Sr, Ca, Eu) host compounds were quickly also discovered, where both hole doping on the M-site and electron doping on the Fe-site, as well as more recently P doping on the As-site, succeeded in causing superconductivity, see Table 2 for a complete listing.  Clearly, the variety of dopants that achieve superconductivity in the 122's is quite large.  An exception is doping with Cu (Canfield and Bud'ko, 2010), three columns to the right of Fe in the periodic table, or Cr (Sefat et al., 2009), two columns to the left of Fe, which do not induce superconductivity in $BaFe_2As_2$.  In addition to doping-induced superconductivity, three Fe-containing 122 compounds superconduct without doping, $KFe_2As_2$ ($T_c$~3.8 K, Rotter et al., 2008a), $RbFe_2As_2$ ($T_c$~2.6 K, Bukowski et al., 2010) and $CsFe_2As_2$ ($T_c$=2.6 K, Sasmal et al., 2008).  $KFe_2As_2$ has been shown to be quite interesting in its properties, including evidence for nodal superconductivity, see section IV, although according to the specific heat discontinuity at $T_c$, $\Delta C(T_c)$, $KFe_2As_2$ does not appear to belong with the other FePn/Ch (section IIIB3).

   The so-called 'isoelectronic' doping (substitution of P for As or Ru for Fe) in $MFe_2As_2$ causing quite respectable $T_c$'s raises the issue of charge doping vs other effects. Since P is smaller than As, one might conclude that the $T_c$ in $MFe_2As_{2-z}P_z$ is at least partly due to 'chemical' pressure, analogous to the physical pressure discussed below in section IIC.  However, Ru is larger than Fe (although as Ru replaces Fe in $BaFe_2As_2$, the



a-axis grows as the c-axis shrinks – Sharma et al., 2010). Wadati, Elfimov and Sawatzky (2010) using DFT calculations have proposed that the transition metals Co and Ni when substituted for Fe in $BaFe_2As_2$ (as well as in FeSe) behave essentially isovalent with Fe, with their effect on superconductivity primarily due to their impurity/scattering nature affecting the Fermiology – "washing out" parts of the Fermi surface. Thus, rather than a rigid band shift due to adding electrons as would come from a naïve picture, the main effect is calculated to be an impurity-scattering-caused washing out of the more flat band contributions to the total Fermi surface. As stated already in this section, $T_c$ is strongly influenced by the structural properties of tetrahedron angle (Lee et al., 2008) and pnictogen height (Kuroki et al., 2009). Rotter, Hieke and Johrendt (2010) conclude by a careful study of the crystal structure in $BaFe_2As_{2-z}P_z$ that P-doping causes a slight reorganization of the crystal structure (not solely a change in the pnictogen height) that influences $T_c$ via its effect on the bandwidth. Klintberg et al. (2010) compare the effect of pressure and P-doping on the superconducting phase diagram of $BaFe_2As_2$, including the effect of pressure on $BaFe_2As_{2-z}P_z$, and conclude from the similarities between P-doping and pressure that impurity scattering is not limiting $T_c$ in the doped samples.

Thus, there are clearly important details involved not only with the 'isoelectronic' doping, but also with the other doping species. The simple 'atomic' picture - where doping is described as simply adding or subtracting electrons, or isoelectronic doping with essentially no expected change - is definitely oversimplified.



## Table 2:  $T_c$ vs Composition in $M_{1-x}A_xFe_{2-y}TM_yAs_{2-z}P_z$
### $T_c$'s given are the maxima vs composition/Only one site is doped at a time

| Material | M-site dopant | $T_c(K)/x$ $y=z=0$ | Ref. | Fe-site dopant | $T_c(K)/y$ $x=z=0$ | Ref. | As-site dopant | $T_c(K)/z$ $x=y=0$ | Ref. |
|---|---|---|---|---|---|---|---|---|---|
| $BaFe_2As_2$ | K | 38/0.4 | 1 | Co | 22/0.2 | 9 | P | 30/0.7 | 22 |
|  | Rb | 23/0.1 | 2 | Ni | 20.5/0.1 | 10 |  |  |  |
|  |  |  |  | Pd | 19/0.11 | 11 |  |  |  |
|  |  |  |  | Rh | 24/0.11 | 11 |  |  |  |
|  |  |  |  | Ru | 21/0.9 | 12 |  |  |  |
|  |  |  |  | Pt | 25/0.1 | 13 |  |  |  |
| $SrFe_2As_2$ | K | 36.5/0.5 | 3 | Co | 20/0.2 | 14 | P | 27/0.7 | 23 |
|  | Na | 35/0.5 | 4 | Ni | 10/0.15 | 15 |  |  |  |
|  | Cs | 37/0.5 | 3 | Pd | 9/0.15 | 16 |  |  |  |
|  | La | 22/0.4 | 5 | Rh | 22/0.25 | 16 |  |  |  |
|  |  |  |  | Ru | 13.5/0.7 | 17 |  |  |  |
|  |  |  |  | Ir | 22/0.5 | 16 |  |  |  |
|  |  |  |  | Pt | 16/0.16 | 18 |  |  |  |
| $CaFe_2As_2$ | Na | 33/0.66 | 6 | Co | 17/0.06 | 19 | P | 13/0.3 | 23 |
|  |  |  |  | Ni | 15/0.06 | 20 |  |  |  |
|  |  |  |  | Rh | 18/0.1 | 21 |  |  |  |
| $EuFe_2As_2$ | K | 32/0.5 | 7 |  |  |  | P | 26/0.6 | 24 |
|  | Na | 35/0.3 | 8 |  |  |  |  |  |  |

Note: Cu substituted for Fe in $BaFe_2As_2$ suppresses $T_S$ and $T_{SDW}$ but does not induce superconductivity (Canfield et al., 2009) while Mn substituted for Fe in $SrFe_2As_2$ up to x=0.3 is relatively ineffective in suppressing $T_S$ and $T_{SDW}$ (Kasinathan et al., 2009).

**b.) Correlation between $T_c$, $T_S$ and $T_{SDW}$:** In order to make the large set of numerical data of $T_c$, $T_S$ and $T_{SDW}$ vs doping level in the 122's more understandable, phase diagrams are shown here for selected dopants. Despite the hole doped $Ba_{1-x}K_xFe_2As_2$ being the discovery superconductor in the 122's (Rotter, Tegel and Johrendt, 2008), this phase diagram shown in Fig. 11 has received much less attention – perhaps due to K homogeneity issues (Ni et al., 2008a; Johrendt and Poettgen, 2009), where the concentration varies by ±5 % so that 'Ba$_{0.6}$K$_{0.4}$Fe$_2$As$_2$' has K concentrations between 0.35 and 0.45. Within the resolution of the early neutron scattering determinations of $T_S$ and $T_{SDW}$ (H. Chen et al., 2009) and of the x-ray/Mössbauer determinations of $T_S/T_{SDW}$ (Rotter et al., 2009), the structural and magnetic transitions remained at the same temperature (see Fig. 11) until both transitions are suppressed in $Ba_{1-x}K_xFe_2As_2$. However, more recent measurements (Urbano et al. 2010) have found that there is clear evidence (distinct anomalies in both $d\rho/dT$ and specific heat)

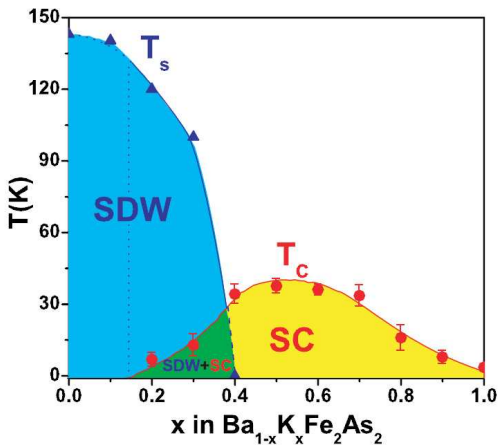

Fig. 11 (color online) H. Chen et al. (2009), $T_S$ and $T_{SDW}$ stay equal vs x. Johrendt and Pöttgen (2009) find that $T_{SDW}$ is suppressed at x=0.3, however both groups find that $T_{SDW}$ does not join the superconducting dome.

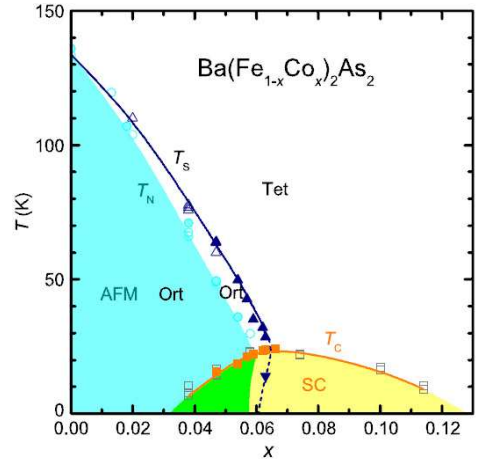

Fig. 12 (color online) Nandi et al. (2010). Note the factor of two between x in their notation vs the y used here and that $T_S$ and $T_{SDW}$ indeed intersect the super-conducting dome.



for splitting of $T_S$ and $T_{SDW}$ in an underdoped single crystal of $Ba_{0.86}K_{0.14}Fe_2As_2$, $T_c \approx 20$ K and RRR~8.5, with $T_S=110$ K and $T_{SDW}=102$ K.   Although this sample was grown using Sn flux, Urbano et al. argue that improved methods have resulted in a high quality sample with little or no effect from Sn-flux inclusion.  This is an important result since, as will now be discussed, 122's in general (with one case – $BaFe_{2-x}Ru_xAs_2$ – still under debate) *all* show such separation with doping.  The exception for K-doped $BaFe_2As_2$ was an important anomaly that needed clarification.   For completeness it should be mentioned that this continues to be a subject of debate, with recent neutron scattering measurements (Avci et al., 2011) on self flux grown samples of $Ba_{1-x}K_xFe_2As_2$ finding no separation at all dopings where $T_S$ and $T_{SDW}$ exist.  The sample from Avci et al. that has the most comparable properties to the sample from the work of Urbano et al. (2010) has a nominal composition of x=0.21 and a similar $T_c \approx 20$ K and $\Delta T_c$ as determined from susceptibility, i. e. the sample seems to be of comparable quality.  Although Avci et al. find no separation in $T_S$ and $T_{SDW}$, their apparent uncertainty in temperature seems to be at least 5 K due to the steep rise of the magnetic moment below $T_{SDW} \approx 80$ K.   These samples should have their magnetic and structural transitions measured by some technique with a higher temperature resolution.

There still remain homogeneity issues in the K-doped $BaFe_2As_2$ samples.  For example, although superconducting samples achieved by doping on both the Fe and As sites (discussed just below) in the 122's show clear specific heat anomalies, $\Delta C$, at $T_c$ (see section IIIB3) for the whole superconducting dome, as yet only samples near optimal doping (x~0.4) show a measurable $\Delta C$ in $Ba_{1-x}K_xFe_2As_2$.  For the Urbano et al. (2010)



data on x=0.14 and in the work of Rotter et al. (2009) for x=0.2, no anomaly in the specific heat is observable in $T_c$ (20 and 23.6 K respectively).

Surprisingly, there are no other studies of doping on the M site in $MFe_2As_2$ (see Table 2 for a summary) that investigate the question of potential splitting of $T_S$ and $T_{SDW,}$ or the presence or absence of finite $\Delta C$ away from optimal doping.

In Fig. 12, the phase diagram for Co-doped $BaFe_2As_2$ is shown, based on resistivity, magnetization, and specific heat measurements. A common feature of doping the $MFe_2As_2$ materials on the Fe-site has been the separation for finite doping of $T_S$ from $T_{SDW}$ (see results similar to those for Co-doping for $T_S/T_{SDW}$ splitting upon doping with TM=Ni and Rh in $BaFe_{2-y}TM_yAs_2$ by Canfield and Bud'ko, 2010). However, Thaler et al. (2010), in single crystal work, report for isoelectronic Ru doping on the Fe site that no splitting is observable, using rather careful consideration of $d\rho/dT$ through the transition. In contradiction to this, another single crystal $BaFe_{2-x}Ru_xAs_2$ work (Rullier-Albenque et al., 2010) claim to see features in their $d\rho/dT$ data indicative of two transitions (95 and 88 K respectively) at x=0.3. This discrepancy deserves further investigation.

The order of the structural phase transition in $BaFe_{1.906}Co_{0.094}As_2$ ($T_S$=60 K) in the neutron scattering study of Pratt et al. (2009a), although there was slight hysteresis, could not be determined with certainty. However, the magnetic transition at $T_{SDW}$ = 47 K is clearly second order. Ni et al. (2009) in their study of $BaFe_2As_2$ doped with Rh and Pd on the Fe-site point out several comparisons in these $BaFe_{2-y}TM_yAs_2$ phase diagrams. Their $T_c$ vs y for Rh falls on the same dome as shown in Fig. 12 for Co, which is isoelectronic with Rh. Their $T_c$ vs y for Pd forms a narrower dome ($T_c$ for Pd-doping is finite for y=0.04 to 0.16 vs 0.06 to 0.24 for Co) that only rises up to $T_c^{max}$ of 19 K, but



again coincides with the $T_c$ vs y data of Ni (Canfield et al., 2009), isoelectronic to Pd. Doping with Cu suppresses $T_S$ and $T_{SDW}$, but does not induce superconductivity (Canfield et al., 2009).

An interesting feature of the phase diagram in Fig. 12 for $Ba(Fe_{1-x}Co_x)_2As_2$ is the *reversal* of the phase boundary upon cooling through the superconducting dome at x~0.063 (see similar work in Rh-doped $BaFe_2As_2$, Kreyssig et al., 2010). Thus, the sample at this composition transforms from orthorhombic back to tetragonal upon cooling below $T_c$. Nandi et al. (2010) discuss this (see also following paragraph) in terms of a magnetoelastic coupling between nematic magnetic fluctuations (no static order is present at this composition) and the lattice. The magnetic fluctuations are weakened by the superconductivity which competes with the magnetic order (Pratt et al., 2009a), thus allowing reentry into the tetragonal lattice structure. In fact, a neutron scattering work (Fernandes et al., 2010a) for the magnetic composition x=0.059 finds not only a weakening of the magnetism by the superconductivity but actually a *reversal* from magnetically ordered back into the paramagnetic state below $T_c$. This reentrant behavior has been used as an argument by Fernandes and Schmalian (2010) that the magnetic order in at least $Ba(Fe_{1-x}Co_x)_2As_2$ must be partly itinerant in nature as discussed in Section IIB above when the question of itinerant vs localized order was considered. INS studies (Lumsden et al., 2009) of near optimally doped $BaFe_{1.84}Co_{0.16}As_2$ show that the anisotropic 3D magnetic interactions in the ordered undoped $BaFe_2As_2$ become much more 2D with doping.

As an introduction to their work on the reentrant behavior around x≈0.06 in $Ba(Fe_{1-x}Co_x)_2As_2$, Nandi et al. (2010) discuss the link between magnetic fluctuations



above $T_{SDW}$, i. e. for x<0.06, and the orthorhombic lattice distortion.  In their description, two antiferromagnetic sublattices have magnetizations $\mathbf{m}_1$ and $\mathbf{m}_2$ which are weakly coupled due to frustration caused by large next nearest neighbor interactions (see Chandra, Coleman and Larkin, 1990, for a discussion.)  Below the magnetic ordering temperature, the time averaged order parameter $<\psi>$, where $\psi=\mathbf{m}_1\cdot\mathbf{m}_2$, and the time averaged sub-lattice magnetizations $<\mathbf{m}_1>$ and $<\mathbf{m}_2>$ are all finite, leading to static magnetic order.  On the other hand, above $T_S$ the time averaged order parameter $<\psi>$, as well as $<\mathbf{m}_1>$ and $<\mathbf{m}_2>$, are zero, while nematic (but not static) ordering (where $\mathbf{m}_1$ and $\mathbf{m}_2$, which still time average to zero, are coupled to give a finite $<\psi>$) sets in at $T_S$ but still above $T_{SDW}$.  Thus, in the view of Nandi et al. (2010), the nematic order above the magnetic transition (and even in the case where the magnetism is totally suppressed) drives the structural distortion.  The relative importance of electronic nematic order, which breaks the tetragonal basal plane a-b axis symmetry, and its possible role in mediating the superconductivity in the FePn/Ch is a subject of significant interest, see also Fernandes et al. (2010b), Chuang et al. (2010), Chu et al. (2010), Park et al. (2010) and Harriger et al. (2010).

Phase diagrams for other $MFe_{2-y}TM_yAs_2$ than M=Ba are less thoroughly studied. Leithe-Jasper et al. (2008) studied $SrFe_{2-x}Co_xAs_2$ and found no superconductivity down to 1.8 K for x≤0.15 and x≥0.5, with $T_c^{max}$ =19.2 K at x=0.2.  Resistive indications of $T_S/T_{SDW}$ were absent for x>0.15.  What is different in this $SrFe_{2-x}Co_xAs_2$ system from the M=Ba data in Fig. 12 is the lack of the gradual ramp up of $T_c$ on the underdoped side of the phase diagram for M=Sr.  F. Han et al. (2009) report phase diagrams based on the measurement of resistivity (i. e. they were unable to distinguish separation of $T_S$ and



$T_{SDW}$) for $SrFe_{2-x}TM_xAs_2$ for TM=Rh, Ir, Pd. Shown in Fig. 13 is the diagram for Rh, isoelectronic to Co just discussed. The behavior shown in Fig. 13 is similar to that seen for $BaFe_{2-x}TM_xAs_2$ discussed above. As shown in Table 2, the $T_c^{max}$ for TM=Ir, isoelectronic to Co and Rh, in $SrFe_{2-x}TM_xAs_2$ found by F. Han et al. (2009) is similar to that for Rh and Co, while that for TM=Pd is significantly lower. Kasinathan et al. (2009) report only weak suppression of $T_S$ in $SrFe_{2-x}Mn_xAs_2$ up to x=0.3, and no superconductivity.

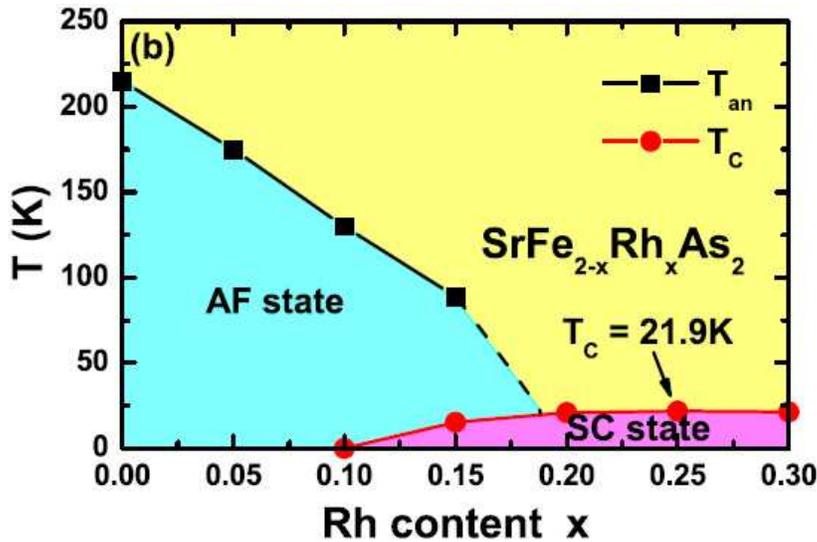

Fig. 13 (color online) The temperature of the anomaly in the resistivity, $T_{an}$ is taken as $T_{SDW}$ by F. Han et al. (2009). The dashed line connecting the last measured $T_{an}$, at x=0.15, to the superconducting dome is a guide to the eye. Note that no data for x>0.3 are reported.

In $CaFe_{2-x}TM_xAs_2$, Kumar et al. (2009a) studied TM=Ni and found superconductivity only for x=0.053 and 0.06, with $T_c$=15 K and both the structural and magnetic transitions suppressed. Drops in the resistivity at 15 K (but not full transitions) were seen at x=0.027, 0.030 and 0.075. This is a much narrower region of



superconductivity with doping than the other Fe-site dopings in M=Ba and Sr discussed above.

Finally, an example of a phase diagram for P-doping is shown in Fig. 14, where data for BaFe$_2$As$_{2-x}$P$_x$ from Kasahara et al. (2010) are shown. Although the shading around x=0.3 is drawn to indicate a gradual fall in T$_S$ and T$_{SDW}$, the data suggest that in fact, just as seen for K doping in BaFe$_2$As$_2$ and Rh and Ir doping in SrFe$_2$As$_2$ (F. Han et al., 2009), there is a region at the top of the superconducting dome where the T$_S$ and T$_{SDW}$ phase boundaries do not join the T$_c$ dome phase boundary. This is also the case for the phase diagram (not shown) of Shi et al. (2009) for SrFe$_2$As$_{2-x}$P$_x$, where T$_c$ becomes finite at x=0.5 while T$_{SDW}$ is still 140 K and disappears for higher P-doping. For EuFe$_2$As$_{2-x}$P$_x$ (Jeevan et al., 2011), the antiferromagnetic ordering in the Fe is suppressed before superconductivity occurs at x=0.4; however, the superconductivity at x=0.4 does coexist with the Eu antiferromagnetism. Such coexistence of antiferromagnetism and superconductivity in

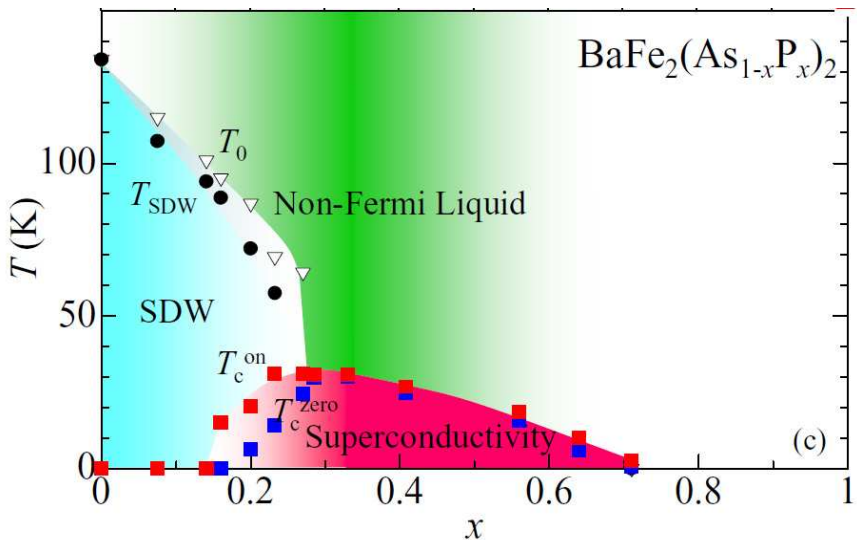

Fig. 14 (color online) Data from Kasahara et al. (2010) for P-doped BaFe$_2$As$_2$. See Jiang et al. (2009) for a similar phase diagram. The open upside down triangles denote T$_S$, while the filled black circles denote T$_{SDW}$ determined from resistivity. Two superconducting T$_c$'s are shown, the upper points are the onset of the resistive transition,



the lower ones are where $\rho \to 0$.  Note the presence of non-Fermi liquid behavior in the resistivity for this compound, discussed in section III.

electrons from different atoms (in this case Eu and Fe) is well known in a variety of compounds, see, e. g., the review on the rare earth borocarbides by Gupta (2006).

### 3.  111, 11, 21311, and 122* Structures:

Relatively fewer data exist for these structures, due to their more recent discovery and, in the case at least of the 11 structure, fewer possibilities for substitution.

#### a.)  $T_c$ vs doping

**111:**  Hole doping in LiFeAs by introducing Li vacancies has been calculated by Singh (2008).  Experimentally, Pitcher et al. (2010) in polycrystalline material find that $T_c$ falls rapidly with increasing Li deficiency in $Li_{1-y}Fe_{1+y}As$.  Pitcher et al. also find that $T_c$ falls with electron doping in $LiFe_{1-x}(Co,Ni)_xAs$, by approximately 10 K for every 0.1 doped electron independent of whether Co (one electron each) or Ni(two electrons each) is used as the dopant.   This agrees fairly well with the $T_c$ suppression measured in single crystal $LiFe_{0.95}Co_{0.05}As$, $T_c \approx 8$ K, reported by Lee et al. (2011).  Based on the Fermiology reported by ARPES (section IVA2), where there is no nesting in LiFeAs because the electron pockets are smaller than the hole pockets, it would be expected that electron doping in LiFeAs might improve the nesting and, if nesting were important for $T_c$ in the 111, therefore $T_c$.  The fact that the opposite effect is observed (especially since Co doping of the Fe site in $BaFe_{2-x}Co_xAs_2$ enhances $T_c$) may be confirmation that nesting is indeed not critical for the superconductivity in LiFeAs – see discussion of the theory in section IVA2.



Before the doping in $Na_{1-\delta}FeAs$ is presented, the question of the superconductivity in the parent compound deserves discussion.  In the early work on polycrystalline $Na_{1-\delta}FeAs$ material, Parker et al. (2009) reported only 10% diamagnetic shielding, i. e. not the more stringent field-cooled Meissner effect expulsion which is generally only a few percent at most due to pinning in the FePn/Ch superconductors.   This 10% fraction of shielding, which is small compared to the typical behavior ($\approx$100 %) of the other FePn/Ch superconductors, in general argues for a small volume fraction of bulk superconductivity, perhaps a sheath of superconducting material or filaments.   Other workers (Chu et al., 2009) reported similarly weak shielding in polycrystalline material.   Then self-flux-grown single crystals of $Na_{1-\delta}FeAs$ were characterized by G. F. Chen et al. (2009) via specific heat, and the lack of a $\Delta C$ anomaly at $T_c$ was attributed to a small superconducting volume fraction.  All of these works estimate a Na deficiency $\delta$ of 1-2 %, which is a kind of "self-doping."

In light of this discussion of the parent compound, the results of doping with Co in either polycrystalline or single crystal material are germane to understanding superconductivity in $Na_{1-\delta}FeAs$.   Parker et al. (2010) doped Co and Ni into polycrystalline $Na_{1-\delta}FeAs$, again with 1-2% Na deficiencies.  The fraction of diamagnetic shielding (zero field cooled susceptibility) grows from 5-10% of full shielding for no Co doping (i. e. not bulk superconductivity), to 60% diamagnetic shielding for $Na_{1-\delta}Fe_{0.99}Co_{0.01}As$ to 100% diamagnetic shielding for $Na_{1-\delta}Fe_{0.975}Co_{0.025}$, $T_c$=21 K.  The superconducting dome ends at 10% Co-doping.  Within the error bar in the $\mu$SR measurement, the magnetism is suppressed at the 2.5% Co-doping as is, determined via neutron scattering, the structural phase transformation (Parker et al., 2010).   Therefore  it



appears that, at least as thus far prepared, undoped $Na_{1-\delta}FeAs$ – presumably due to defects – is not a bulk superconductor but that slight electron doping brings it back to being equivalent to undoped 111 LiFeAs, $T_c$=18 K.   Since Li and Na are isoelectronic, comparable $T_c$'s - as seen for the doped 122's in Table 2 above - are expected.   Xia et al. (2010) have prepared single crystal $Na_{1-\delta}Fe_{0.95}Co_{0.05}As$ ($T_c$=19 K) and $Na_{1-\delta}FeAs_{0.8}P_{0.2}$ ($T_c$=33 K, a record high for P-doping of an As pnictide superconductor), with resistive transition widths for both samples ~ 0.5 K.   The resistivity measured up to room temperature in both compounds has no anomalies above $T_c$, confirming in the case of the Co-doping the reported suppression of the magnetic transition by Parker et al. (2010).

**11:**  McQueen et al. (2009a) performed a careful study of $T_c$ in $Fe_{1+\delta}Se$ with Fe content variation and found that 'stoichiometric' $Fe_{1+\delta}Se$, when made single phase, has $\delta$=0.01 and $T_c$=8.5 K, while for $\delta$=0.03, $T_c$ is below 0.6 K.  Mizuguchi et al. (2009) have studied FeSe doping with Te and S on the Se site and Co and Ni on the iron site.  $T_c$ rises from the initial ~ 8 K up to about 20% doping for both the S and Te, while Ni and Co both suppress $T_c$ by 10% substitution.   Replacing 10% of the Te in $Fe_{1+\delta}Te$ with S results in a depression of the magnetic transition from 72 K to ~ 30 K and $T_c$~8.5 K, i. e. coexistent magnetism and superconductivity (Hu et al., 2009).

**21311:**  As discussed above in the beginning of section IIA, replacing Sc by V and P by As in $Sr_2ScO_3FeP$, $T_c$=17 K, gives $T_c$= 37 K in $Sr_2VO_3FeAs$ (Zhu et al., 2009b). Replacing V by $Mg_{0.2}Ti_{0.8}$ increases $T_c$ up to 39 K (Sato et al., 2010), with a c-axis spacing of 15.95 Å.  A derivative structure of the 21311 is the 2(1.5)411 - doubled to preserve integer ratios, known as the '43822' structure (N. Kawaguchi et al., 2010).  This 43822 extension of the 21311 structure follows the idea (see, e. g., Ogino et al., 2010a) of



inserting or 'doping' more layers between the FeAs planes to expand the c-axis, based on the correlation that $T_c$ and c-axis spacing scale in the first four structures: $FeSe_{1-y}$ ($T_c$=8 K, 5.49 Å), LiFeAs ($T_c$=18 K, 6.36 Å), $Ba_{0.6}K_{0.4}Fe_2As_2$ ($T_c$ = 38 K, 6.65 Å), $SmFeAsO_{1-x}F_x$ ($T_c$ = 55 K, 8.44 Å). (Note that, within a given structure, $T_c$ does not scale with c-axis spacing, e. g. 1111 $LaFeAsO_{1-x}F_x$ has $T_c$=26 K and c=8.73 Å.) Ogino et al. (2010c) reported $Ca_2(Mg_{0.25}Ti_{0.75})_{1.5}O_{-4}FeAs$ to have $T_c^{mid}$ = 47 K, with a c-axis spacing of 33.37 Å. This related structure is still tetragonal, but has space group I4/mmm, i. e. the same as the 122 structure which has an atom in the body center of the unit cell, and can be further expanded according to the formula $Ca_{n+1}(M,Ti)_nO_{-3n-1}Fe_2As_2$, M=Sc, Mg (Ogino et al., 2010a; Shimizu et al., 2010), with 'n' equal to the number of intercalated layers. As yet, only the discovery works discuss this further progression of seeking higher $T_c$ by stretching the c-axis and the distance between the FePn/Ch layers so that understanding the 21311 and derivative structures is still a work in progress.

**122\*:** The discovery of superconductivity in this structure, before the correct stoichiometry as it presently is understood ($K_{0.8}Fe_{1.6}Se_2$) was worked out, was in the nominal composition $K_{0.8}Fe_2Se_2$ by Guo et al. (2010), with a $T_c^{onset}$ determined resistively in polycrystalline material of 30 K. Within 2 ½ weeks of Guo et al.'s publication, Krzton-Maziopa et al. (2011) reported superconductivity at $T_c$=27.4 K in single crystals of $Cs_{0.8}Fe_2Se_2$. Fang et al. (2011b) then reported $T_c$=20 K in $TlFe_{1.7}Se_2$ (nominal composition), and also – in order to affect the known (Zabel and Range, 1984) Fe-sublattice deficiency in the $TlFe_2Se_2$ compound – prepared single crystals of $Tl_{1-y}K_yFe_xSe_2$ ($1.50 \leq x \leq 1.88$, $0.14 \leq y \leq 0.57$) where the compositions were determined using energy dispersive x-ray (EDX) spectrometry. For $1.78 \leq x \leq 1.88$, Fang et al. observe



superconductivity in their samples, sometimes with multiple dips in ρ starting already at 40 K with decreasing temperature, with $T_c$ (ρ→0)≈30 K. It is interesting to note that Zhang and Singh (2009) *predicted* $TlFe_2Se_2$ as a possible parent compound for superconductivity. Rounding out the list of discovery of superconductivity in $A_{0.8}Fe_{1.6}Se_2$ (A=K, Rb, Cs, Tl), C.-H. Li et al. (2011) reported superconductivity at $T_c^{onset}$=31 K in single crystals of $Rb_{0.8}Fe_2As_2$ (nominal composition.)

Although the 122* structure is relatively new, some $T_c$ vs doping information is available. The most important parameter for superconductivity is not the addition of an element to the parent compound (as is necessary for most of the FePn/Ch and particularly the 1111 and the 122), but rather – as mentioned in Section IIA when the structure of 122* was first discussed – insuring the order of the Fe vacancies peculiar to the 122* structure. Bao et al. (2011b) (see also Ye et al., 2011) report that the metallic behavior (and the superconductivity) in these materials is centered at the composition $K_{0.8}Fe_{1.6}Se_2$ (or $A_2Fe_4Se_5$) where the Fe2 sites (see Fig. 6, 16 per unit cell) can be completely occupied and the Fe1 sites (Fig. 6, 4 per unit cell) completely empty. In a contrasting work, F. Han et al. (2011) argue that their data are consistent with disorder being critical for the superconductivity, although they measure a degradation of superconductivity for samples left at room temperature over a time period of days that is unreported by others. Also, Z. Wang et al. (2011), in a transmission electron microscopy study of $K_{0.8}Fe_xSe_2$, conclude that the superconducting samples have Fe vacancy disorder. This question continues to be of central interest in the 122* materials.

Partially substituting the smaller S (i. e. effectively 'chemical pressure') for Se in $K_{0.8}Fe_{1.7}SSe$, Guo et al. (2011a) find $T_c$(ρ→0)=24.8 K, while both L. Li et al. (2011) and



Wang, Lei, and Petrovic (2011b) find essentially no suppression in $T_c$ when only 20% of the Se is replaced by S. $T_c$ is fully suppressed by 80% substitution of Se by S (Lei et al., 2011). Zhou et al. (2011) in a series of Co dopings in crystalline material found that $T_c$ was suppressed below their lowest temperature of measurement (5 K) already in $K_{0.8}Fe_{1.70}Co_{0.01}Se_2$ (composition determined by inductively coupled plasma atomic emission spectroscopy). This result, if it withstands scrutiny concerning possible alteration of the important-for-superconductivity Fe-sublattice vacancy ordering, would be a record in the FePn/Ch for change of $T_c$ with Co-for-Fe substitution.

   **b.) Correlation between $T_c$, $T_S$ and $T_{SDW}$:** Phase diagrams of $T_c$, $T_S$ and $T_{SDW}$ do not exist in either the 111 or the 21311 structures, since there are not enough data (e. g. only one indication of magnetism in the 21311's so far, Sefat et al., 2010). A phase diagram for $FeSe_xTe_{1-x}$ has been produced (Martinelli et al., 2010) using neutron diffraction to determine the structural and magnetic transitions. $T_S$ and $T_{SDW}$ remain coincident and finite with increasing Se-doping for x≤0.075 – decreasing from 72 K at x=0 down to 43 K at x=0.075, whereas superconductivity is induced increasing Se for x≥0.05, i. e. there is a range of Se composition where long range magnetism and superconductivity coexist. Katayama et al. (2010) offer a competing phase diagram for $FeSe_xTe_{1-x}$, with spin glass behavior for $0.15 \leq x \leq 0.3$, with no range of Se composition with coexistence of long range magnetism and superconductivity. Further, these $FeSe_xTe_{1-x}$ phase diagrams are like those of K-doped $BaFe_2As_2$ (Fig. 11), Ir- and Rh-doped $SrFe_2As_2$ (Fig. 13) and P-doped $BaFe_2As_2$ (Fig. 14) and $SrFe_2As_2$ in that $T_{SDW}$ does not coincide with/smoothly join $T_c$ in the phase diagram. In the 122* structure, Bao et al. (2011b) present a phase diagram for $K_xFe_{2-x/2}Se_2$ in which the magnetic transition vs



x varies between ≈520 K determined by $\chi$ (559 K from neutron scattering) for x≈0.8 down to ≈475 K for x≈1.0, while $T_c$ remains constant at around 30 K for 0.77≤x≤0.86 and becomes abruptly 0 (insulating phase) for x>0.86. The only structural transition in the 122* materials is the ordering of the Fe atoms on the two sublattices (Fe1 and Fe2, see Fig. 6), changing the structure from the disordered tetragonal 122 structure (I4/mmm symmetry) at high temperature with random defect occupation of the Fe1 and Fe2 sublattices to the ordered defect tetragonal 122* structure (Fig. 6, I4/m symmetry) where the vacancies are preferably on the Fe1 site, below $T_S$. Zavalij et al. (2011) give an occupation of the Fe1 site in their ordered superconducting $K_{0.8}Fe_{1.6}Se_2$ and $Cs_{0.8}Fe_{1.6}Se_2$ of 3.2-7.8 % and hold open the possibility that this Fe1 site occupation is only in isolated small domains. According to Bao et al. (2011b) the Fe-defect ordering transition occurs at 578 K for x=0.82 and ≈500 K for x=0.99. Liu et al. (2011), using resistivity and susceptibility measurements, find that the transition they associate with the vacancy ordering transition $T_S$ is generally 10-20 K higher than $T_N$ (see Table 1), just as observed by Bao et al. (2011b), in all of the $A_{0.8}Fe_{1.6}Se_2$ systems they studied with the lowest $T_S$ = 512 K for $A_{0.8}=Tl_{0.4}Rb_{0.4}$.

## C. Coexistence of Magnetism and Superconductivity in the FePn/Ch Superconductors:

From the discussion above, experimentally it is clear that superconductivity coexists with magnetism in a number of FePn/Ch superconductors, including $Ba_{1-x}M_xFe_2As_2$ (Fig. 11), a large number of different transition metal dopants (see Table 2) in $BaFe_{2-y}TM_yAs_2$ (Fig. 12 for TM=Co), $SrFe_{2-y}TM_yAs_2$ (TM= Rh – Fig. 13, Ir, Pd), $MFe_2As_{2-z}P_z$ (M=Ba – Fig. 14, Sr), $Na_{1-\delta}FeAs$, $FeTe_{1-x}Se_x$ and the ordered defect 122*



structure $A_{0.8}Fe_{1.6}Se_2$ (A= K, Rb, Cs, Tl). Certainly other doped systems, e. g. the Ca and Eu 122's, would likely show coexistence as well, when sufficient phase diagram data are gathered. On the other hand, it is equally clear that magnetism is suppressed by doping *before* the appearance of superconductivity in systems like $LnFeAsO_{1-x}F_x$ (Ln=Pr - Fig. 9, La, Ce – Fig. 10, Nd, and possibly Sm).

The issue that researchers have considered is: when coexistence is indicated in the phase diagram, do magnetism and superconductivity evolve from the *same* conduction electrons on a microscopic scale?

Coexistent magnetism and superconductivity evolving from *different* bands, as is the case for example (see Gupta, 2006) in the quaternary borocarbides $RENi_2B_2C$, where RE is a rare earth, is simply magnetic ordering independent of (uncoupled from) the superconductivity, although the magnetically aligned spins can cause pairbreaking and thus the superconductivity is coupled (in a deleterious fashion) to the magnetism. Interestingly, this kind of negative influence of the magnetic rare earth ions on the superconductivity seen in the borocarbides has one comparison example in the FePn/Ch - in $EuFe_2As_2$ under pressure - due to the antiferromagnetism on the Eu sublattice affecting the superconductivity on the Fe sublattice. In $HoNi_2B_2C$ with decreasing temperature in an applied field of 0.2 T (Gupta, 2006) the resistivity, $\rho$, with decreasing temperature first goes to 0 at $T_c \approx 7.6$ K, followed by a finite value of $\rho$ at somewhat lower temperature $\approx 5$ K where the magnetic Ho rare earth ions undergo an ordering transition followed by *reentrance* into the superconducting state again below 4.4 K. In $EuFe_2As_2$ under 3.1 GPa (Kurita et al., 2011), $\rho \to 0$ at $T_c \approx 28$ K, then $\rho$ reenters the normal state around the antiferromagnetic ordering temperature of $T_N = 23$ K, followed by $\rho \to 0$ again below 18 K.



However to reiterate, this is the interaction of the *Eu* magnetic spins on the superconducting Fe electrons, i. e. not the sometimes observed positive interaction discussed in this review between the magnetism and superconductivity on the *same* Fe electrons (see in particular Section IVA1 on the spin resonance in INS below $T_c$). Thus, the question in the FePn/Ch is whether there is coupling between the (antiferro-) magnetic and superconducting order parameters, i. e. unconventional superconductivity.

Certainly some theories (see section IV) suggest that the answer to this question is yes. There is also strong evidence experimentally for microscopic coexistence coming from the same Fe 3d electrons, particularly in Co-doped $BaFe_2As_2$ which has excellent sample homogeneity. Prozorov et al. (2009), using magneto-optic imaging of Meissner screening, find homogeneous superconductivity on a scale of 2-4 μm in $BaFe_{2-x}Co_xAs_2$ over the whole superconducting dome. Laplace et al. (2010), using NMR, find lack of electronic homogeneity down to the nanometer scale in underdoped $BaFe_{1.88}Co_{0.12}As_2$. Pratt et al. (2009a) find in their neutron scattering work that the integrated antiferromagnetic intensity in the underdoped, coexistent FePn superconductor $BaFe_{1.906}Co_{0.094}As_2$ is "substantially" reduced when superconductivity sets in at 17 K. This implies a direct coupling between the superconductivity and the magnetism, as seen in, for example, the unconventional heavy Fermion superconductor $UPt_3$ (Aeppli et al., 1988) and is consistent with (although not direct evidence of) microscopic homogeneity like reported by Prozovov et al. (2009) and inferred from thermodynamic and transport measurements (Ni et al., 2008b).

However, there are contrary data. Shen et al. (2011) argue for phase separation (islands of superconductivity) in their single crystals of 122* $K_{0.8}Fe_{1.6}Se_2$ (approximate



composition), although Shermadini et al. (2011) present μSR data arguing for microscopic coexistence of superconductivity and magnetism in single crystal $Cs_{0.8}Fe_2Se_2$. There is certainly discussion about coexistence of superconductivity and magnetism for K-doped $BaFe_2As_2$ where, as mentioned above in section IIB2b, there are sample homogeneity issues. For example, Park et al. (2009), using magnetic force microscopy and μSR measurements on $Ba_{1-x}K_xFe_2As_2$, find the magnetic and superconducting regions to be mesoscopically separated, on a scale of ~ 65 nm. Using point contact Andreev reflection spectroscopy, Lu et al. (2009) in both K-doped and Co-doped $BaFe_2As_2$ find their results also consistent with mesoscopic-scale phase separation, and no true microscopic coexistence of magnetism and superconductivity in the same electrons.

Lu et al., however raise the issue of whether this phase separation in K-doped $BaFe_2As_2$ could be due to crystalline inhomogeneity. This is the conclusion of Rotter et al. (2009) in the case of underdoped $Ba_{1-x}K_xFe_2As_2$ (which, as discussed elsewhere in this review, is known to have ±5 % K inhomogeneity). Using Mössbauer spectra of their underdoped $Ba_{1-x}K_xFe_2As_2$, Rotter et al., find at lower temperatures that all the domains in the sample are antiferromagnetically ordered. Thus, the theorists' proposals (section IV), that superconductivity in the FePn/Ch's is intimately connected with magnetism/spin fluctuations, find at least partial support from experimental measurements. Sample quality issues (see section V), particularly in the defect structure 122*'s, still need to be resolved however to draw clear conclusions on this coexistence question.

**D. $T_c$ and $T_S/T_{SDW}$ vs Pressure:**



As discussed in the Introduction, the pressure dependence of the $T_c$ of these FePn/Ch materials can be quite significant, and of interest for understanding the relative importance of various factors, e. g. lattice spacing or tetrahedral angle, that affect superconductivity. For example, as discussed above (see Fig. 7) $T_c$ scales with the a-axis spacing in REFeAsO$_{1-x}$. Thus, pressurizing REFeAsO$_{1-x}$ for the smaller rare earths Sm and Nd (which are at or below the peak in $T_c$ vs increasing a-axis lattice parameter in Fig. 7), results in a monotonic decrease in $T_c$ with increasing pressure as shown in Fig. 15. For the larger rare earths in REFeAsO$_{1-x}$ like La that are to the right of the Fig. 7 peak in $T_c$ with increasing a-axis, pressure first increases $T_c$, followed thereafter by a decrease, see Fig. 15 and Fig. 16, which focuses on $T_c$ vs pressure for LaFeAsO$_{1-x}$F$_x$. Thus far there is no evidence for pressure suppressing magnetism just at the point that superconductivity appears in those samples - such as the undoped 1111's and 122's - where pressure induces $T_c$ in a non-superconducting parent compound. In fact, several of the underdoped 1111's and undoped SrFe$_2$As$_2$ show evidence under pressure for coexistence of magnetism and superconductivity.

Technically, pressure is typically applied in the 10 to 20 kbar range (1 to 2 GPa) via a metal (often BeCu alloy) clamp arrangement, while higher pressures use some form of diamond anvil cell. The metal clamp or diamond cell contains some liquid pressure transmission medium (e. g. Daphne oil) that remains liquid (i. e. continues to give approximately hydrostatic conditions) to ~ 1 GPa upon application of pressure at room temperature. When the pressure medium solidifies upon cooling or at room temperature at higher pressures, then shear strains can occur causing possible non-reproducibility of properties in samples where shear (see discussion of CaFe$_2$As$_2$ below)



is important.  For a comparison of the effects of pressure media on the effect of $T_c$ vs P

in $BaFe_2As_2$, see Duncan et al. (2010).

    **1.) 1111 Structure:**  The pressure response of $T_c$ in electron doped $LaFeAsO_{1-x}F_x$ is positive, irregardless if the sample is underdoped, optimally doped, or overdoped as

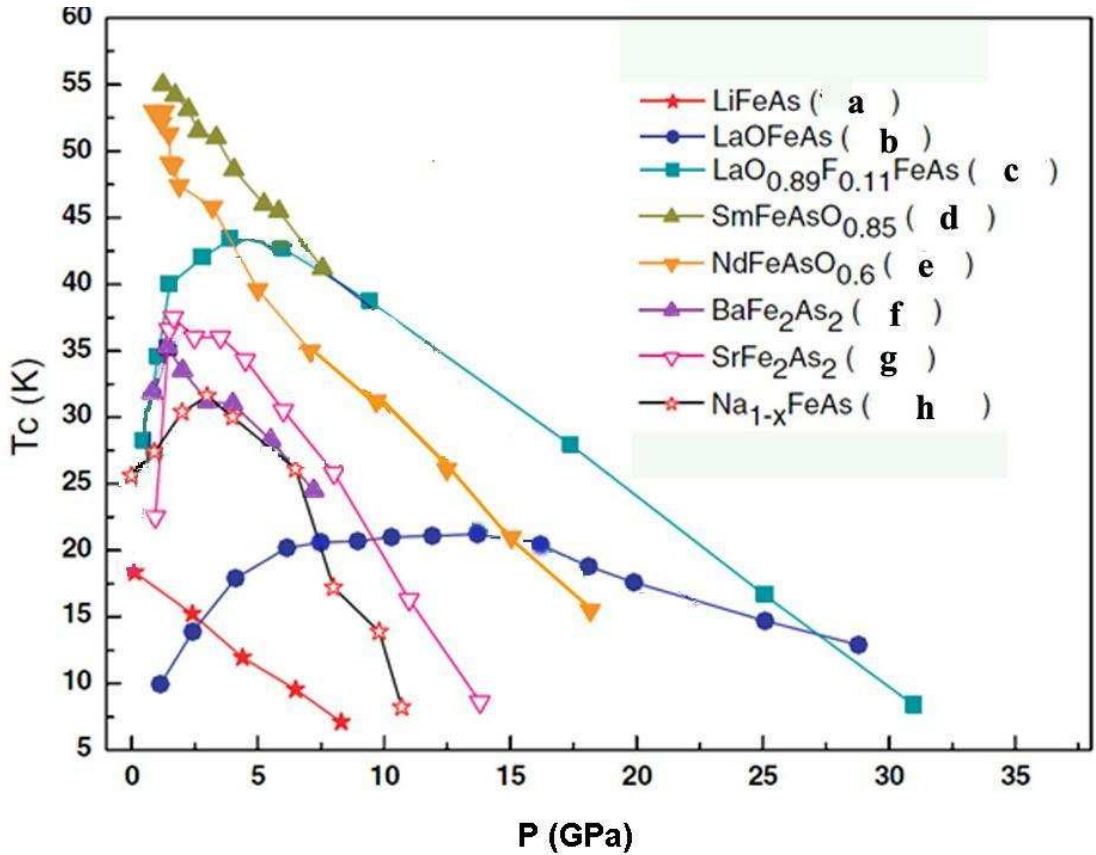

Fig. 15 (color online)  $T_c$ vs pressure in representative FePn/Ch superconductors.  As shown, while some systems undergo an initial $T_c$ increase vs pressure because pressure optimizes some controlling parameter (see discussion), a number of systems are already at their maximum $T_c$ at zero pressure.  Note the difference in the two 111 compounds. The basis for this figure is from S. J. Zhang et al. (2009a), whose data for $Na_{1-\delta}FeAs$ are shown (ref. h). The other references are a (S. J. Zhang et al., 2009b), b (Okada et al., 2008), c (Takahashi et al., 2008a), d (Yi et al., 2008), e (Takeshita et al., 2008), f (Mani et al., 2009) and g (Igawa et al., 2009).  Note that for $BaFe_2As_2$ and $SrFe_2As_2$ $T_c$ is zero until finite pressure.  For an early review of the effect of pressure on the FePn/Ch's, see Chu and Lorenz (2009).  The effects of non-hydrostatic pressure can be quite significant, see discussions of $BaFe_2As_2$ and $CaFe_2As_2$.



shown in Fig. 16. The initial slope $dT_c/dP|_{P=0} = +2$ K/GPa for x=0.05 (Takahashi et al., 2008a). For optimally doped LaFeAsO$_{0.89}$F$_{0.11}$, Takahashi et al., measured the behavior of $T_c$ with pressure all the way to 30 GPa (Fig. 15, data set (c), and Fig. 16): initially $T_c$ goes up to 43 K at 4 GPa as mentioned in the Introduction, with $dT_c/dP|_{P=0} = +3$ K/GPa, and then decreases monotonically to 9 K at the highest pressure. In a follow up work, Takahashi et al. (2008b) completed the $T_c - P$ phase diagram, Fig. 16, showing that overdoped LaFeAsO$_{0.86}$F$_{0.14}$ behaves similarly to optimally doped material, while the pressure variation of $T_c$ in undoped LaFeAsO is similar in sign but smaller in magnitude.

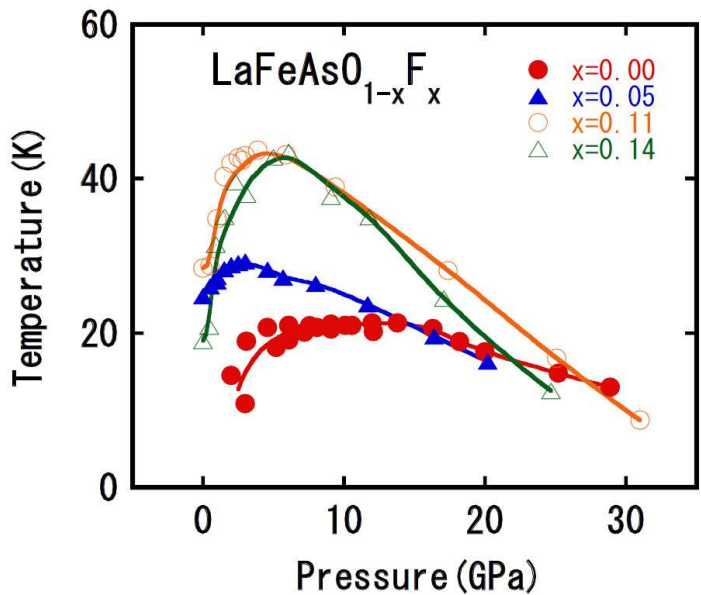

Fig. 16 (color online) $T_c$ is plotted on the y-axis vs pressure for electron doped LaFeAsO$_{1-x}$F$_x$ for various x (Takahashi et al., 2008b.) The data for x=0.0 and 0.11 are reproduced in Fig. 15 for comparison with the other FePn/Ch superconductors.

$T_c$ vs P measurements for other 1111's have returned varied results. Lorenz et al. (2008) measured SmFeAsO$_{1-x}$F$_x$ up to 1.7 GPa and found, contrary to the behavior shown in Fig. 16 for the La analog, that $T_c$ increases with pressure for undoped material, and decreases with pressure for an overdoped composition. Lorenz et al., also found that $T_S$/



$T_{SDW}$ decreases from ~ 100 K at an initial rate of 3.7 K/GPa (i. e. for a total suppression of only 6 K in the pressure range of measurement) in the underdoped $SmFeAsO_{0.95}F_{0.05}$. This is comparable to work on oxygen-deficient $NdFeAsO_{1-x}$ by Takeshita et al. (2008), where for an underdoped x=0.15 sample they find that $T_S/T_{SDW}$ decreases from ~140 K at an initial rate of 5 K/GPa. This decrease in $T_S/T_{SDW}$ in $NdFeAsO_{0.85}$, which is difficult experimentally to determine from the resistivity measured under pressure for higher pressures, is not at a high enough rate to imply suppression of $T_S/T_{SDW}$ by the time that an applied pressure of 10 GPa gives a drop in $\rho$ (but not completely to 0) at around 15 K in this material. Thus, the question of whether pressure suppresses $T_S/T_{SDW}$ in the 1111's before superconductivity appears is answered in the negative, at least in these two underdoped cases where $T_{SDW}$ could be measured.

Takeshita et al.,'s work on optimally doped $NdFeAsO_{0.6}$ shows (see Fig. 15, data set (e)) a monotonic decrease in $T_c(P=0)=53$ K with increasing pressure up to their maximum pressure of 18 GPa since, as already discussed, Nd is a smaller rare earth, vs Takahashi et al.'s (2008b) result (Fig. 16) of initial increase in $T_c$ with applied pressure for the large La in optimally doped $LaFeAsO_{0.89}F_{0.11}$. Further, Takeshita et al., find that $T_c$ for underdoped $NdFeAsO_{0.8}$ decreases from $T_c(P=0)=41$ K monotonically with increasing pressure, contrary to Lorenz et al.'s (2008) result that pressure increases $T_c$ in underdoped $SmFeAsO_{1-x}F_x$ even though Sm is smaller than Nd.

**2.) 122:** Interestingly, the inducement of superconductivity via application of pressure in the undoped $MFe_2As_2$ mother compounds revealed important differences between M=Ca and the other $MFe_2As_2$ – yet one further example of the richness and



variety of behavior in the FePn/Ch - that would perhaps have remained unknown without the application of pressure.

BaFe$_2$As$_2$ was reported (Alireza et al., 2009) to become superconducting with T$_c^{max}$ at ~ 29 K at P=4.5 GPa with no superconductivity below 2.8 GPa vs Mani et al., see Fig. 15, who report T$_c^{max}$ ~ 35 K at 1.5 GPa.  Both works involved single crystals. Kimber et al. (2009) report T$_c$=31 K for P=5.5 GPa, in somewhat better agreement with Mani et al.   Interestingly, Kimber et al. (2009) find – using neutron powder diffractometry - that, just as Lee et al. (2008) point out at *zero* pressure for 1111 and 122 FePn superconductors as a function of doping, that the maximum T$_c$ in their pressure work on BaFe$_2$As$_2$ corresponds to the pressure where the FeAs$_4$ tetrahedra are regular, with an angle of 109.47 $^o$.  At zero pressure, the irregular tetrahedra in undoped BaFe$_2$As$_2$ have a As-Fe-As bond angle of 108.5 $^o$.   Kimber et al., note that the structural phase transition in BaFe$_2$As$_2$ appears to be suppressed with increasing pressure at ~1.3 GPa *before* superconductivity appears around 2.2 GPa.  However, Fukazawa et al. (2008), using NMR measurements on polycrystalline material up to 2.5 GPa and resistivity measurements up to 9 GPa, argue that T$_{SDW}$ is suppressed only slowly with pressure, about -6.7 K/GPa, and is still finite (> 70 K) over the entire pressure region (2.2-6 GPa) of Kimber et al.'s superconducting dome.

Thus, due to the difficulty of the experimental technique, pressure measurements sometimes return conflicting results.  In the case of BaFe$_2$As$_2$ (see also the discussion of CaFe$_2$As$_2$ below), Yamazaki et al. (2010) use a quite hydrostatic cubic anvil apparatus up to 14 GPa on single crystals.  They argue that the earlier results (including the data shown in Fig. 15, data set (f)) were strongly affected by a small uniaxial stress along the c-axis



under non-hydrostatic conditions, stabilizing islands of tetragonal phase and causing filamentary superconductivity. They find no coexistence of magnetism and superconductivity, and state that $T_{SDW}$ is suppressed only at 10 GPa (consistent with the NMR results of Fukazawa et al., 2008), with superconductivity occurring between 11 and 14 GPa and $T_c^{max}$=13 K (not > 30 K) at 11.5 GPa.

Alireza et al. (2009) further report $T_c^{max}$ ~ 27 K at P=3.2 GPa for $SrFe_2As_2$, while Takahashi et al. (2008b) found $T_c$(~4 GPa) for $SrFe_2As_2$ to be 34 K, in agreement with Kotegawa, Sugawara and Tou (2009) and Igawa et al. (2009), the latter data being displayed in Fig. 15, data set (g). Kotegawa, Sugawara and Tou (2009) were able – unlike most pressure works - to measure a fairly complete set of $T_S/T_{SDW}$ values vs pressure and formed a phase diagram vs pressure where $T_S/T_{SDW}$ was still finite (at ~105 K) after superconductivity was already induced at around 3.6 GPa. Thus, their phase diagram was similar to those with doping discussed above (e. g. $Ba_{1-x}K_xFe_2As_2$ or $SrFe_{2-x}Rh_xAs_2$, Figs. 11 and 13 respectively) where $T_S/T_{SDW}$ does not join or intersect the superconducting dome, and provides another example of coexistence of magnetism and superconductivity.

Uhoya et al. (2010) report $T_c$ vs pressure for $EuFe_2As_2$, with $T_c$=22 K at 2 GPa rising up to $T_c^{max}$ = 41 K at 10 GPa, the highest pressure-induced $T_c$ of any of the undoped 122 parent compounds. Note that Eu undergoes a valence change to non-magnetic $Eu^{3+}$ between 3 and 9 GPa, the pressure region where $T_c$ rises monotonically with increasing pressure.

Concerning $CaFe_2As_2$, first reports (Torikachvili et al., 2008) for $T_c$(P) for $CaFe_2As_2$ showed a superconducting dome that started at the much lower pressure,



compared to M=Ba,Sr, of 0.23 GPa, with a peak at only $T_c$=12 K at 0.5 GPa. In addition, this pressure work on CaFe$_2$As$_2$ found a new, additional transition (identified later, Kreyssig, et al., 2008, as a "collapsed" tetragonal structure) at ~100 K that appeared at 0.55 GPa and moved to higher temperature with increasing pressure. Park et al. (2008) also found superconductivity in CaFe$_2$As$_2$, with $T_c$~13 K at 0.69 GPa. After significant further work, the sensitivity of the structural transitions to different pressure conditions was solved (Yu et al., 2009) using helium gas as a more nearly perfect hydrostatic pressure medium (cf. the discussion of $T_c$(P) in BaFe$_2$As$_2$ above). The result is that, under improved hydrostatic conditions, there is actually no superconductivity in CaFe$_2$As$_2$ under pressure up to 0.6 GPa, i. e. the previous observations of superconductivity were due to shear stress from the pressure medium. The new structural phase transition (found in the hydrostatic helium to be at 0.4 GPa rather than the originally reported 0.55 GPa) is hysteretic in both temperature and pressure.

**3.) 111:** Gooch et al. (2009) report a monotonic decrease of $T_c$ with increasing pressure in LiFeAs at a rate of 1.5 K/GPa, in agreement with the data shown in Fig. 15 from S. J. Zhang et al. (2009b), data set (a). In Na$_{1-\delta}$FeAs, S. J. Zhang et al. (2009a) report an increase of $T_c$ from 26 K up to 31 K at 3 GPa, followed by a sharp decrease down to $T_c$=8 K by 11 GPa, Fig. 15, data set (h). Presumably LiFeAs under pressure behaves differently from Na$_{1-\delta}$FeAs due to the smaller ionic radius of Li vs Na, i. e. LiFeAs is already "pre-compressed" (S. J. Zhang et al., 2009b). The $T_c$=6 K phosphorous analog of LiFeAs, LiFeP, discovered by Deng et al. (2009) has been studied under pressures up to 2.75 GPa by Mydeen et al. (2010). $T_c$ declines monotonically with increasing pressure at a rate of 1.2 K/GPa, similar to the result for LiFeAs.



**11:**  As mentioned in the Introduction, Margadonna et al. (2009b) found that 7 GPa increases the $T_c$ of FeSe from 8 K at zero pressure up to 37 K, with $T_c$ already 27 K at 2.6 GPa, followed by a decrease down to 6 K as pressure increases to 14 GPa.   FeSe has a much larger compressibility (~twice that of LiFeAs, ~three times that of the 1111's) than the other FePn/Ch superconductors, at least partially explaining the large response of $T_c$ to pressure.  However, the explanation of Kimber et al. (2009) for their observed maximum in $T_c$ vs pressure for $BaFe_2As_2$ (which was called into question because of implied non-hydrostatic effects by Yamazaki et al., 2010) – that the tetrahedral bonding angle approached the optimal $109.47^o$ at that pressure – does not hold for the work of Margadonna et al., on FeSe.  They observe rather that the tetrahedral bonding angle in FeSe, which starts around $111.5^o$, *increases* monotonically with pressure, leaving changes in the band structure with the much changed interatomic spacing with pressure (the c-axis contracts by 7.3% at 7.5 GPa vs 4% at 6 GPa in $BaFe_2As_2$, Kimber et al., 2009) as a possible  explanation.  Another possible explanation for the enhanced $T_c$ with pressure of FeSe was pointed out by Imai et al. (2009), who found in an NMR study that applied pressure enhances spin fluctuations (proportional to $1/T_1T$) above $T_c$.

A positive enhancement of $T_c$ with increasing pressure has also been found in $FeSe_{1-x}Te_x$ for x=0.43 (Gresty et al., 2009) and 0.50 (Horigane et al., 2009) with an increase of $T_c$ from ~ 15 K at zero pressure up to ~25 K at 2 GPa, while for x=0.75 (Mizuguchi et al., 2010b) the $T_c$ enhancement at 1 GPa is only ~ 1.5 K, (see Mizuguchi and Takano, 2010 for an overview of the FeCh).

**21311:**  Sato et al. (2010) found that pressure monotonically increased the $T_c$ of $Sr_2Mg_{0.3}Ti_{0.7}O_3FeAs$ from 37 K at P=0 up to 43 K at 4.2 GPa.  Kotegawa et al. (2009)



showed that 4 GPa increased $T_c$ of $Sr_2VO_3FeAs$ from 36 K (P=0) to 46 K, while the same pressure decreased the $T_c$ of $Sr_2ScO_3FeP$ from 16 K (P=0) to 5 K. The authors discuss this difference in pressure effect as being due to the height of the pnictogen, as discussed in the theory of Kuroki et al. (2009) discussed above in Section IIA.

**122\*:** Guo et al. (2011b) report that $T_c$ in $K_{0.8}Fe_{1.7}Se_2$ remains constant with pressure at $\approx$32 K up to 1 GPa, and then falls monotonically to 0 at around 9.2 GPa. Seyfarth et al. (2011) report that $T_c$ in $Cs_{0.8}Fe_2Se_2$ is approximately constant at $\approx$30 K also up to 1 GPa, and then falls monotonically to $T_c^{onset} \approx$12 K at 7.5 GPa.

### E. $T_c$ vs Magnetic Field:

Measuring the upper critical field of a superconductor, $H_{c2}(T)$, has impact not only on potential applications, but also helps the understanding of the superconductivity. The upward curvature of $H_{c2}(T) \parallel$ c-axis with temperature in both the 1111 and 122 FePn superconductors has been interpreted as consistent with the existence of two superconducting gaps, while the size of $H_{c2}(T \rightarrow 0)$ (60-400 T in the 1111's, depending on sample and crystal orientation) is consistent with strong coupling (Jo et al., 2009), see following discussion. Two straightforward models are commonly used to fit the $H_{c2}$ data and extract qualitative conclusions, sometimes followed by more intricate analysis involving, e. g., two band models and more adjustable parameters. The weak coupling Werthamer, Helfand, and Hohenberg (1966), WHH, model assumes that $H_{c2}$ is limited at higher fields and lower temperatures by spin orbit pair breaking in addition to spin paramagnetic effects (where alignment of the spins in the applied field breaks the pairs.) Contrary wise, when spin paramagnetism pair breaking effects dominate those from spin orbit coupling, then the Pauli paramagnetic limiting model is used. Qualitatively (see, e.



g., the original paper by WHH), Pauli paramagnetic limiting being the dominant mechanism over spin orbit effects causes saturation ("flattening") of the upper critical field at lower temperatures/higher fields ($T_c(H)/T_c(H=0) \leq 0.2$-$0.4$. Because paramagnetic limiting is isotropic, a stronger effect is found in the higher critical field direction (H in plane in the FePn/Ch) which reduces the anisotropy in the two field directions at lower temperatures (Putti et al., 2010). As discussed below, this reduction in the $H_{c2}(\parallel ab)/H_{c2}(\perp ab)$ anisotropy at higher fields/lower temperatures is indeed often found in the FePn/Ch. When the upper critical field data qualitatively shows such saturation, but $H_{c2}(T=0)$ exceeds the weak coupling BCS paramagnetic limit ($\mu_0 H_p^{BCS}=1.84\ T_c$, where $H_p^{BCS}$ is in units of T and $T_c$ has units of K) – which for the observed high values of $H_{c2}(0)$ in the FePn/Ch is often the case, then enhancements of the weak coupling BCS paramagnetic limit due to strong coupling effects (proportional to $1+\lambda$, where $\lambda$ is the strength of the coupling) can be considered (Schlossmann and Carbotte, 1989). Thus, measurements of $H_{c2}(0)$ are often used as evidence for strong coupling effects being present (see e. g. Jo et al., 2009).

A more difficult measurement, that of the temperature and orientation dependence of the lower critical field (where flux first penetrates the superconductor), $H_{c1}(T)$ (~ 10 mT as T→0)), of an underdoped, oxygen deficient single crystal of $PrFeAsO_{0.9}$, $T_c = 35$ K, also was interpreted as consistent with multiple gap superconductivity (Shibauchi et al., 2009).

**1.) 1111 Structue:** The excitement of the discovery of high $T_c$'s in $LnFeAsO_{1-x}F_x$, where Ln started with La and then progressed rapidly to the smaller rare earths like Sm and Nd, was fed by the early measurements of very high upper critical fields, $H_{c2}(T)$,



required to extinguish superconductivity in these compounds. Using DC fields of up to 45 T, Jaroszynski et al. (2008) reported $H_{c2}(T)$ data for optimally doped polycrystalline LaFeO$_{0.89}$F$_{0.11}$ ($T_c$=28 K), SmFeAsO$_{0.85}$ ($T_c$=53.5 K) and NdFeAsO$_{0.94}$F$_{0.06}$ (50.5 K), finding already $H_{c2}(0)$ of 60 T for the lowest $T_c$ sample. Jia et al. (2008), measuring single crystal NdFeAsO$_{0.82}$F$_{0.18}$, $T_c$=52 K, at low (up to 9 T) fields found -d$H_{c2}$(T)/dT|$_{T=Tc}$ = 9 T/K for field in the ab-plane, and 1.85 T/K for field in the c-axis direction, i. e. an anisotropy of only about 5. Using the WHH formula ($H_{c2}(0)$ = - 0.69*$T_c$* d$H_{c2}$(T)/dT|$_{T=Tc}$) Jia et al., calculated $H_{c2}(0)$ in the two field directions of ~300 and 66 T respectively. Using data up to 45 T on a similar crystal (NdFeAsO$_{0.7}$F$_{0.3}$, $T_c^{mid}$ = 47.4 K), Putti et al. (2010) find the critical field slopes at $T_c$ (10.1 and 2.1 T/K for H⊥c and H∥c respectively) to give a similar anisotropy, and calculate the coherence lengths in the ab plane and c-axis directions to be 1.8 and 0.45 nm respectively. These are quite short compared to the penetration depth - determined from various methods (see, e. g., Luan et al., 2010) to be in the 100's of nm.

  **2.) 122 Structure:** Critical field studies on single crystal Ba$_{0.6}$K$_{0.4}$Fe$_2$As$_2$ ($T_c$=29 K) up to 45 T by Jo et al. (2009), on single crystal Ba$_{0.6}$K$_{0.4}$Fe$_2$As$_2$ ($T_c$=28 K) up to 60 T by Yuan et al. (2009), on single crystal BaFe$_{2-x}$Co$_x$As$_2$ (x=0.076, 0.094, 0.116, 0.148, 0.20, 0.228; $T_c$=7, 15, 23, 22, 17, 8 K) up to 35 T by Ni et al. (2008b) and on single crystal BaFe$_{2-x}$Co$_x$As$_2$ (x=0.20; $T_c$=22 K) up to 35 T by Putti et al. (2010) allow several conclusions. Unlike the 1111's but like the 11's discussed below, the anisotropy for $H_{c2}(T)$ for the 122's is only about 2-3 near $T_c$ and essentially vanishes as T→0. The possible reasons for such isotropic $H_{c2}(0)$ values, which are in strong contrast to the cuprates, is still under discussion but include band warping in the cylindrical Fermi



surfaces (see section IVA2 which discusses ARPES meaurements of the Fermiology) or multiband effects (Khim et al, 2010). Also unlike the 1111's, whose resistive transitions broaden significantly with field presumably due to vortex depinning/dissipation, the transition widths in 122's remain fairly narrow with increasing field and merely shift downwards in temperature with increasing field. A comparison of $H_{c2}(T)$ graphs for 1111 $NdFeAsO_{0.7}F_{0.3}$ and 122 Co-doped $BaFe_2As_2$ (Putti et al., 2010) shown in Fig. 17 makes this latter comparison visually very clear. The critical fields extrapolated to T=0, whether via the WHH formula or via $H_{c2}(T)=H_{c2}(0)(1-(T/T_c)^2)$, for the 122's just as for the 1111's exceed the weak-coupling Pauli paramagnetic limiting field, $H_P=1.84k_BT_c$. Thus, the pairing breaking effect of the magnetic field is qualitatively more dominated by orbital effects (WHH model) than by spin alignment effects (Pauli limit), although consideration of the detailed interplay of the two scaled by the Maki parameter $(\alpha=\sqrt{2}H_{c2}^{WHH}(0)/H_P)$ can bring more quantitative understanding (see, e. g., Kida et al., 2009.)

**3.) 111 Structure:** Song et al. (2010) measured the critical fields up to 9 T in single crystal LiFeAs, $T_c=19.7$ K, and found via the WHH formula $H_{c2}(0)=83$ and 72 T for field in the ab-plane and c-axis directions respectively. In addition to this small anisotropy, they found a lack of curvature in the measured $H_{c2}(T)$ curves where, as discussed above, curvature in $H_{c2}(T)$ was discussed as consistent with multi-gap superconductivity. They also found significant broadening of the transition with increasing field, consistent with vortex dissipation. Sasmal et al. (2010), in their measurements of $H_{c1}(T)$ for single crystal LiFeAs found, on the other hand, evidence for a two band gap scenario. G. F. Chen et al. (2009) measured the critical fields up to 14 T



in a single crystal of Na$_{1-\delta}$FeAs, although T$_c$ was only 15 K and there was no measurable ΔC anomaly at T$_c$. Using the WHH formula they found H$_{c2}$(0)=60 and 30 T for field in the ab-plane and c-axis directions respectively.

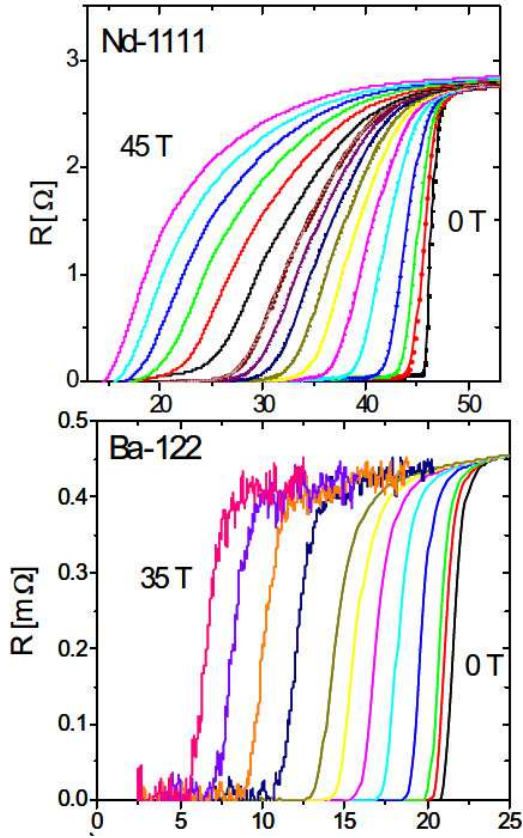

Fig. 17 (color online) Transitions into the superconducting state as measured by the resistivity as a function of field for single crystals of NdFeAsO$_{0.7}$F$_{0.3}$ (upper panel) and BaFe$_{1.8}$Co$_{0.2}$As$_2$, measured with field along the c-axis direction (Putti et al., 2010).

**4.) 11 Structure:** Putti et al. (2010) measured H$_{c2}$(T) of a single crystal of FeSe$_{0.5}$Te$_{0.5}$, T$_c$=15 K, up to 32 T. Broadening of the transition with field was observed, not as severe as in the 1111's (Fig. 17) but much more than seen in the 122's. The critical field slope at T$_c$, -dH$_{c2}$(T)/dT|$_{T=Tc}$, was found to be the very high value of 25 T/K for field in the ab-plane, and 14 T/K for field in the c-axis direction, giving an anisotropy



of 2 close to $T_c$. This anisotropy decreases at 32 T ($T_c \sim 11$ K) to ~1 due to downward (not concave upwards) curvature for the ab-plane field direction. Due to these high slopes, despite the lower $T_c$, the WHH formula gives $H_{c2}(0)=260/145$ T for field in the ab-plane/c-axis, again exceeding the weak-coupling Pauli paramagnetic limiting field as just discussed for the 122's. $H_{c2}(T)$ measurements (Braithwaite et al., 2010) of a single crystal of $FeSe_{0.48}Te_{0.52}$, $T_c=15$ K, in pulsed fields up to 46 T confirmed the decreasing anisotropy reported by Putti et al. (2010). In fact, the two curves for ab-plane/c-axis cross at about 41 T (4 K) and $H_{c2}(T\rightarrow0)$ for field in the ab-plane is in fact *smaller* than for in the c-axis direction. This crossing of the $H_{c2}$ curves for the ab- and c-directions was confirmed in DC measurements to 45 T in a single crystal of $FeSe_{0.4}Te_{0.6}$ by Khim et al. (2010).

**5.) 21311 Structure:** Measurements (Sefat et al., 2010) up to 10 T in $Sr_2VO_3FeAs$, $T_c \sim 33$ K, give a value of $dH_{c2}/dT|_{Tc}$ = -9 T/K. By using the WHH formula, this gives $H_{c2}(0)\approx200$ T, comparable with values by Zhu et al. (2009b).

**6.) 122* Structure:** C.-H. Li (2010) in single crystal $Rb_{0.8}Fe_2Se_2$, $T_c\approx31$ K, report $-dH_{c2}/dT|_{Tc}$ values of 6.78 T/K for field in the ab-plane and 1.98 T/K for field along the c-axis, resulting (using the WHH formula) in $H_{c2}(0)$ values of 145 and 42 T respectively.



## III. Structural and Electronic Properties, Part Two – Normal State ρ, χ, C down to T_c

The present section focuses on the normal state from which the superconducting state forms, with the magnetic and structural transitions already discussed in Section II. Measurement of the resistivity and susceptibility, and to a lesser extent (due to its greater difficulty) the specific heat, is often used to indicate, via anomalies in these parameters, the progression with doping of the structural and magnetic anomalies discussed in section II, as shown here in Fig. 18. Such measurements allow a more rapid estimate of the part of the phase diagram of particular interest in a given study, which can then be further examined with more microscopic measurement techniques (e. g. x-ray diffraction, neutron diffraction, μSR, Mössbauer.)

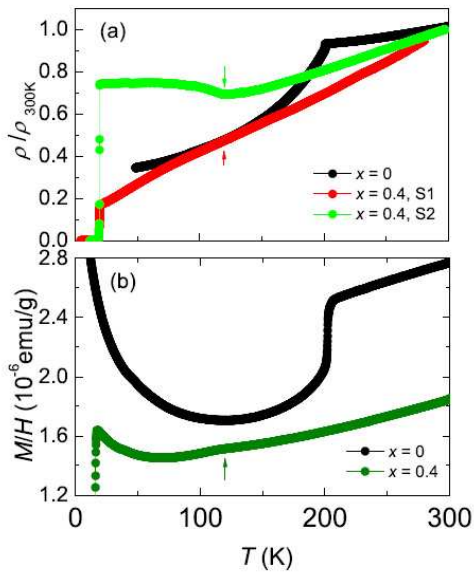

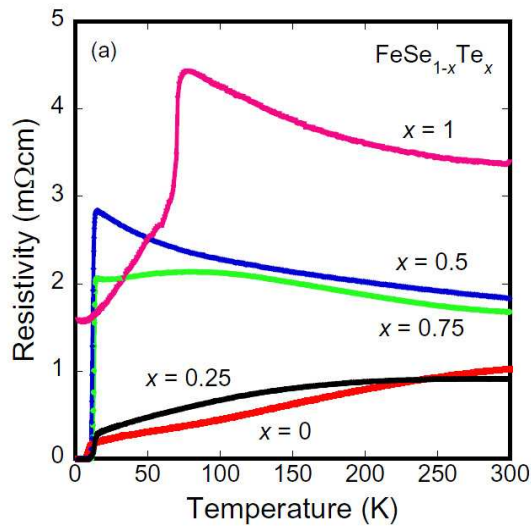

Fig. 18 (color online) Resistivity (upper panel) and magnetic susceptibility (lower panel) of single crystal SrFe$_{2-x}$Co$_x$, x=0 and 0.4 (J. S. Kim et al., 2009b). Arrows mark anomalies for x=0.4. $2*10^{-6}$ emu/g is 0.7 memu/mole. Note the sample dependence in ρ for x=0.4, samples S1 and S2.

Fig. 19 (color online) Resistivity vs temperature of polycrystalline FeSe$_{1-x}$Te$_x$ (Mizuguchi et al., 2009). Note the anomaly at 72 K in pure FeTe (upper curve) at T$_S$/T$_{SDW}$.



In addition, the residual resistivity ratio (RRR), defined as $\rho(300\text{ K})/\rho_0$

($\rho_0 \equiv \rho(T \to 0)$, where the extrapolation to T=0 is from the normal state above $T_c$ if the

sample is superconducting), serves as an important indicator of sample quality since

scattering from impurities increases the residual resistivity $\rho_0$. Similarly, the sharpness

and size of the specific heat anomaly at the superconducting transition, $\Delta C(T_c)$,

(discussed below in Section IIIB3) also serves as a commonly used indicator of the

quality of a sample.

A third use for these normal state measurements is that their temperature

dependence can provide insights useful for understanding the superconductivity. For

example, the temperature dependence of the resistivity in the normal state has been used

in the study of the FePn/Ch superconductors to determine nearness to quantum criticality

in so far as $\rho$ does not follow Fermi liquid behavior. Landau Fermi liquid behavior is

$\rho = \rho_0 + AT^2$, with 'A' a constant. Quantum critical behavior can occur (see Stewart,

2001; Stewart, 2006; von Löhneysen et al., 2007) at (or near) the point in a phase

diagram where a second order phase transition, e. g. antiferromagnetism, has been

suppressed to T=0. In the case of the FePn/Ch, $T_{SDW}$ being suppressed to T=0 by doping

(section IIB) is an obvious pathway to such quantum critical behavior, with the associated

non-Fermi liquid behavior at finite temperatures – including long range magnetic

fluctuations potentially important for understanding superconductivity.

**A. Resistivity and Susceptibility**

Some representative examples of the measurements are offered here to give an idea of

the common behavior. The references given in Section II in the discussion of materials



and their phase diagrams can also be followed to learn more about the various normal state properties of a particular compound.

In general, the resistivities of the FePn/Ch superconductors are metallic in their temperature dependence ($d\rho/dT > 0$) as seen in Fig. 18 for pure and Co-doped $SrFe_2As_2$ and for FeSe in Fig. 19, although $FeSe_{1-x}Te_x$, x>0.25, provides counterexamples to this metallic behavior.  Also, as a function of composition in the 122*, there can - depending on whether the composition is optimized for metallic and superconducting behavior - be a 'hump' in the resistivity peaked at around 150 K, where $\rho$ rises over a large maximum when cooling from room temperature to $T_c$.  For samples in the insulating composition range of the phase diagram in the 122*'s, $\rho$ continues to rise with decreasing temperature, while optimized samples near in composition to $A_{0.8}Fe_{1.6}Se_2$ show $\rho$ decreasing monotonically (i. e. no 'hump') between room temperature and $T_c$ (Bao et al., 2011b), with decent RRR values (>40, D. Wang et al., 2011; ≈20, Luo et al., 2011).

In all cases the absolute values at room temperature for the FePh/Ch are high, ≈ 1 m$\Omega$-cm (≥50 m$\Omega$-cm for the 122*), where for a good metal (e. g. Cu or Ag) $\rho$ ~ 1 $\mu\Omega$-cm. A band structure calculation (Singh, 2009) for FeSe results in small Fermi surface sections, resulting in a semi-metallic classification, although in general the FePn/Ch are called metallic.   In the beginning of the study of the FePn/Ch the question of itinerant metal vs localized insulator (weak Coulomb repulsion U vs strong) was important for deciding how to understand the physics of these materials, see a discussion by Tesanovic (2009).  Rather early on, the xray measurements of Yang et al. (2009b), as discussed in the Introduction, indicated that the 1111 and 122 FePn/Ch are actually similar to Fe metal, with relatively (compared to the bandwidth) small Coulomb correlation U, an even



smaller Hund's coupling (diminishing the tendency to form large local moments), Fe 3d

hybridized bands, and metallic behavior. Singh points out in general that for the FePh/Ch

materials, the small carrier density (which gives the high values of ρ), does not imply a

small density of states, N(0) (in units of states/eV-atom), at the Fermi energy, which in

fact turns out (see discussion of the specific heat γ, proportional to N(0), below) to be

relatively high compared to, e. g., the cuprates. This affects the scaling of ΔC/$T_c$, see

discussion in section IIIB3.

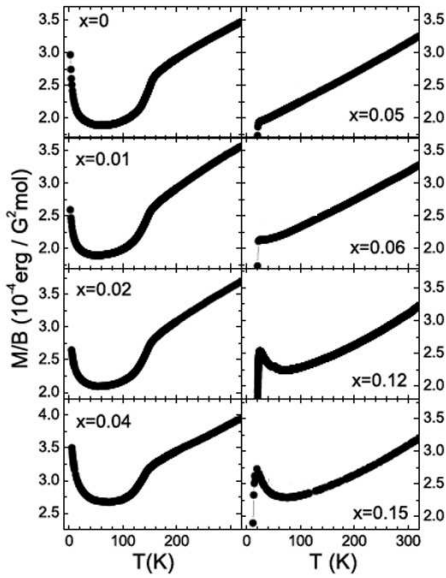
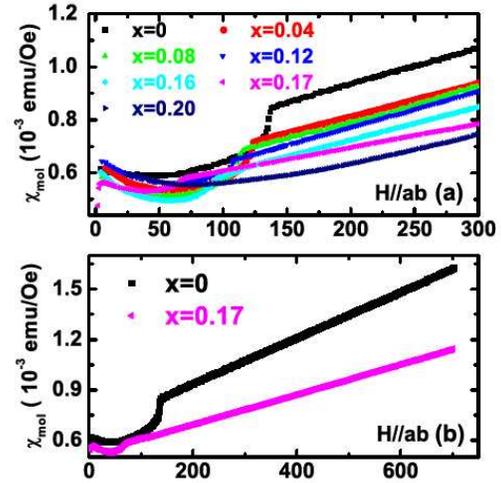

Fig. 20. Magnetic susceptibility for LaFeAsO$_{1-x}$F$_x$, 0 ≤ x ≤ 0.15, by Klingeler et al. (2010). Note the large anomalies at $T_{SDW}$ up to x=0.04.

Fig. 21 (color online) Magnetic susceptibility for BaFe$_{2-x}$Co$_x$As$_2$, X. F. Wang et al. (2009b). Note that the linearity with T, χ ~ T, disappears abruptly for x=0.20.

The magnetic susceptibility, χ, shows a large anomaly at $T_{SDW}$ in the FePn/Ch

structures (see Figs. 18, 20-21 for examples) where this transition exists (the 1111's, the

122's, the 122*'s and some of the 11's). Perhaps more importantly, χ data when taken

above $T_{SDW}$ (not yet the case in the 122* with their > 500 K ordering temperature) give

an idea about the magnetic fluctuations. As was discussed in Section II, the structural



transition occurs at *higher* temperature than $T_{SDW}$ in the 1111's, and in the 122's after doping on the Fe or Pn sites splits the two transitions (Figs. 9-10, 12, 14). However, a number of early theories (Yildirim, 2008; Mazin and Johannes, 2009) suggest that the lower transition temperature magnetism causes the structural transition through a fluctuating magnetic state without long range order (see Singh, 2009 for a discussion.) Profiting from development in understanding of the magnetic state, this argument was later refined (see, e. g., Nandi et al. (2010), Fernandes et al. (2010b)) to argue that the structural transition is caused by nematic magnetic fluctuations which break the tetragonal a-b axis symmetry as described above in Section IIB2b where the reversal of the phase boundary in Ba(Fe$_{1-x}$Co$_x$)$_2$As$_2$, x$\approx$0.06, was presented. Cano et al. (2010) discuss experiments from the point of view of their Ginzburg-Landau theory to further investigate whether magnetic fluctuations drive the structural transformation.

Instead of the above explanation for the cause of $T_S$, a number of theories (see, e. g., Lv, Wu and Phillips, 2009; Turner, Wang and Vishwanath, 2009; Lee, Yin and Ku, 2009) propose instead that orbital ordering plays an important role for understanding the structural order. The five Fe d-orbitals include two (the d$_{xz}$ and d$_{yz}$) in directions that are asymmetric in the xy plane and thus could play a role in the tetragonal-orthorhombic distortion. If the orbitals in either of these two directions order, then the total energy is lowered, thus inducing the structural phase transition. ARPES (Shimojima et al., 2010) and optical experiments (Akrap et al., 2009; Dusza et al., 2010) have been cited as consistent with orbital ordering below the magnetic transition. Yet another explanation for the structural transition involves a local Fe-moment picture described as the 'Hund's rule correlation' model (see Ji, Yan, and Lu, 2011 and references therein.)



The other temperature range where the magnetic susceptibility and its temperature dependence might shed light on the underlying physics would be at low temperatures, where the resistivity for some systems indicates quantum criticality. There are known (Stewart, 2001; Stewart, 2006; von Löhneysen et al., 2007) temperature dependences in $\chi$ below $\approx$20 K that would be worthwhile to compare to the $\rho$ data. Unfortunately, samples of the FePn/Ch appear almost uniformly to have at least some trace impurity phases that are magnetic, e. g. FeAs, $Fe_3O_4$, Fe (all of which also affect the low temperature specific heat discussed in Section IIB3b below), which prevent the detailed analysis of the intrinsic low temperature temperature dependence of $\chi$.

**1.) 1111 Structure:** Kamihara et al. (2008) in their discovery of superconductivity in $LaFeAsO_{1-x}F_x$ reported that the undoped LaFeAsO resistivity was approximately temperature independent at 5 m$\Omega$-cm, with an anomaly at 150 K and an upturn below 100 K. Upon fluorine doping, the upturn in $\rho$ below 100 K decreases and by x=0.11 resistivity falls uniformly from room temperature (metallic behavior) with an RRR of ~5.

Kamihara et al. (2008) report that the susceptibility of LaFeAsO is about 0.4 memu/mole and roughly temperature independent below room temperature except for the 150 K anomaly and an upturn below ~25 K. McGuire et al. (2008), with an expanded set of data for $\chi$ of LaFeAsO, show that $\chi$ increases with increasing temperature above the anomaly up to room temperature by about 30%. Klingeler et al. (2010) extend $\chi$ for LaFeAsO up to 500 K, showing that $\chi$ continues to rise almost linearly up to the highest temperature of measurement. Klingeler et al., also find the same general behavior of $\chi$ increasing monotonically (see Fig. 20) starting at either $T_{SDW}$ (x<0.05) or $T_c$ (x$\geq$0.05) up



to 300 K for seven additional compositions of $LaFeAsO_{1-x}F_x$, $0 \leq x \leq 0.15$. Note however that, as shown in Fig. 20, the linearity of $\chi$ with temperature does not hold for x=0.12 and 0.15.

G. M. Zhang et al. (2009) consider the data in Fig. 20, along with similar data for $MFe_2As_2$ (M=Sr – see Fig. 18, Ca - G. Wu et al., 2008a, and Ba) above the respective $T_{SDW}$'s up to 300 K (the linearity of $\chi$ vs T for M=Sr extends up to 600 K – Mandrus et al., 2010, and for M=Ba up to 700 K as shown in Fig. 21 above), as evidence for a "universal" $\chi \sim T$ dependence in the FePn/Ch. These authors compare these results to theory for a Heisenberg antiferromagnet (Chubukov and Sachdev, 1993) and to $\chi$ data for Cr which are approximately linear with T from 300-900 K (Fawcett et al., 1994) as evidence for strong (antiferro-) magnetic fluctuations above $T_{SDW}$ (and indeed, as seen in Fig. 20 for $LaFeAsO_{1-x}F_x$, above $T_c$ even after $T_{SDW}$ is suppressed for x>0.04) in the FePn/Ch superconductors. Corroborating evidence for the "universal" $\chi \sim T$ behavior proposed by G. M. Zhang et al. (2009), but not remarked on by them, is the close to linear-in-temperature behavior of $\chi$ between $T_{SDW}$=180 K and room temperature reported by Tegel et al. (2008a) in SrFeAsF. For further discussion of this $\chi \sim T$ behavior, see Korshunov et al. (2009) and Sales et al. (2010).

For oxygen deficient $LnFeAsO_{1-x}$ polycrystalline samples prepared under high pressure (Miyazawa et al., 2009), again $d\rho/dT$ is positive (metallic behavior), $\rho(300 \text{ K}) \sim$ 2 m$\Omega$-cm, and the RRR values range from ~9 for La and ~5 for Ce, to over 20 for Sm, Gd, Pr, and Nd. For high-pressure-prepared single crystal $PrFeAsO_{0.7}$, $T_c$=35 K, Kashiwaya et al. (2010) report an anisotropy $\rho_c/\rho_{ab}$ = 120 at 50 K, which is comparable to



the transport anisotropies discussed below for single crystals of the other structures. Hole doped $La_{1-x}Sr_xFeAsO$ shows metallic behavior in $\rho$ vs T below 200 K, RRR ~ 5 (Mu et al., 2008a). Polycrystalline $Gd_{0.8}Th_{0.2}FeAsO$, $T_c$=56 K, has RRR~5 and a magnetic susceptibility that increases below room temperature up to ~0.27 emu/mole at $T_c$ (C. Wang et al., 2008).

**2.) 122 Structure:** Measurements of polycrystalline $BaFe_2As_2$ (Rotter, Tegel, and Johrendt, 2008) gave essentially constant $\rho$ vs T from room temperature down to the $T_S/T_{SDW}$ transition, followed by a monotonic fall off of $\rho$ to lower temperatures with an RRR~5, while $Ba_{0.6}K_{0.4}Fe_2As_2$ is metallic in behavior ($d\rho/dT$>0) down to $T_c$, with RRR~17. The same qualitative resistivity vs temperature behavior as seen in undoped $BaFe_2As_2$ is also seen in $SrFe_2As_2$, RRR=3 (Saha et al., 2009b) – see Fig. 18 - and $EuFe_2As_2$ RRR=3 (Jeevan et al., 2008a). With single crystals, the anisotropy $\rho_c/\rho_{ab}$ at 300 K in the $MFe_2As_2$ for M=Ba, Sr, Ca, and Eu has been determined to be 150 (X. F. Wang et al., 2009a), 80 (G. F. Chen et al., 2008), $\geq$50 (Ronning et al., 2008), and 7 (D. Wu et al., 2009) respectively. Sample quality also plays an important role, Krellner et al. (2008) report RRR=32 for $SrFe_2As_2$ and Rotundu et al. report RRR=36 in single crystal $BaFe_2As_2$ after 30 days of annealing at 700 $^o$C, vs the usual RRR=5 for the unannealed sample. Krellner et al. (2008) also report an increasing $\chi$ with decreasing temperature in their high quality $SrFe_2As_2$ as seen in the 1111's, see also Fig. 18. Undoped single crystal $KFe_2As_2$ has a wide range of RRR reported, see Fukazawa et al. (2009a) for RRR=67, Dong and Li (2010) for RRR=265, J. S. Kim et al. (2011c) for RRR=650 and Hashimoto et al. (2010a) for RRR>1200. (Samples without doping can in general be prepared with larger RRR due to the lack of any dopant-atom scattering.)



Dong et al. (2010b) report that their $KFe_2As_2$ samples (RRR≈90) show non-Fermi liquid behavior, $\rho=\rho_0 + AT^{1.5}$ above $T_c \approx 3.5$ K up to 15 K in zero field or in a 5 T applied field to suppress $T_c$ between 0.05 and 15 K.  In contrast, Hashimoto et al. (2010a) report that their (RRR>1200) $KFe_2As_2$ sample shows Fermi liquid behavior, $\rho=\rho_0 + AT^2$, above $T_c$ up to 10 K and Terashima et al. (2009), in $KFe_2As_2$ with RRR≈90, report $\rho=\rho_0 + AT^2$ between 4 and 45 K.   Specific heat in field on a RRR=650 crystal shows (J. S. Kim et al., 2011c) a decreasing γ with decreasing temperature, i. e. consistent with Fermi liquid behavior.  This controversy remains unresolved, although the non-Fermi liquid result of Dong et al. is often cited as one proof of such behavior in the FePn/Ch.  Where a quantum critical point would be in the phase diagram of $KFe_2As_2$ to cause non-Fermi liquid behavior is unclear, but there seems to be general agreement that $KFe_2As_2$ exhibits unconventional superconductivity (Dong et al., 2010b; Hashimoto et al., 2010a; Fukazawa et al., 2009a).

Another interesting resistivity behavior seen in the undoped $MFe_2As_2$ is that, for certain samples of M=Ba (J. S. Kim et al., 2009a) and Sr (Saha et al., 2009b), $\rho\rightarrow0$ at $T_c^{\rho}$~22 K but with no bulk indications of superconductivity (although Saha et al., see diamagnetic zero-field-cooled shielding of 15% in one sample).  Partial transitions in $\rho$ at ~ 10K are seen in $CaFe_2As_2$ (Torikachvili et al., 2009).  The explanation for these resistive transitions to superconductivity (including possible filamentary or planar defects) is still under investigation.

Considering now doped $MFe_2As_2$, Ahilan et al. (2008) point out that $\rho=\rho_0 + AT^1$ above $T_c$=22 K in $BaFe_{1.8}Co_{0.2}As_2$ up to 100 K, a significant range of non-Fermi liquid



behavior. The authors discuss the nearness of $BaFe_{1.8}Co_{0.2}As_2$ to a magnetic instability and the possibility of this being linked to the superconductivity. Interestingly, X. F. Wang et al. (2009b) find that, as long as there is an SDW anomaly in the sample ($x \leq 0.17$, $T_{SDW} \sim 70$ K for x=0.17), that $\chi$ for $BaFe_{2-x}Co_xAs_x$ rises linearly with increasing T up to their highest temperature of measurement, generally 300 K (see Fig. 21). Ronning et al. (2008) report $\chi \sim T^1$ for field both in the ab-plane and in the c-axis directions up to 350 K in $CaFe_2As_2$. Klingeler et al. (2010) report the magnetic susceptibility for $CaFe_{2-x}Co_x As_2$ $0 \leq x \leq 0.25$ increases above $T_{SDW}$ up to room temperature for all six compositions studied, with $\chi \sim T$ for as long as $T_{SDW}$ remains finite (up to x=0.056). For undoped $BaFe_2As_2$ and $BaFe_{1.83}Co_{0.17}As_2$ X. F. Wang et al., extend their range of measurement up to 700 K and $\chi$ is seen (Fig. 21) to rise linearly with increasing temperature for $T_{SDW} < T$ $\leq 700$ K. These data are consistent with the arguments of G. M. Zhang, et al. (2009), discussed above, for the existence of strong antiferromagnetic fluctuations above $T_{SDW}$ and up to high temperature in these Co-doped $MFe_2As_2$ alloys, M=Ba and Ca. Note that the linearity in $\chi$ with T disappears when $T_{SDW}$ is suppressed for x=0.20/0.065 in the Co-doped Ba (Fig. 21a)/$CaFe_2As_2$, while the $\chi \sim T$ survives in $LaFeAsO_{1-x}F_x$, after $T_{SDW}$ is suppressed for x=0.05 and (approximately) for x=0.06, Fig. 20. Presumably this implies stronger fluctuations surviving in the $LaFeAsO_{1-x}F_x$ after the magnetic transition is suppressed than in Co-doped $BaFe_2As_2$, a point of potential interest for theorists and for neutron scattering (see section IVA1) and NMR investigation of the fluctuations.

For further evidence for non-Fermi liquid behavior in the resistivity of the doped 122's, $\rho = \rho_0 + AT^1$ above $T_c$=21 K in $SrFe_{1.7}Rh_{0.3}As_2$ and in $SrFe_{1-x}Ir_xAs_2$, x>0.4, up to



300 K (F. Han et al., 2009). Kasahara et al. (2010) find $\rho = \rho_0 + AT^1$ in single crystal BaFe$_2$As$_{1.4}$P$_{0.6}$ above $T_c$=30 K up to 150 K, while Jiang et al. (2009) report $\rho \sim T^1$ up to 300 K above $T_c$ for BaFe$_2$As$_{2-x}$P$_x$, $0.6 \leq x \leq 0.9$.

**3.) 111 Structure:** Song et al. (2010) report metallic behavior in the $\rho$ of single crystal LiFeAs, with RRR ~ 35 and $\chi$ approximately (to within 10%) temperature independent from $T_c$ to room temperature. G. F. Chen et al. (2009) also report d$\rho$/dT>0 for single crystal Na$_{1-\delta}$FeAs, but RRR was only 1.8. G. F. Chen et al. further report that $\chi$ increases by about 40% approximately linearly with increasing temperature between 40 and 300 K, i. e. this would be consistent with the universal behavior proposal for $\chi \sim T$ of G. M. Zhang et al. (2009)..

**4.) 11 Structure:** As an example of how measurements of resistivity offer a good overview of a phase diagram, Fig. 19 shows $\rho$ up to room temperature of polycrystalline samples of FeSe$_{1-x}$Te$_x$ (Mizuguchi et al., 2009). With the later advent of single crystals of FeSe the absolute value of $\rho$ decreased by approximately a factor of two (Braithwaite et al., 2009), but the temperature dependence (metallic, with rounding towards room temperature) remains qualitatively the same. An expanded view of $\rho$ vs T in FeSe showed a linear temperature dependence (i. e. non-Fermi liquid behavior as has been discussed above for the 122's) from $T_c$ ~ 8 K up to almost 50 K (Sidorov, Tsvyashchenko and Sadykov, 2009; Masaki et al., 2009). Unlike the $\chi \sim T$ behavior reported above $T_{SDW}$ for the 1111's, the 122's, and Na$_{1-\delta}$FeAs, $\chi$ for single crystal FeSe – which has no magnetic transition - increases faster than linearly with temperature by a factor of three



above $T_c$ up to ~180 K, at which point $\chi$ falls by about 20% by room temperature (Braithwaite et al., 2009).

The magnetic susceptibility in $FeTe_{0.92}$ above $T_{SDW}$ ~ 70 K *decreases* linearly with increasing temperature up to about 240 K (Iikubo et al., 2009). Upon S-doping (Hu et al., 2009), this $\chi$ ~ -T behavior persists above the depressed $T_{SDW}$ (~ 30 K for $Fe_{1+\delta}Te_{0.9}S_{0.1}$) up to room temperature. In $FeSe_{0.5}Te_{0.5}$, $T_c$ ~ 14 K, $\chi$ increases linearly with temperature by about 15% between 100 and 250 K (highest temperature of measurement) (Sales et al., 2009).

**5.) 21311 Structure:** Resistivity of polycrystalline $Sr_2Mg_{0.5}Ti_{0.5}O_3Fe_{1-x}Co_xAs$ shows metallic behavior from room temperature down to low temperature for x=0, with RRR~6, while $\rho$ vs T shows a slight upturn in $\rho$ with decreasing temperature above $T_c$ caused by the Co-doping (Sato et al., 2010). Resistivity of polycrystalline $Sr_2VO_3FeAs$, $T_c$ ~ 33 K (the superconductivity is sample dependent) is also metallic in behavior from 300 K down to $T_c$, with an extrapolated RRR of ~ 10. There appears to be no evidence for a structural ordering anomaly up to 300 K in this class of material. In the undoped parent compound $Sr_2CrO_3FeAs$, Cr orders antiferromagnetically at 31 K (Tegel et al., 2009), while in $Sr_2VO_3FeAs$ there is evidence in $\chi$ and specific heat (Sefat et al., 2010) and $\rho$ (Cao et al., 2010) for a weak (~ 0.1 $\mu_B$) magnetic transition at ~ 155 K. The magnetic susceptibility for $Sr_2VO_3FeAs$ shows a definite anomaly at this temperature, as does the specific heat, while only the derivative of the resistivity reveals an anomaly. Investigations have not been reported above 50 K for any other of the superconducting



examples of 21311 (Sr$_2$ScO$_3$FeP, T$_c$=17 K; Sr$_2$Mg$_{0.2}$Ti$_{0.8}$O$_3$FeAs, T$_c$=39 K) nor in the recently discovered example of 43822 (Ca$_2$(Mg$_{0.25}$Ti$_{0.75}$)$_{1.5}$O$_{-4}$FeAs, T$_c$=47 K).

**6.) 122* Structure:** As shown below in Fig. 22 (data from Bao et al., 2011b), there is a wide range of resistivity behavior in K$_x$Fe$_y$Se$_2$, depending on the exact composition, which is thought to be caused by the effects of disorder on the iron Fe1 and Fe2 sublattices. However, the magnetic susceptibility is relativity insensitive to these small variations in composition as shown in Fig. 22. Susceptibility data by Liu et al. (2011) for a similar composition as those shown in Fig. 22 are about 10% larger and show about the same ≈40 % decrease below T$_N$. Bao et al. also report data (not shown here) for the insulating compositions K$_{0.87}$Fe$_{1.57}$Se$_2$, K$_{0.94}$Fe$_{1.54}$Se$_2$, and K$_{0.99}$Fe$_{1.48}$Se$_2$ where with decreasing Fe content the resistivity climbs more and more steeply with decreasing temperature below 300 K. For the last composition, ρ can be fit to an exponential activation form, exp(-δ/k$_B$T), with the energy gap δ≈85 meV.

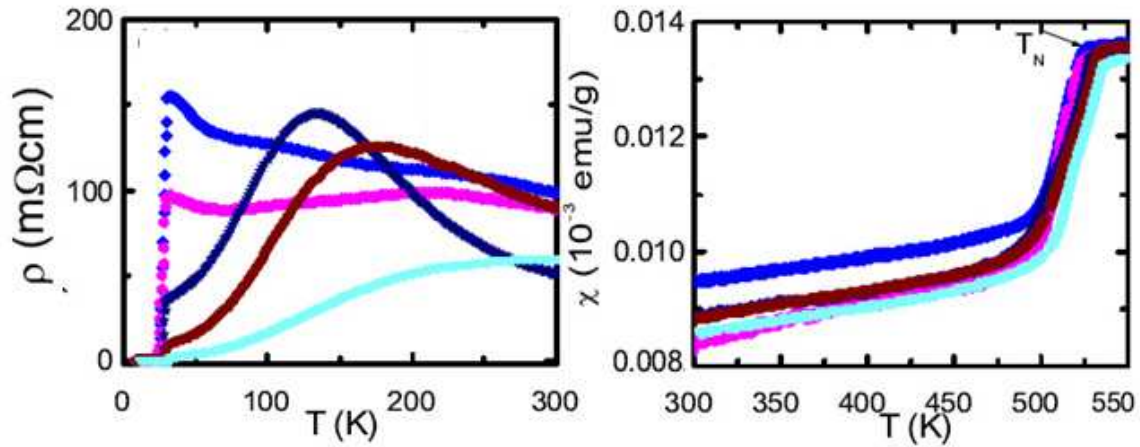

Fig. 22 (color online) Resistivity (left panel) and magnetic susceptibility (right panel) from Bao et al. (2011b) for metallic (dρ/dT>0) K$_{0.82}$Fe$_{1.63}$Se$_2$ (light blue), as well as samples with 'humps' in ρ - K$_{0.86}$Fe$_{1.62}$Se$_2$ (maroon) and K$_{0.84}$Fe$_{1.58}$Se$_2$ (dark blue) and samples even closer to insulating behavior: K$_{0.77}$Fe$_{1.60}$Se$_2$ (pink hexagons) and



$K_{0.77}Fe_{1.58}Se_2$ (blue diamonds). Although all five compositions show a superconducting transition $\rho \rightarrow 0$, the latter two show only about 75% of a zero-field-cooled diamagnetic shielding effect (not shown) while the first three compositions show a full shielding effect. Note the similarity for all five compositions of the high temperature $\chi$ data.

## B. Specific Heat

Measurements of the specific heat of superconductors in the normal state are generally of use to show higher temperature transitions, such as $T_S$ and $T_{SDW}$ in the FePn/Ch superconductors. If the $T_c$ is low enough or if enough magnetic field can be applied to suppress $T_c$ appreciably, C/T extrapolated to T=0 from normal state data gives $C^{normal}/T|_{T \rightarrow 0} = \gamma_n$. The parameter $\gamma_n$ is proportional to the renormalized (by $1+\lambda$, where $\lambda$ can be a combination of electron-phonon and electron-electron interactions) bare electronic density of states at the Fermi energy N(0), i. e. $\gamma_n \sim (1+\lambda)N(0)$. The parameter $\gamma_n$ is a useful parameter for various discussions including those of band structure calculations of N(0) and dHvA measurements of the effective masses, m*, of the various Fermi surface orbits since $\gamma_n \propto m*$. Although there have been a few cases in the new FePn/Ch superconductors where $\gamma_n$ has been either measured or estimated, extrapolating $\gamma_n$ from above a superconducting transition of 10 K or higher is problematic. If the phonon contribution to the specific heat below $T_c$ can be accurately estimated, e. g. via a neighboring composition (fortunately for this purpose Co-doping of Fe involves almost the same molar mass) that is not superconducting, one can attempt to extrapolate the electronic specific heat below $T_c$ by using the second order nature of the superconducting transition and matching entropies. Thus, the *measured* superconducting state specific heat, $C_{sc}$, gives the superconducting state entropy at $T_c$, $S_{sc}(T_c)=\int(C_{sc}/T)dT$, by integrating the superconducting state data from T=0 to $T_c$. Then, if the phonon contribution to the entropy (which is large) can be subtracted or accurately estimated, the extrapolated



normal state electronic contribution $C^{el}_{normal}/T$ must give, for a second order phase transition, a matching $S_{normal}(T_c)$ by integrating $\int (C^{el}_{normal}/T)dT$ and adding in the phonon contribution.  Another possibility is if C/T in the superconducting state is proportional to $H^1$ (from nodeless superconductivity, discussed below in section IV), then measurements of C/T up to some fraction of the upper critical field $H/H_{c2}(T\rightarrow 0)$ will give C(H)/T in the superconducting mixed state equal to the product $\gamma_n*H/H_{c2}(T\rightarrow 0)$.  However, this is so far a rather rare measurement, since $H_{c2}(T\rightarrow 0)$ values are quite high, and this method of estimating $\gamma_n$ is dependent on rather high applied fields to be of any accuracy.

**1.) $\gamma_n$ (experiment):**  A short list of those superconducting FePn/Ch materials for which estimated $\gamma_n$ values in the normal ($T>T_c$) state exist consists of the following.  Due to the higher $T_c$'s and sample quality issues, most 1111 materials have unknown $\gamma_n$ values.  Kant et al. (2010) estimate $\gamma_n$ for $Ba_{1-x}K_xFe_2As_2$ to be in the range 50-65 mJ/moleK$^2$ for x between 0 and 0.6.   Popovich et al. (2010) find $\gamma_n$=50 mJ/moleK$^2$ for $Ba_{0.68}K_{0.32}Fe_2As_2$, $T_c$=38.5 K.  Using 9 T C/T data which are proportional to $H^1$ and extrapolating $\gamma$ up to $H_{c2}(T\rightarrow 0)$ of 100 T (such a long extrapolation involves a large potential error), Mu et al. (2009a) estimate $\gamma_n$ for $Ba_{0.6}K_{0.4}Fe_2As_2$ to be 63 mJ/moleK$^2$.

**Table 3:  Specific heat $\gamma_n$ and $T_c$ for unannealed and annealed* $BaFe_{2-x}Co_xAs_2$**

| x= | $T_c$(K) | $\gamma_n$(mJ/moleK$^2$) | reference |
|---|---|---|---|
| 0.08/0.09 | 5.8/5.6,8.0* | 14.9/13.7,14* | a/b |
| 0.10 | 19.5 | 17.2 | a |
| 0.11 | 21.5 | 19 | a |



| 0.115 | 24.3 | 21.3 | a |
| 0.15/0.16 | 22.9/20,25* | 22.1/18,22* | a/b |
| 0.18 | 20.7 | 20 | a |
| 0.22/0.21 | 11.1/11,17.2* | 17/23.2,20* | a/b |
| 0.24 | 5.1 | 14.6 | a |
| 0.31 | 0 | 16 | a |

Ref. a:  Hardy et al., 2010 a;  Ref. b:  Gofryk et al. (2011a,b), annealed values are with *

In a thorough study of the specific heat of $BaFe_{2-x}Co_xAs_2$ over the whole superconducting dome (see Fig. 12), Hardy et al. (2010a) (see also Hardy et al., 2010b) reported unannealed $\gamma_n$ and $T_c$ values vs composition (Table 3), while values for three compositions (x=0.09, 0.16, 0.21) of both unannealed and annealed (1 week, 800 °C) material were reported by Gofryk et al. (2011 a,b).  There is relatively good agreement between the annealed and unannealed $\gamma_n$ values for comparable compositions (although note the differences in $T_c$'s, discussed with $\Delta C/T_c$ later in Section IIIB4).

J. S. Kim et al. (2010a) and Y. Wang et al. (2011), using superconducting state data to 15/35 T on a collage of single crystal $BaFe_2As_{1.4}P_{0.6}$, $T_c$=30 K, estimate $\gamma_n$ to be 16 mJ/mole$K^2$ by extrapolating to $H_{c2}(T\rightarrow0)$ of 52 T.   Zeng et al. (2011), using data to 9 T in $K_{0.8}Fe_{1.6}Se_2$ ($H_{c2}(0)\approx48$ T, $T_c$=32 K) offer the rough estimate that $\gamma_n$ is roughly 6 mJ/mole$K^2$, or significantly smaller than found for the other FePn/Ch with comparable $T_c$'s.



Low $T_c$ compounds, such as FeSe$_{0.88}$, $T_c \approx 8$ K, LaFePO, $T_c \approx 5$-6 K, and KFe$_2$As$_2$, $T_c$=3.4 K, have $\gamma_n$'s that are more easily determined.   For a polycrystalline sample of FeSe$_{0.88}$ Hsu et al. (2008) find that $C_{normal}$ /T =$\gamma_n$ + $\beta T^2$ with $\gamma_n$ =9.2 mJ/moleK$^2$.   For a mosaic of single crystals of LaFePO, Analytis et al. (2008) found that $\gamma_n$=7 mJ/moleK$^2$, while Fukazawa et al. (2009a) found $\gamma_n$ =69 mJ/moleK$^2$ for polycrystalline KFe$_2$As$_2$, RRR=67.  In a later work on KFe$_2$As$_2$, RRR>1200, Hashimoto et al. (2010a) reference an unpublished result for $\gamma_n$ of 93 mJ/moleK$^2$, and J. S. Kim et al. (2011c) report $\gamma_n$=102 mJ/moleK$^2$ for single crystal KFe$_2$As$_2$ with RRR=650, so clearly there is sample dependence of $\gamma_n$ in KFe$_2$As$_2$ (and presumably in other FePn/Ch compounds).

**2.) $\gamma_n$ (calculated):**  It is also interesting to compare, where possible, the measured $\gamma_n$ values to those calculated from band structure calculations.  The normal state specific heat $\gamma_n$ can be related to the calculated bare density of states, N(0), at the Fermi energy by $\gamma_n$=1/3 $\pi^2 k_B^2$N(0)(1+$\lambda$), where $k_B$ is the Boltzmann constant and $\lambda$ is the sum of the electron-phonon as well as the electron-electron coupling parameters, $\lambda_{el-ph}$ and $\lambda_{el-el}$.  If $\gamma_n$ is in units of mJ/moleK$^2$ and N(0) is in units of states/eV-atom, then – by combining the constants 1/3 $\pi^2 k_B^2$ – we get N(0)(1+$\lambda$)= 0.42$\gamma_n$/n.  Usually the scaling between "mole" and "atom" is that the mole contains "n" atoms, e. g. n=5 in the case of the 122's, without regard to whether the atoms are greater or lesser contributors to N(0), i. e. a mole of 122 is not considered to consist of just the two Fe atoms even though band structure calculations tell us that N(0) comes mostly from the Fe bands.  Most band structure calculations have been on the undoped parent compounds, which in the case of the 1111's (with the exception of LaFePO) and the 122's (excepting KFe$_2$As$_2$, RbFe$_2$As$_2$ and



CsFe$_2$As$_2$) are not superconducting and thus not the focus here. Further, the 1111 and the 122 parent compounds all undergo a spin density wave transition (which typically lowers N(0)) around 100-200 K, while in the 21311 there is at least indication of magnetic order in Sr$_2$VO$_3$FeAs at 155 K (Sefat et al., 2010) and the 122* have magnetic order above 500 K. Therefore the measured low temperature $\gamma_n$ will have a lower value than the calculations (which do not take into account the reduction in N(0) due to magnetic order) predict in any case. Thus, in order to compare band structure calculations with experimental $\gamma_n$ values, what is needed is either such a calculation on a non-magnetic doped system, or to compare the calculated and measured $\gamma_n$ on a non-magnetic 111 or 11 compound. We present here three disparate examples.

For FeSe, T$_c$=8 K, Subedi et al. (2008) calculate N(0)=0.95 states/eV-atom. Based on the measured specific heat $\gamma_n$ of Hsu et al. (2008), this implies, using n=2, a 1+$\lambda$ of 2.05. A number of calculations exists for N(0) in LaFePO, T$_c$≈5-6, see e. g. Lu et al. (2008), Lebegue (2007), and Skornyakov et al. (2010). Using $\gamma_n$=10 mJ/moleK$^2$ from Suzuki et al. (2009), the consensus for 1+$\lambda$ is 1.7. Considering these two values of 1+$\lambda$, it is interesting to note that the authors of such band structure calculations themselves note that their calculated band structures need to be a factor of ~ 2 narrower to correspond to the measured angle resolved photoemission spectroscopy (ARPES), e. g. Lu et al. (2008) renormalize their DFT band structure by narrowing it a factor of 2.2 to fit their ARPES data. Shein and Ivanovskii (2009c) calculate N(0)=1.11 states/eV-atom for Ba$_{0.5}$K$_{0.5}$Fe$_2$As$_2$, T$_c$=38 K. However, they note that the Fermi energy in the calculation lies on the slope of a sharp peak in the density of states, so that small changes in the Fermi energy would have a large effect on N(0). Using the $\gamma_n$ for this composition from



Kant et al. (2010) of 54 mJ/moleK$^2$, and n=5, leads to a 1+λ of 4.1, clearly far larger than any possible 1+λ$_{el-ph}$ and perhaps indicative indeed that N(0) has been underestimated.

At the present juncture of theoretical understanding of the pairing mechanism (see also the discussion of the isotope effect in section IVA and the discussion of spin fluctuations below T$_c$ in the discussion of inelastic neutron scattering in section IVA1), it is clear that the pairing mechanism for the superconductivity in the FePn/Ch is not electron phonon coupling (Boeri, Dolgov and Golubov, 2009; Subedi et al., 2008), but some other interaction that is presumably electronic, perhaps spin fluctuations.

*If* the so-called mass renormalization (~ 1+λ) *were* due to electron phonon coupling in FeSe$_{1-x}$ or LaFePO, a standard estimate (e. g. the McMillan, 1968 formula) in the BCS formalism would in fact, for T$_c$=8/6 K and the lattice stiffness of FeSe/LaFePO as reported by Hsu et al. (2008)/Suzuki et al. (2009), require λ$_{el-ph}$~0.8/0.6. This is not inconsistent with the 1+λ values of 2.05/1.7 derived from the ratio of the measured specific heat γ$_n$ and calculated N(0) discussed in the previous paragraph. However, Subedi et al. (2008) calculate λ$_{el-ph}$=0.17 for FeSe, making it clear (see also the inelastic neutron scattering detected spin fluctuations below T$_c$ discussed in section IVA1 below) that even this low T$_c$ FeCh is not an electron-phonon pairing superconductor.

Thus, it should be stressed that the ratio between measured γ$_n$ values and band structure calculations for N(0) – even for such low T$_c$ materials as FeSe – is giving values for 1+λ that either involve large contributions to λ from electron-electron mass renormalization or indicate errors in the calculations. For the higher T$_c$ Ba$_{0.5}$K$_{0.5}$Fe$_2$As$_2$ it is clear that the derived 1+λ of 4.1 implies a problem with the calculated N(0). Such



strong electron-electron interactions, if present, should strongly affect other measurements, for example the low temperature resistivity.

**3.) $\Delta C/T_c$:** A very interesting correlation between $\Delta C$ and $T_c$ has been proposed by Bud'ko, Ni and Canfield (2009) (hereafter 'BNC'), namely that for 14 samples of various doped $BaFe_2As_2$ superconductors (including Co and Ni on the Fe site and K on the Ba site) $\Delta C/T_c = aT_c^2$ (see Fig. 23), where analyzing their graph gives a~0.056 mJ/moleK$^4$. Zaanen (2009) has proposed that this $\Delta C/T_c \sim T_c^2$ scaling behavior argues against a Fermi liquid picture, and instead discusses the idea that the superconductivity could be forming from a non-Fermi liquid quantum critical metal. Rather than the usual quantum critical *point* in a phase diagram (see Stewart, 2001, Stewart, 2006, and von Löhneysen et al., 2007), Zaanen argues for a quantum critical *region* over some fraction of the superconducting dome in composition space. To explain the observed BNC scaling Kogan (2009, 2010) considers instead that the FePn/Ch superconductors are weak coupled Fermi liquids with strong pair breaking, with the observed $\Delta C$'s and $T_c$'s much reduced from those in hypothetical clean material. A third theory (Vavilov, Chubukov, and Vorontsov, 2011) calculates that $\Delta C/T_c \approx T_c^2$ below optimal doping in the FePn/Ch for part of the underdoped dome as $T_c \to 0$ due to the coexistence of SDW magnetism and $s_{+-}$ superconductivity. However, above optimal doping in the absence of coexistent magnetism their work discusses a return to BCS behavior.

### a.) Possible errors in determining the intrinsic $\Delta C/T_c$

Before discussing this scaling of the discontinuity in the specific heat at $T_c$, a discussion of the determination of $\Delta C$ will help to establish the source of possible errors.



Due to sample quality (disorder/strain) issues, these transitions can be quite broadened in temperature. One way to analyze and intercompare such broadened transitions is the so-called 'equal area construction', sketched in Fig. 24. In this method, the low temperature superconducting state data up to the initial bend over in $C/T$ at $T_c^{low}$ are extrapolated linearly further as $C_{sc}^{ex}/T$; likewise, the normal state data are extrapolated linearly as $C_n^{ex}/T$ to lower temperature. Then an ideally narrow discontinuity $\Delta C$ is constructed at a temperature approximately midway between $T_c^{onset}$ and $T_c^{low}$ at $T_c^{mid}$ with the area (which is an entropy) between the linearly extrapolated $C_{sc}^{ex}/T$ and the actual measured data below $T_c^{mid}$ equal to the area (entropy) between the measured data above $T_c^{mid}$ and the extrapolated $C_n^{ex}/T$ from above $T_c^{onset}$. This then preserves the correct measured value of the superconducting state entropy at $T_c$ in the new, idealized transition. Sometimes, however, the transition is so broad (for example in $Sr(Fe_{0.925}Ni_{0.075})_2As_2$, $T_c^{onset}$=8.5 K, $\Delta T_c \approx 3.5$ K, Saha et al., 2009a) or even non-existent (e. g. in underdoped $Ba_{1-x}K_xFe_2As_2$, Urbano et al., 2010 and Rotter et al., 2009, as discussed in Section IIB2b or in $Ca_{0.5}Na_{0.5}Fe_2As_2$, $T_c$=18 K, Dong et al., 2008b) that the equal area construction fails.

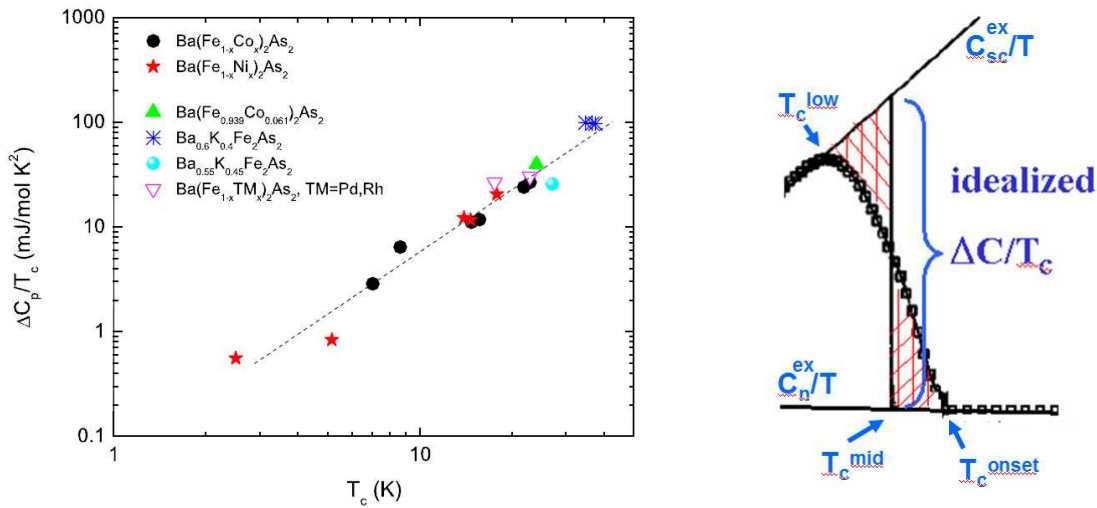



Fig. 23 (color online) Discontinuity in the specific heat $\Delta C$ at the superconducting transition in doped $BaFe_2As_2$ on a log-log plot showing $\Delta C/T_c$ proportional to $T_c^2$ (Bud'ko, Ni and Canfield, 2009).

Fig. 24 (color online) Sketch of the equal area construction method for determining $\Delta C/T_c$ in a broadened transition. Data points are denoted by squares. Red cross-hatching marks the equal areas, which are entropies, discussed in the text.

Further complicating the determination of $\Delta C/T_c$, for many samples of the FePn/Ch superconductors there is a finite $\gamma$ in the superconducting state that is likely not intrinsic. How to distinguish if this residual $\gamma_r$ is a sign of a part of the sample being non-superconducting (thus decreasing $\Delta C/T_c$ but not affecting $T_c$) or a sign of defects and gapless behavior (with both $\Delta C/T_c$ and $T_c$ decreased, while the transition width $\Delta T_c$ is broadened) will now be discussed using examples from the FePn/Ch.

In $KFe_2As_2$, where $\gamma_n$ extrapolated from above $T_c$ is 69 mJ/moleK$^2$ in the data of Fukazawa et al. (2009a) for an RRR=67 sample as already mentioned, C/T in the superconducting state as T→0, $\gamma_r$, is ≈40 mJ/moleK$^2$ while in the data of J. S. Kim et al. (2011c) down to 0.08 K for an RRR=650 sample, $\gamma_n$=102 mJ/moleK$^2$ and $\gamma_r$≈0. The fact that the sums of $\gamma_n$ and $\gamma_r$ in both samples are approximately the same gives credence to the idea that $\gamma_r$ in the Fukazawa et al. sample is simply from a non-superconducting fraction. Further, if one continues this logic, then the Fukazawa et al. sample would, using their values for $\gamma_n$ and $\gamma_r$, be approximately $\gamma_n/(\gamma_n+\gamma_r)$ (=63%) superconducting, and one would expect in this sample only this fraction of the $\Delta C/T_c$ observed in the fully superconducting ($\gamma_r$≈0) sample of J. S. Kim et al. (2011c), or $\Delta C_{partially\ super}/T_c$ =$[\gamma_n/(\gamma_n+\gamma_r)]*\Delta C_{fully\ super}/T_c$. This is, within the error bars, borne out, since $\Delta C/T_c \approx 23$



mJ/moleK$^2$ for the Fukazawa et al. (2009a), RRR=67 sample, or 56% of the $\Delta$C/T$_c \approx 41$

mJ/moleK$^2$ for the J. S. Kim et al. (2011c), RRR=650 sample with $\gamma_r \approx 0$.

In Suzuki et al.'s (2009) data for LaFePO, $\gamma_n$ extrapolated from above T$_c$=5.8 K is

10.1 mJ/moleK$^2$ whereas C/T extrapolated to T=0 from their superconducting state data

below T$_c$ (between 2 and 4 K) gives a residual $\gamma_r$~7.5 mJ/moleK$^2$ – seemingly similar to

the results for KFe$_2$As$_2$.

Thus, in KFe$_2$As$_2$ and possibly in LaFePO a reasonable explanation is that only part

of the sample is superconducting (since only part of the normal state $\gamma_n$ is removed below

T$_c$) and therefore for an ideal, 100% superconducting sample $\Delta$C/T$_c$ would be

proportionately larger.   Thus, in general, without high quality ($\Leftrightarrow$low $\gamma_r$) samples it can

be difficult comparing $\Delta$C/T$_c$ values and care must be taken.

As an aside, it should be stressed that such a large residual $\gamma_r$ in the superconducting

state as found in LaFePO, in early, low RRR samples of KFe$_2$As$_2$ or in unannealed non-

optimally doped BaFe$_{2-x}$Co$_x$As$_2$ (where $\gamma_r$>10 mJ/moleK$^2$ or roughly ½ of $\gamma_n$) is a sample

quality issue (see Section V), not a sign of nodal behavior.  Since specific heat is a bulk

measurement (vs resistivity and thermal conductivity which can be dominated by one

dimensional pathways), even line nodes on a Fermi surface – if unsmeared due to defects

– will have only a miniscule amount of normal Fermi surface electronic density of states

contribution to $\gamma_r$.  Whether the extrinsic behavior is due to normal regions (as the

conservation of $\gamma_r$+$\gamma_n$ in KFe$_2$As$_2$ with improving sample quality with no change in T$_c$ but

an increase in $\Delta$C/T$_c$ would imply), or defects on a microscopic, approximately



homogeneous scale causing gapless behavior (where annealing of, e. g., $BaFe_{2-x}Co_xAs_2$ – Gofryk et al., 2011a,b - decreases $\gamma_r$ markedly, down to 0.25 mJ/moleK$^2$ on one sample of optimally doped x=0.16,  and increases $T_c$ while leaving $\gamma_n$ – see Table 3 and discussion - approximately unchanged) has to be determined on a case by case basis.   In any case, nodal behavior (line or point nodes) in a single crystalline (although no real material is ideal) superconductor cannot lead to over 30% of a Fermi surface being gapless and causing the large $\gamma_r$ seen, e. g., in $KFe_2As_2$ and LaFePO.  As an example of a known d-wave superconductor with line nodes, $YBa_2Cu_3O_{6.99}$ has $\gamma_r$ in a high quality sample (but presumably still with some defect broadening of the line nodes at the Fermi surface, as well as possible other contributions to $\gamma_r$) equal to 1.2 mJ/moleK$^2$ and $\gamma_n \approx 20$ mJ/moleK$^2$ (Moler et al., 1994).  Further optimization of the YBCO samples could decrease $\gamma_r$ even further, but the ratio 1.2/20 or 6% provides a useful 'upper bound' estimate for the effect of nodal superconductivity on $\gamma_r$ in well ordered single crystals.

If there are sufficient defects on a quasi-homogeneous microscopic scale (rather than normal regions) to make a large $\gamma_r$, then $T_c$ should be strongly affected (cf.  Kogan, 2009, 2010).  Although this is not the case in $KFe_2As_2$ ($T_c$ seems to be fairly constant as a function of sample quality measured via RRR), in the annealing studies of $BaFe_{2-x}Co_xAs_2$, $T_c$ increases with annealing approximately by $\approx 50\%$ for the non-optimally doped samples (Gofryk et al, 2011a,b) shown above in Table 3.  How the $\Delta C/T_c$ results for the annealed and unannealed samples of $BaFe_{2-x}Co_xAs_2$ compare on the BNC plot will be discussed below.



In $Sr_2VO_3FeAs$, the status of the sample quality is that as yet *no* anomaly at $T_c$ is visible in the specific heat (Sefat et al., 2010), while the residual gamma in the superconducting state, $T<<T_c$, is 25 mJ/moleK$^2$ for the sample with the largest fraction of superconductivity in the susceptibility ($\approx$10% Meissner fraction, $\approx$50% shielding) and $\gamma_n$= 60 mJ/moleK$^2$ for the non-superconducting sample. In the defect 122* superconductors, determinations of $\Delta C/T_c$ give about 10 mJ/moleK$^2$ (Luo et al., 2011) to 12 mJ/moleK$^2$ (Zeng et al., 2011), with $T_c\approx$31 K, which is small compared to the BNC plot value expected for this $T_c$ of about $\Delta C/T_c\approx$50 mJ/moleK$^2$. Although $\gamma_r$ was reported by Zeng et al. to be small compared to $\gamma_n$ (0.4 vs 6 mJ/moleK$^2$ respectively), another work (Shen et al., 2011) by the same group on improved samples reported the possibility that these materials were made up of superconducting islands surrounded by insulating (i. e. $\gamma$=0) material. Thus, for the 122* samples evaluation of $\Delta C/T_c$ awaits homogeneous, single phase samples.

Now that potential sources of error in $\Delta C/T_c$ values in the FePn/Ch have been discussed, it is interesting to examine the error bars for several samples, both with large and small disagreements from the BNC scaling plot, shown in Fig. 23. First, $BaFe_{2-x-}Ni_xAs_2$, x=0.144 and $T_c\sim$5 K, has a very broad, small and hard to analyze transition in the specific heat, and the $\Delta C/T_c$ shown in Fig. 23 is likely underestimated – which would bring that point closer to the BNC fitted line. Another point which also lies too low vs the BNC $\Delta C/T_c \sim T_c^2$ trend ($Ba_{0.55}K_{0.45}Fe_2As_2$, $T_c\sim$28 K) had $\Delta C/T_c\sim$25 mJ/moleK$^2$ (vs 44 mJ/moleK$^2$ expected from the plot) estimated from a very broad, $\Delta T_c\sim$3 K, transition in a Sn-flux grown single crystal (Ni et al., 2008a), RRR~3. The sample quality as well as



the width of the transition again contribute to the possible error bar.  Considering now a data point that lies on the BNC line, $\Delta C/T_c$ of a self-flux grown single crystal $Ba_{0.6}K_{0.4}Fe_2As_2$ ($\Delta C/T_c$=100 mJ/moleK$^2$ at $T_c^{mid}$=34.7 K), was idealized (Welp et al., 2009) from a $\Delta T_c$~1 K broad transition, rather high quality (RRR~15) sample (Luo et al., 2008).  A more recent measurement on $Ba_{0.68}K_{0.32}Fe_2As_2$, $T_c$=38.5 K and a $\Delta T_c$~0.4 K broad transition – not plotted in the original BNC plot in Fig. 23 - found $\Delta C/T_c$=125 mJ/moleK$^2$ (Popovich et al., 2010).   Based on the square of the ratios of $T_c$ ($[38.5/34.7]^2$), this $\Delta C/T_c$ value of Popovich et al. matches the BNC plot equally as well as the Welp et al. value.  Thus, it seems reasonable to conclude that the BNC scaling law fit – which was conceived for doped 122 FePn's only – seems reasonably robust.

In order to supplement the BNC plot with data (and structures) not in the original version, as well as to introduce data that perhaps speak to the proposed theories, J. S. Kim et al. (2011a) considered $\Delta C/T_c$ values for several other FePn/Ch materials.  In addition, they added $\Delta C/T_c$ data for conventional electron-phonon coupled superconductors (elements with $T_c$>1 K and A-15 superconductors) and for several unconventional heavy Fermion superconductors.   This revised BNC plot, with $\Delta C/T_c \approx 0.083 T_c^{1.89}$ is shown in Fig. 25 and discussed here.

**b.)  Some additional examples of $\Delta C/T_c$ to discuss with respect to the BNC plot:**

**$KFe_2As_2$:**  The disputed report of non-Fermi liquid behavior in the resistivity (Dong et al., 2010b) of the 3.4 K superconductor $KFe_2As_2$ discussed above in section IIIA2 makes this material perhaps germane for the quantum critical picture of Zaanen.  The values for



$T_c$ and $\Delta C/T_c$ for $KFe_2As_2$ (Fukazawa, 2009a) are $T_c$=3.4 K and $\Delta C/T_c$=20-24 mJ/moleK$^2$, in a sample with RRR=67. The lower value quoted for $\Delta C/T_c$ is from simply taking $\Delta C$ at the maximum in $C_{sc}/T$ and the higher value is from the equal area construction method discussed above. This value for $\Delta C/T_c$ for an undoped 122 compound is approximately a factor of 40 larger than the 0.65 mJ/moleK$^2$ calculated from $\Delta C/T_c = aT_c^2$, Fig. 23. Also, as discussed in the preceding subsection above, due to the large value of $C/T$ as $T \rightarrow 0$ in the superconducting state, $\Delta C/T_c$ for an improved sample (such as the RRR=650 sample reported by J. S. Kim et al., 2011c) of $KFe_2As_2$ is even larger, $\approx$41 mJ/moleK$^2$. J. S. Kim et al. (2011a) then concluded, in their updated BNC plot discussion, that this large positive discrepancy with $\Delta C/T_c \propto T_c^2$ is an indication that $KFe_2As_2$ does not belong to the class of superconductor represented by the BNC plot. Although not discussed by J. S. Kim et al. (2011a), $RbFe_2As_2$ with $T_c$=2.6 K (Bukowski et al., 2010), $\gamma_n \approx$110 mJ/moleK$^2$ and $\Delta C/T_c$=55 mJ/moleK$^2$ (Kanter, et al., 2011) is presumably also more comparable to a conventional, electron-phonon coupled superconductor.

**$BaFe_2(As_{0.7}P_{0.3})_2$ / annealed $Ba(Fe_{0.92}Co_{0.08})_2As_2$ / $Sr(Fe_{0.82}Pt_{0.08})_2As_2$ / $Eu_{0.5}K_{0.5}Fe_2As_2$ / $Ba(Fe_{0.95}Pt_{0.05})_2As_2$:**

Five additional 122 superconductors have been measured since the original BNC plot, and are included in the updated BNC plot, Fig. 25. J. S. Kim et al. (2011a) measured $\Delta C/T_c$ in a collage of single crystals of $BaFe_2(As_{0.7}P_{0.3})_2$ and found a 1K wide transition, $\Delta T_c$, at $T_c^{mid}$=28.2 K and $\Delta C/T_c$=38.5 mJ/moleK$^2$.



Since the original BNC plot, Gofryk et al. (2011a,b) have been the first to report specific heat on annealed (800 $^o$C, 1 week) single crystals of Co-doped BaFe$_2$As$_2$. For optimally doped Ba(Fe$_{0.92}$Co$_{0.08}$)$_2$As$_2$, T$_c$=25 K, Gofryk et al. (2011a) report $\Delta$C/T$_c$=33.6 mJ/moleK$^2$ for $\Delta$T$_c$~1 K, vs values for unannealed samples of approximately the same composition of $\approx$24 mJ/moleK$^2$, T$_c$=22 K (Fukazawa et al., 2009a) and T$_c$=20 K (Gofryk et al., 2011a,b). As can be seen in Fig. 25, this T$_c$=25 K point fits well with the other Co-doped points of BNC to the general trend. For the other two compositions (x=0.09, T$_c$$\approx$8 K and x=0.21, T$_c$$\approx$17.2 K) annealed by Gofryk et al., the $\Delta$C/T$_c$ values of $\approx$8.4 and 14 mJ/moleK$^2$ respectively (not shown in Fig. 25) match fairly well values already in the original BNC plot, Fig. 23. Annealing single crystal BaFe$_{2-x}$Co$_x$As$_2$ showed that annealing reduced the 'residual' $\gamma_r$ in the superconducting state by large amounts (from 10.5 to 1.3 mJ/moleK$^2$ for x=0.09 and from 14.6 to 3.8 mJ/moleK$^2$ for x=0.21) in the non-optimally doped samples, vs a smaller reduction (from 3.6 to 1.3/0.25 mJ/moleK$^2$) for optimally doped, x=0.16 (Gofryk et al., 2011a,b). (Values for $\gamma_r$ in the unannealed samples of BaFe$_{2-x}$Co$_x$As$_2$ of Hardy et al., 2010a, are 9.8, 2.9 and 7.9 mJ/moleK$^2$ for the comparable compositions x=0.08, 0.15, and 0.22, i. e. the $\gamma_r$ values are – except for the overdoped case – in good agreement.) In contrast to the large changes in $\gamma_r$ with annealing in BaFe$_{2-x}$Co$_x$As$_2$, Gofryk et al. (2011b) found (see Table 3 above) that $\gamma_n$ changed only by +0.3, +4, and -3.2 mJ/moleK$^2$ for their samples of x=0.09, 0.16, and 0.21 respectively. Thus, in terms of the previous discussion about errors in determining $\Delta$C/T$_c$, the non-optimally doped BaFe$_{2-x}$Co$_x$As$_2$ samples show a marked *decrease* in $\gamma_r$ with $\gamma_n$ approximately unchanged in comparison. This, along with the $\approx$50% increase in T$_c$ with annealing (Table 3) and rather broad transition widths ($\Delta$T$_c$$\approx$0.2T$_c$) even after



annealing for these two samples, x=0.09 and 0.21, seems more consistent with defects and gapless behavior (cf. Kogan, 2009, 2010) rather than non-superconducting regions. However, the optimally doped annealed sample of Gofryk et al., even though $T_c$ increases 25% with annealing, has the same $\Delta T_c$ as the unannealed sample, as well as relatively small changes in $\gamma_r$ – properties that are less consistent with a defect/gapless picture.

Kirschenbaum et al. (2010) reported $\Delta C/T_c$=17 mJ/moleK$^2$, $T_c$=14.5 K, and $\Delta T_c$~0.8 K for their single crystal Sr(Fe$_{0.92}$Pt$_{0.08}$)$_2$As$_2$. Jeevan and Gegenwart (2010) reported $\Delta C/T_c$=70 mJ/moleK$^2$, $T_c$=32 K, and $\Delta T_c$~3 K for their polycrystalline Eu$_{0.5}$K$_{0.5}$Fe$_2$As$_2$. Finally, Saha et al. (2010b) reported $\Delta C/T_c \approx$20 mJ/moleK$^2$, $T_c^{mid}$=20 K in Ba(Fe$_{0.95}$Pt$_{0.05}$)$_2$As$_2$ for an addition to the original BNC (Fig. 23) Ba(Fe$_{1-x}$TM$_x$)$_2$As$_2$, TM=Pd,Rh points.

As may be seen in the updated BNC plot in Fig. 25, all five of these added 122 $\Delta C/T_c$ values agree rather well with the original BNC fit and support the robustness of their observation of $\Delta C/T_c \propto T_c^2$ for a broader range of 122's.

**LiFeAs/LiFeP:** These 111 structure superconductors have been well characterized by specific heat, and were not included in the original BNC plot. In particular, there are a number of works on the higher $T_c$ LiFeAs - Wei et al. (2010), Chu et al. (2009), Lee et al. (2010a), and Stockert et al. (2010) - and one on the $T_c$ ~ 6 K LiFeP (Deng et al., 2009). Although the transition of Stockert et al. in their self-flux-grown crystal is sharp and their residual $\gamma_r$ is essentially zero, their $\Delta C/T_c$ is only 12.4 mJ/moleK$^2$, $T_c$=14.7 K, while the broader transition of Lee et al. in their Sn-flux-grown gives $\Delta C/T_c$ ~ 20 mJ/moleK$^2$, $T_c$=16.8 K. The sample of Lee et al. has a residual gamma over half of the extrapolated



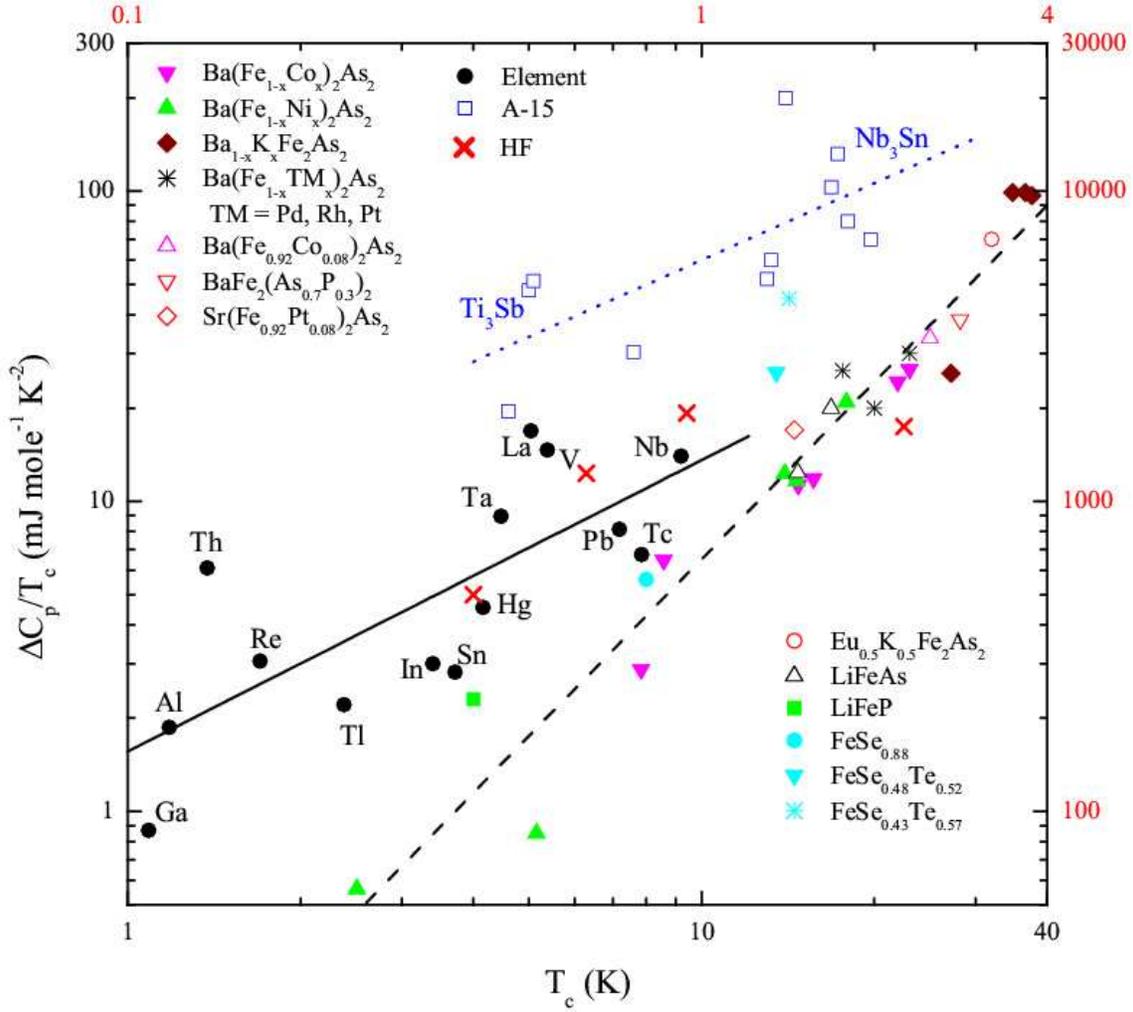

Fig. 25 (color online) Expanded BNC plot based on the work by J. S. Kim et al. (2011a) with additional FePn/Ch data as discussed in the text, along with $\Delta C/T_c$ data for the elemental superconductors with $T_c > 1$ K as well as a selection of A-15 superconductors – both conventional, electron-phonon coupled, superconducting families. In these two kinds of superconductors the $\Delta C/\gamma_n T_c$ values, while they may deviate from the weak-coupling BCS value of 1.43, are generally between 1.3 (Re) and 2.7 (Pb), i. e. fairly constant compared to the wide range of $\Delta C/T_c$. Thus, the two groups of conventional superconductors lie at different places on the y-axis in this $\Delta C/T_c$ plot since the $\gamma_n$ values which would normalize the higher $\gamma_n$ A-15's into rough agreement with the elements are not considered. In addition, four heavy Fermion superconductors are shown. These materials, CeIrIn$_5$ ($T_c$=0.4 K), CeCu$_2$Si$_2$ ($T_c$=0.63 K), UBe$_{13}$ ($T_c$=0.94 K) and CeCoIn$_5$ ($T_c$=2.25 K), due to the different scale of their $\Delta C$ values, are plotted against the upper and right hand (red) axes; all other points are plotted vs the left and lower axes. The slope of the black elemental superconductor line gives $\Delta C/T_c \sim T_c^{0.94}$ and for the A-15 superconductors (which show a large spread in $\Delta C/T_c$ at the higher $T_c$ end due to sample quality issues) the blue best fit line gives $\Delta C/T_c \sim T_c^{0.75}$. The heavy Fermion



superconductors, which are presumably non-conventional, surprisingly show $\Delta C/T_c$ vs $T_c$ behavior similar to the conventional superconductors. Numerical values for $T_c$ and $\Delta C/T_c$ for most of the plotted points are given in J. S. Kim et al. (2011a), while the others are given here in the text.

normal state $\gamma_n$ which, following the discussion above for $KFe_2As_2$, implies a larger $\Delta C/T_c$ in a sample where $\gamma_r$ could be reduced. For LiFeP, Deng et al. (2009) find a broad transition, with $\Delta C/T_c \sim 2.3$ mJ/moleK$^2$ at a midpoint $T_c$ of 4 K. These values are plotted in the updated BNC plot in Fig. 25, and agree well with the trend of the 122 superconductors, $\Delta C/T_c \propto T_c^2$. Due to the lack of magnetism in these 111 samples (see also FeSe$_{0.88}$ below), the theory of Vavilov, Chubukov, and Vorontsov (2011) is not applicable to the comparison of these data with the BNC trend.

**FeSe$_{0.88}$:** Hsu et al. (2008) fit their normal state data above $T_c \sim 8$ K to a straight line on a $C/T$ vs $T^2$ plot and arrive at $C^{ex}_n/T = 9.17 + 0.522 \, T^2$ (units of mJ/moleK$^2$) and $\Delta C/T_c$ of 5.6 mJ/moleK$^2$, which is somewhat large compared to the BNC plot value of 3.6 mJ/moleK$^2$, see Fig. 25. The superconducting $C/T$ $(T \rightarrow 0) \approx 0$, implying a clean sample.

**FeSe$_{0.48}$Te$_{0.52}$:** For this doped 11 compound, Braithwaite et al. (2010) find in single crystal material $T_c^{mid} = 13.5$ K, transition width $\Delta T_c \approx 3$ K, and $\Delta C/T_c = 20\text{-}26$ mJ/moleK$^2$ (where the larger value is from an equal area construction). In a later work (after J. S. Kim et al.'s, 2011a, revised BNC plot) with improved single crystals of FeSe$_{0.43}$Te$_{0.57}$, $T_c^{mid} = 14.2$ K and $\Delta T_c \approx 2$ K, Hu et al. (2011) report the much larger value of $\Delta C/T_c = 40\text{-}51$ mJ/moleK$^2$, with the upper value again from an idealized, sharp transition. In the Hu et al. sample there is an upturn above $T_c$ in the normal state $C/T$ (fit to a Schottky anomaly in comparable data by Tsurkan et al., 2011) which makes the correct determination of $\Delta C/T_c$ more difficult. In any case, these values for $\Delta C/T_c$ for FeSe$_{0.48}$Te$_{0.52}$/FeSe$_{0.43}$Te$_{0.57}$



lie well above the modified BNC fit value in Fig. 25 of $\Delta C/T_c$ for $T_c$=14 K of 12

mJ/moleK$^2$.   The C/T data of Braithwaite et al. below 2.5 K show an upturn, as has been

seen in the specific heat of other FePn/Ch superconductors (Kim, Kim, and Stewart,

2009).   However, this upturn is likely due to some magnetic impurity rather than a

fraction of the sample being normal, since C/T from above 2.5 K appears to extrapolate

to approximately zero in this sample.   The data of Hu et al. show $\gamma_r \approx$2.3 mJ/moleK$^2$ vs

$\gamma_n \approx$27 mJ/moleK$^2$.   Therefore, both values of $\Delta C/T_c$ for $FeSe_{1-x}Te_x$ should **a priori** be

approximately correct for intrinsic material.  Why the two values are so disparate does

not seem to be based on some obvious issue of sample quality.

In summary, most of the five additional 122 samples, two 111 examples, and two

11 examples, which are neither quantum critical nor show strong signs of pair breaking,

seem  approximately comparable to the 14 superconductors assembled by BNC for their

proposed correlation between $\Delta C/T_c$  and $T_c{}^2$.   However, the Hu et al. (2011) result for

$FeSe_{0.43}Te_{0.57}$, like that for $KFe_2As_2$, lies well above the BNC trend.

One question that J. S. Kim et al. (2011a) addressed is how such a plot of $\Delta C/T_c$

vs $T_c$ looks for conventional superconductors.  The answer is not simply $\Delta C/\gamma_n T_c \sim$

constant, therefore $\Delta C/T_c$ is also just a constant, independent of $T_c$.  Such a plot,

conventional superconductors together with the FePn/Ch data discussed above, was put

forward by J. S. Kim et al. (2011a) and – together with the additional data for FePn/Ch -

is the basis for Fig. 25.  All the superconducting elements with $T_c$>1 K are shown, as well

as representative A-15 superconductors, in order to provide $T_c$ values up to 20 K.  The

gamma values for the elemental superconductors are bounded by around 10 mJ/moleK$^2$



(V and La) (Stewart, 1983), while $\gamma_n$ values for the A-15's are several times larger (see references in J. S. Kim et al., 2011a). The slopes of the two $\Delta C/T_c$ vs $T_c$ sets of data for the conventional superconductors are clearly quite close, and in strong contrast to that for the FePn/Ch.

Thus, this modified BNC plot from J. S. Kim et al. (2011a) makes clear that whatever the pairing mechanism in the superconducting state in the FePn/Ch is, that this superconductivity is different in a fundamental fashion from conventional superconductivity. Broadly speaking, the electron-phonon coupled elemental and A-15 superconductors have a $\Delta C/T_c$ that is dependent on three factors: the electronic density of states at the Fermi energy, $N(0)$, the spectral density $\alpha^2 F(\omega)$ and the Coulomb pseudopotential $\mu^*$ (Carbotte, 1990). This dependence, using the slopes of the fits of $\Delta C/T_c$ to $T_c^\alpha$ in Fig. 25, says that for these superconductors a.) these three factors combine to give $\Delta C/T_c \sim T_c$ and b.) since in these superconductors $\Delta C/T_c$ roughly varies as $\gamma_n$, $T_c$ then (again broadly speaking) must vary as $\gamma_n$ ($\propto N(0)(1+\lambda_{el\text{-}ph})$). (In a less approximate fashion, in weak coupling BCS theory, $T_c \propto \exp(-1/N(0)V)$), where $(1+\lambda_{el\text{-}ph})N(0) \propto \gamma_n$.) This dependence of $T_c$ on the renormalized density of states in BCS superconductors derivable from Fig. 25 is of course the paradigm that drove the search for higher $T_c$ in the A-15 superconductors, with some success. It is also the paradigm that Bednorz and Mueller ignored to discover high $T_c$ superconductivity in the cuprates.

Now, the BNC plot suggests another paradigm, namely that whatever instead of (or in addition to) $N(0)$, $\alpha^2 F(\omega)$, and $\mu^*$ determines $\Delta C/T_c$ for the FePn/Ch, the result is that $\Delta C/T_c$ varies as $T_c^2$. As will be discussed below in the next subsection, even for the



FePn/Ch, $\Delta C/T_c$ – in so far as $\gamma_n$ values are known - remains approximately proportional to $\gamma_n$. Also, the measured $\gamma_n$'s (see section IIIB above) combined with calculations imply that $\gamma_n$ for the FePn/Ch comes primarily from $N(0)(1+\lambda_{el-el})$ since $\lambda_{el-ph}$ is negligible. Thus, since for the FePn/Ch $\Delta C/T_c \propto T_c^2$ and $\Delta C/T_c \propto \gamma_n \propto N(0)(1+\lambda_{el-el})$, the BNC plot has implications for how the superconducting transition temperature $T_c$ depends on the electron-electron interactions that are presumably involved in the superconducting pairing.

It is also interesting to note that, according to the quick look by J.S. Kim et al. (2011a) in Fig. 25 at the behavior for the heavy Fermion superconductors $CeIrIn_5$, $T_c$=0.4 K and $\Delta C/T_c$=500 mJ/moleK$^2$, $CeCoIn_5$, $T_c$=2.25 K and $\Delta C/T_c$=1740 mJ/moleK$^2$ as well as $CeCu_2Si_2$ and $UBe_{13}$ - which include non-Fermi liquid systems and unconventional superconductivity (d-wave gap for $CeCoIn_5$), see Pfleiderer (2009) - the FePn/Ch present *another kind* of unconventional superconductivity than the heavy Fermion superconductors. The further question – what about $\Delta C/T_c$ vs $T_c$ for the cuprates – runs into two difficulties in the cuprates: a.) $\Delta C$ is not easy to measure at such high transition temperatures due to the large phonon contribution to the total specific heat (e. g. $\Delta C$ in YBCO is just ≈1% of $C_{Total}(T_c)$), just as is the case for the FePn/Ch and b.) determining $\Delta C$ is complicated by the pseudogap behavior for some compositions that affects the specific heat above $T_c$. If however one considers $\Delta C/T_c$ vs $T_c$ for $La_{1-x}Sr_xCuO_4$, x=0.17, 0.22, 0.24, $T_c$'s from 17-25 K (other compositions can have similar $T_c$'s and much different $\Delta C$'s) and $YBa_2(Cu_{0.98}Zn_{0.02})_3O_7$, $T_c$=65 K (Loram et al., 2001), YBCO ($T_c$=91 K, Junod et al., 1997), $HgBa_2Ca_2Cu_3O_8$ ($T_c$=133K, Calemczuk et al., 1994), and



$Bi_{1.74}Sr_{1.88}Pb_{0.38}CuO_6$ ($T_c$=9.4, Wen et al., 2010a), then for this choice of cuprate systems $\Delta C/T_c \sim T_c^{1.05}$. Again, the FePn/Ch seem quite different in the behavior of $\Delta C$ with $T_c$.

In summary, the BNC plot provides a simple but insightful method for organizing data on the specific heat discontinuities at $T_c$. In addition, the BNC plot, vis a vis the discussion of $KFe_2As_2$, provides a simple test as to whether a material belongs to the FePn/Ch (magnetism/fluctuation dominated) class of superconductors. As with all the comparisons offered in this review, sample quality (e. g. in the 122*'s) is definitely an issue for reaching correct conclusions. Whether the different dependence of $\Delta C/T_c$ with $T_c$ for the FePn/Ch vs that of elemental and A-15 superconductors ($T_c^2$ vs $T_c$) can provide a link between the superconductivity and related parameters such as $\lambda_{el\text{-}el}$ might be an interesting path for theoretical investigation.

**4.) $\Delta C/\gamma_n T_c$:** In weak coupling BCS theory $\Delta C/\gamma_n T_c = 1.43$ and serves as a traditional method to estimate the coupling strength of a superconductor, with larger values implying stronger coupling. In a d-wave superconductor, $\Delta C/\gamma_n T_c$ is (in the calculation of Won and Maki, 1994) about 0.9. For superconductors with multiple gaps (which ARPES data – see section IVA2 below, as well as penetration depth, NMR, specific heat, tunneling, optical data, and a host of other measures, reveal for many of the FePn/Ch), $\Delta C/\gamma_n T_c$ can be a wide variety of values from above 1.43 to significantly below. For example, in the canonical two gap electron-phonon mediated superconductor $MgB_2$, the normalized discontinuity at $T_c$=38.7/37 K is $\Delta C/\gamma_n T_c$=1.3/0.9 (Bouquet et al., 2001/Wang et al., 2003), where the disagreement is apparently due to sample differences with the higher $T_c$ and $\Delta C/\gamma_n T_c$ coming from the sample with narrower $\Delta T_c$.



Now that both $\Delta C/T_c$ and $\gamma_n$ are accurately known for several FePn/Ch (believed to be unconventional) superconductors, with understood error bars, this ratio can be discussed in these specific cases. For $Ba_{0.6}K_{0.4}Fe_2As_2$, $T_c^{onset} = 37$ K, Kant et al. (2010) determine $\gamma_n = 49$ mJ/moleK$^2$ while Welp et al. (2009), with a sample with comparable $T_c^{onset}$ (35.5 K) determine $\Delta C/T_c = 100$ mJ/moleK$^2$. Thus, for $Ba_{0.6}K_{0.4}Fe_2As_2$, $\Delta C/\gamma_n T_c = 2.04$. Using the value of $\Delta C/T_c = 125$ mJ/moleK$^2$ from Popovich et al. (2010) for $Ba_{0.68}K_{0.32}Fe_2As_2$ and the appropriate $\gamma_n$ from Kant et al. (2010) of 53 mJ/moleK$^2$, this value of $\Delta C/\gamma_n T_c$ rises to 2.36, indicative of even stronger coupling. As will be discussed below in section IV, numerous measurement techniques (ARPES, penetration depth, NMR, tunneling and others) imply that K-doped $BaFe_2As_2$ has multiple superconducting energy gaps, i. e. a large value for $\Delta C/\gamma_n T_c$ is not a contraindication for multiple gaps in the FePn/Ch.

For annealed optimally doped $BaFe_{1.85}Co_{0.16}As_2$, Gofryk et al. (2011a,b) determine $\gamma_n = 22$ mJ/moleK$^2$ and $\Delta C/T_c = 33.6$ mJ/moleK$^2$. This gives ~1.5 for $\Delta C/\gamma_n T_c$, a more weak coupled value and consistent with their fit of their data to a two gap model. Finally, taking $\Delta C/T_c = 24$ mJ/moleK$^2$ for $KFe_2As_2$ from the equal area construction as discussed above, and $\gamma_n = 69$ mJ/moleK$^2$ (Fukazawa et al., 2009a, RRR=67), we obtain $\Delta C/\gamma_n T_c = 0.35$, presumably indicative of sample quality issues. However, a sample of $KFe_2As_2$ with even higher quality (J. S. Kim et al., 2011c, RRR=650) with $\gamma_n = 102$ mJ/moleK$^2$) and $\Delta C/T_c \approx 41$ mJ/moleK$^2$ still only has $\Delta C/\gamma_n T_c \approx 0.40$, arguing perhaps for a two gap model.



## IV. Superconducting pairing mechanism, Theory and Experiment; Symmetry and Structure of the Energy Gap

Approximately 8 years after the discovery of superconductivity in the cuprates (Bednorz and Muller, 1986), Tsuei et al. (1994) were able to show that the pairing symmetry was d-wave.  In less than half that time after the discovery of superconductivity in the iron pnictides (Kamihara, et al., 2008), thanks to the experience amassed studying the cuprates and heavy Fermion superconductors plus significantly improved experimental and theoretical tools, the question of the pairing symmetry is being heavily studied.  There is significant experimental evidence for some version of the so-called $s_\pm$ state, predicted first by Mazin et al. (2008) for the FePn superconductors, although predictions abound for other pairing states which may be dominant (e. g. the proposal for the $s_{++}$ state mediated by orbital fluctuations - see Kontani and Onari, 2010, Yanagi, Yamakawa, and Ono, 2010 and Kontani, Saito, and Onari, 2011) or coexist in the $s_\pm$ materials.  Fernandes and Schmalian (2010) (see also Vorontsov, Vavilov, and Chubukov, 2010) argue that - within their model for the magnetism and superconductivity (where the same electrons that form the superconducting pairs also cause the ordered moment) - the observed coexistence of antiferromagnetism and superconductivity in, e. g., underdoped $BaFe_{2-x}Co_xAs_2$, implies a sign changing $s_{+-}$ state and rules out $s_{++}$ pairing.  The discovery of superconductivity in the 122* materials, with the large local moment (3.3 $\mu_B$/Fe, Bao et al., 2011a) and different Fermi surface (no hole pockets, L. Zhao et al., 2011) seems at present to argue against the $s_\pm$ model being applicable to all the FePn/Ch, but see Mazin (2011) for a discussion.

Predictions for the actual superconducting pairing mechanism are quite broad in scope, with some concentration on spin fluctuations due to, among other reasons, the



nearness (sometimes coexistence) in the phase diagram of magnetism to the superconductivity and the inelastic neutron scattering evidence for at least some linkage between superconductivity and a spin fluctuation resonance peak below $T_c$ (section IVA1). Related ideas have been explored using phenomenological intra- and interband interaction parameters, leading to similar conclusions (Chubukov, 2009 and F. Wang et al., 2009).

## A. Theory of Superconductivity and Some Relevant Experiments in FePn/Ch

A number of authors have pointed out that the electron phonon coupling is too weak (by about a factor of five, Osborn et al., 2009) in these materials to account for the >20 K $T_c$'s. Boeri, Dolgov and Golubov's (2008) calculation of the Eliashberg $\alpha^2F(\omega)$ produces an electron phonon coupling parameter $\lambda_{el-ph}$~0.2, with a followup work in the magnetic state by Boeri et al. (2010) finding $\lambda_{el-ph}$≤0.35. As examples of experimental determinations, Rettig et al. (2010) find in the 122 parent compound $EuFe_2As_2$, using time resolved ARPES, that $\lambda_{el-ph}$<0.5 while Mansart et al. (2010) find in $BaFe_{1.84}Co_{0.16}$, $T_c$=24 K, using transient optical reflectivity that $\lambda_{el-ph}$≈0.12. However, there are several experimental works indicating an isotope effect (in BCS theory, $T_c \propto M^{-\alpha}$, $\alpha$=1/2), indicating some role of the phonons in the superconductivity. In $SmFeAsO_{0.85}F_{0.15}$, $T_c$=41 K, and $Ba_{0.6}K_{0.4}Fe_2As_2$, $T_c$=38 K, Liu et al. (2009b) find a conventional isotope effect, but only for the Fe: substitution of $^{54}$Fe for $^{56}$Fe results in an increase of $T_c$ proportional to $M^{-0.35}$ with essentially no isotope effect due to substitution of $^{18}$O for $^{16}$O. Thus, phonon modes involving the Fe may through a magnetoelastic effect affect the magnetic fluctuations and therefore superconductivity, but the results of Liu et al. argue against an electron-phonon pairing mechanism. Shirage et al. (2010) in oxygen deficient



SmFeAsO$_{1-y}$, T$_c$=54 K, find essentially no isotope effect on the Fe site, with α=0.02. Shirage et al. (2009) in contradiction to Liu et al. (2009b) find an *inverse* Fe-isotope in Ba$_{1-x}$K$_x$Fe$_2$As$_2$, T$_c$=38 K, with T$_c \propto$ M$^{+0.18}$. Khasanov, et al. (2010b) find a conventional Fe-isotope effect in FeSe$_{1-x}$, T$_c$=8.2 K, with – after some involved analysis (half of the change in T$_c$ with $^{54}$Fe isotopic enrichment is assigned to structural changes in the samples) – T$_c \propto$ M$^{-0.4}$. Khasonov, et al. (2010a), following the same analysis as used in their FeSe$_{1-x}$ isotope effect work, argue that – when adjusted for structural changes – Liu et al.'s and Shirage et al.'s results are also consistent with a conventional α≈0.35-0.4. Obviously, the possible partial role of the phonons in superconductivity in these materials is still not entirely decided but the evidence from the isotope measurements to date – with the possible exception of the low T$_c$ FeSe$_{1-x}$ – argues against electron-phonon coupling as the primary pairing mechanism.

Theorists, based on years of experience with the cuprate, heavy Fermion, and other exotic superconductors and on the clear inability of the electron phonon coupling to explain T$_c$, have proposed a number of electronic ("unconventional") pairing schemes (as opposed to the conventional, phononic, pairing) for the FePn/Ch materials. Beyond the short introduction to these ideas given here, the reader is referred to in-depth theoretical reviews, see articles by Mazin and Schmalian (2009), Boeri, Dolgov and Golubov (2009), Kuroki and Aoki (2009), Chubukov (2009), Korshunov, Hirshfeld and Mazin (2011) and references therein. For a discussion of the 122* superconductors, see the discussion by Mazin (2011).

Many of these proposals for the pairing center around the early idea of Mazin et al. (2008), that even if the excitation (e. g. spin fluctuations) being exchanged to produce



the coupling is repulsive it can still lead to attractive pairing if the excitation is being

exchanged between parts of the Fermi surface with opposite signs of the order parameter.

Simply put, if $\Delta_k = - \Delta_{k+Q}$ then a repulsive interaction with wave vector Q (Fig. 26) can be

attractive due to the sign reversal in the order parameter $\Delta$. This is a realization, specific

to the FePn/Ch materials' Fermi surface with several small pockets separated by Q, of the

general spin fluctuation pairing mechanism (Berk and Schrieffer, 1966; Scalapino, 1995).

See section IVA2 below for a discussion of the experimental work on the Fermiology of

the FePn/Ch, which – like the inelastic neutron scattering results discussed just below in

section IVA1 - is mostly consistent with the proposed spin fluctuation, electronic-in-

origin 'pairing glue' picture.  See also supporting evidence from optical conductivity

measurments, e. g. by Yang et al. (2009a).

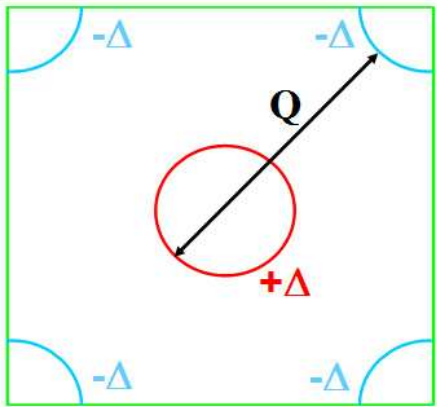

Fig. 26  (color online) Sketch of an idealized
Fermi surface of undoped FePn/Ch with the hole pocket (red) at the $\Gamma$ point (0,0) with
energy gap $+\Delta$, the electron pockets (blue) at the corner M (called 'X' in some works'
notation) points $(\pi,\pi)$ with energy gap $-\Delta$, and the spin density wave momentum wave
vector Q spanning the two nested pockets.  This schematic Brillouin zone (BZ) follows
the two Fe atoms/unit cell 'folded' BZ notation.  For a comparison with the 'unfolded'
BZ, one Fe/unit cell notation, see Chubukov (2009) or Korshunov, Hirschfeld, and Mazin
(2011).

### 1.)  Spin Resonance in INS below $T_c$



Early inelastic neutron scattering experiments in polycrystalline $Ba_{0.6}K_{0.4}Fe_2As_2$ (Christianson et al., 2008) found evidence – a magnetic resonance below $T_c$ – for a sign change (although see Onari, Kontano and Sato, 2010, for an opposing argument) in the superconducting energy gap $\Delta$ on different parts of the Fermi surface. Such a sign change in the order parameter is consistent with the $s_{+-}$ model and the Fermiology of the FePn/Ch sketched above in Fig. 26. For a system like $Ba_{0.6}K_{0.4}Fe_2As_2$, which experiments indicate is nodeless (see section IVB below), d-wave pairing would be ruled out. This type of collective excitation/resonant mode below $T_c$ is found in most of the cuprate superconductors (although with differences in, e. g., Sr-doped 214, see Tranquada et al., 2004) as discussed in the review by Eschrig, 2006 and the experimental work (and references therein) of Dai et al. (2000). In the cuprates, the resonance mode, which is thought to be a triplet excitation of ground state singlet Cooper pairs, is centered in k-space at the antiferromagnetic ordering wave vector and is 2D in behavior.

The first INS work on single crystals of $BaFe_{1.84}Co_{0.16}As_2$ (Lumsden et al., 2009) found that the magnetic fluctuations associated with the resonance were – just as in the cuprates - also 2D in nature. Follow up work on Ni-doped $BaFe_2As_2$ found instead different resonant energies at (1/2, 1/2, L) depending on whether L was even or odd, indicating dispersion along the c-axis (3D behavior). As shown in Table 4, this 3D character survives in overdoped $BaFe_{1.85}Ni_{0.15}As_2$ (M. Wang et al., 2010). As well, Park et al. (2010) have been able to find this dispersive behavior of the resonance fluctuations in Co-doped $BaFe_2As_2$.



Interestingly, INS studies (see Table 4) of $FeSe_{1-x}Te_x$ (Qiu et al., 2009; Wen et al., 2010b; Mook et al., 2009) find the wave vector of the resonance at the in-plane nesting vector between the electron and hole pockets (Fig. 26), or (1/2, 1/2, 0), like in the 122 single crystal work, and not at the 11 structure magnetic ordering wave vector ((1/2, 0, 0) as sketched in section IB, Fig. 8).   The case of non- magnetic LiFeAs, in which ARPES data discussed in the next section (IVA2) indicate that there is – due to the size and shape of the Fermi surface pockets - *no* nesting is also interesting.  Despite this lack of nesting and magnetism, INS studies of polycrystalline LiFeAs (Taylor et al., 2011) also find antiferromagnetic spin fluctuations (although no clear sign of a resonance) in the same (1/2,1/2) wavevector direction.  NMR results also report evidence for antiferromagnetic fluctuations in LiFeAs (polycrystalline work - Jeglic et al., 2010; single crystal work - Ma et al., 2010).

Bao et al. (2010) and others found using unpolarized INS that the resonant spin correlations in $FeSe_{1-x}Te_x$ were quasi-2D, just as Lumsden et al. (2009) reported in the first work on single crystal Co-doped $BaFe_2As_2$, in $BaFe_{1.84}Co_{0.16}As_2$.  Whether this 2D characterization of the 11 FeCh survives further investigation is an open question.

A general feature of the resonance in optimally doped 122 $BaFe_{2-x}(Co,Ni)_xAs_2$ and $FeSe_{0.4}Te_{0.6}$  material is that its spectral weight comes from a spin gap that opens at even lower energy ( $\leq$ 1/2 $E_{resonance}$) as temperature is lowered below $T_c$ (see, e. g., Chi et al., 2009, H.-F. Li et al., 2010 and Qiu et al., 2009).  For underdoped 122 $BaFe_{1.92}Co_{0.08}As_2$, this spin gap is not observed down to 2 meV (Christianson et al., 2009).  Note that in the 122's the underdoped samples all have coexistent magnetism and



# Table 4.  Spin Resonance Energies in the FePn/Ch

With the exception of the initial work and those on the 1111 and P-doped 122 samples, all the experiments have been on single crystals in order to determine the wavevector(s) unambiguously.

| Compound | $T_c$(K) | Resonance Energy(meV) | $E_r/k_B T_c$ | Ref. |
|---|---|---|---|---|
| $BaFe_{2-x}Co_xAs_2$ x=0.08 | 11 | 4.5 | 4.9 | a |
| x=0.094 | 17 | ~ 4.5 | 3.2 | b |
| x=0.13 | 23 | ~ 10 | 5.2 | c |
| x=0.148 | 22.2 | 8.3 | 4.5 | d |
| x=0.15 | 25 <br> 25 | 9.5 <br> 9.6, 10.5* | 4.6 <br> 4.6,5.0 | e <br> f |
| x=0.16 | 22 | 8.6 | 4.7 | g |
| $BaFe_{2-x}Ni_xAs_2$ x=0.075 | 12 | 5, 7* | 5.0, 7.0 | h |
| x=0.09 | 18 | 6.5, 8.8* | 4.3, 5.9 | f |
| x=0.1 | 20 | 7.0, 9.1* | 4.2, 5.5 | i |
| x=0.15 | 14 | 6, 8* | 5.1, 6.9 | h |
| $FeSe_{0.4}Te_{0.6}$ <br> $FeSe_{0.5}Te_{0.5}$ | 14/14.6 <br> 14 | 6.5/7.1 <br> 6/6.5 | 5.6 <br> ~5.6 | j/k <br> l/m |
| $LaFeAsO_{1-x}F_x$ x=0.057/0.082 | 25/29 | 11 | 5.3/4.6 | n |
| $Ba_{0.6}K_{0.4}Fe_2As_2$ | 38 | 14 | 4.4 | o |
| $BaFe_2(As_{0.65}P_{0.35})_2$ | 30 | 12 | 4.8 | p |

* Resonances at *two* wavevectors – ½, ½, 1 and ½, ½, 0 - with different energies.
a.  Christianson et al. (2009)  b.  Pratt et al. (2009a)  c.  Lester et al. (2010)  d.  H.-F. Li et al. (2010)  e.  Inosov et al. (2010a)  f.  Park et al. (2010)  g.  Lumsden et al. (2009)  h. M. Wang et al. (2010)  i.  Chi et al. (2009)  j.  Qiu et al. (2009)  k.  Bao et al. (2010)  l. Wen et al. (2010b)  m.  Mook et al. (2010)  n.  Wakimoto et al. (2010)  o.  Christianson et al. (2008)  p.  Ishikado et al. (2010)



superconductivity (discussed with the phase diagrams in Section IIB2b), while in the optimally and overdoped materials $T_{SDW}$ is suppressed. Indeed, Lumsden and Christianson (2010) point out that the spectral weight for the resonance in underdoped $BaFe_{2-x}Co_xAs_2$ may indeed come from the observed suppression of the spectral weight in the magnetic Bragg peaks below $T_c$.

In agreement with cuprate work, INS studies (see, e. g., Chi et al., 2009, H.-F. Li et al., 2010, Inosov et al., 2010a) of the FePn/Ch superconductors have found that the intensity associated with the spin fluctuation resonance increases with decreasing temperature below $T_c$ similar to the superconducting order parameter itself. Based on these results, one of the possible conclusions is that if the superconducting order parameter and the spin resonance are indeed linked in a causal fashion, then the order parameter – at least in Co- and Ni-doped $BaFe_2As_2$ - is 3D and should depend sensitively on the c-axis wave vectors. See, e. g., M. Wang et al. (2010) for further discussion of this.

When discussing the magnetic resonance in cuprates, it is common to point out that there is an approximately uniform scaling of the resonance energy with $T_c$, implying that the resonance is intimately connected to the superconductivity. In the cuprates, Hüfner et al. (2008) state that $E_{resonance}$ is about $5k_BT_c$. Discussion of this scaling in the FePn/Ch (see Table 4) is complicated by the dispersion of $E_{resonance}$ along the c-axis, as discussed explicitly by M. Wang et al. (2010). As Table 4 makes clear, there is in addition significant scatter in some of the values. This leads to a breadth in quoted values for the average $E_{resonance}/k_BT_c$ (~4.9, Lumsden and Christianson, 2010; ~ 4.3, Park et al., 2010). In any case, the scaling argument made in the cuprates for the resonance appears



to be valid in the FePn/Ch as well, with the caveat that there may be differences between, e. g., the 122's and the 11's.

Another method for investigating the resonance in the superconducting state of the FePn/Ch is to measure its field dependence. If the applied field depresses the intensity and energy of the resonance similarly to its reduction of the superconducting energy gap $\Delta$, this would provide a link between the two like the observed similar temperature dependence. In $BaFe_{1.9}Ni_{0.1}As_2$, $T_c$=20 K, J. Zhao et al. (2010) find that a 14.5 T applied field suppresses $E_{resonance}$ and the associated neutron scattering intensity both by ~ 20%, while $T_c$ is also suppressed by 20% to 16 K. They argue that their data are evidence that the resonance is related to the superconducting $\Delta$. Wang et al. (2011), in a neutron scattering study of underdoped $BaFe_{1.92}Ni_{0.08}As_2$ ($T_c$=17 K, $T_{SDW}$=44 K) in zero and 10 T, find that the intensity of the INS resonance below $T_c$ is reduced by field while the static antiferromagnetic order is enhanced. They argue that therefore the magnetic order competes with the superconducting order, similar to some of the cuprate superconductors.

A further use of magnetic field for probing the magnetic resonance below $T_c$ in the FePn/Ch has been the work of Bao, et al. (2010). They applied 14 T to an optimized set of single crystals of $FeSe_{0.4}Te_{0.6}$ with a smaller mosaic spread than in previous INS works, and succeeded in their high resolution experiment in finding that the resonance peak splits into a set of three equal intensity peaks in field, a signature of a triplet excited state.

In another work that bears on the question of the triplet character of the resonance in the FePn/Ch, Lipscombe et al. (2010), performed a polarized INS experiment



(previous work discussed above in this section has been with unpolarized neutron sources) on a different material, $BaFe_{1.9}Ni_{0.1}As_2$. Their results are inconsistent with the usual understanding of the magnetic resonance in the cuprates (Eschrig, 2006) as being an isotropic triplet excited state of the ground state Cooper pair singlet, since their polarized neutron results are able to resolve an anisotropy in the resonance. In contrast to this, but in agreement with the magnetic field work of Bao, Babkevich et al. (2010) find using polarized INS in $FeSe_{0.5}Te_{0.5}$ (comparable to Bao et al.'s $FeSe_{0.4}Te_{0.6}$ sample) a 'quasi-isotropic' resonance consistent with the triplet excitation scenario.

These INS works on the magnetic resonance in the superconducting state of the FePn/Ch indicate that the iron containing superconductors have fundamental differences in their behavior. Although it is too early to reach a firm conclusion, certainly these resonance studies are of great interest since many theories posit that the FePn/Ch superconductivity is mediated by spin fluctuations/magnetic excitations. In terms of actual calculations of the strength of the INS-detected fluctuation resonances and their wavevector, Maier and Scalapino (2008) calculate for which gap functions and for which wavevectors resonances in the dynamic spin susceptibility occur. They find for Mazin's predicted $s_\pm$ gap a predicted resonance in the (1/2, 1/2) wavevector direction that matches the antiferromagnetic ordering vector, as well as resonances for two triplet p-wave gaps. Maier et al. (2009a), in a following calculation, find in addition to the prediction for the strongest resonance being for q ∥ (1/2, 1/2) and an $s_\pm$ gap, two other weaker possible resonances for a non-sign-changing extended s-wave gap and a $d_{x2-y2}$ gap. They argue for further INS measurements along other wavevectors to distinguish which gap is causing the observed resonance.



As well from the experimental perspective, Wu et al. (2010) – based on a strong similarity between their optical-conductivity-derived $\alpha^2F(\omega)$ electron-boson spectral function and the INS-determined spin excitation spectrum in optimally doped $BaFe_{2-x}Co_xAs_2$ - argue that the charge carriers in these superconductors are strongly coupled to the spin fluctuations.  Thus, thorough studies of this resonance continue to be one of the best approaches (see also experimental determination of the nodal structure below in section IVB) in use to help elucidate the relation between magnetism and superconductivity in these new superconductors.

### 2.) Fermiology in the FePn/Ch:  Theory and Experiment

**Theory:**  The calculated Fermi surfaces of undoped LaFeAsO (Singh and Du, 2008) have two electron cylinders around the tetragonal M point, plus two hole cylinders and a hole pocket around the $\Gamma$ point.  Similar results for the Fermiology of LaFePO - the first reported superconducting ($T_c \approx 5$ K) iron pnictide, Kamihara et al., 2006, - were obtained by Lebegue (2007).  Mazin et al.'s (2008) calculation of the Fermiology for F (electron)-doped superconducting $LaFeAsO_{1-x}F_x$ resulted in a somewhat simplified Fermi surface, with the hole pocket filled.  (See the experimental ARPES determinations of the Fermi surface of K-doped $BaFe_2As_2$ in Figs. 28 and 29.)

Due to the nearness (even, in parts of the phase diagram in some samples, coexistence) of magnetism (section II), Mazin et al. (2008) proposed spin-fluctuation-mediated pairing (weak coupling) for wave vectors connecting the electron and hole cylinders, the so-called "$s_+$" pairing state, while rejecting the other possible spin-fluctuation-induced order parameter, i. e. triplet pairing.



Numerous authors have also discussed spin-fluctuation mediated pairing in the FePn/Ch, with some theories stating that the "$s_\pm$" (also known as "sign-reversing s-wave" or "extended s-wave") is the only pairing symmetry allowed (Y. L. Wang et al. (2009), Maier et al. (2009b), Chubukov et al. (2009)), while some give d-wave pairing as the preferred state for particular values of the parameters chosen (Kuroki et al. (2008), Graser et al. (2009), Kuroki et al. (2009), Thomale et al. (2009), Ikeda et al. (2010)). 'Nesting' between cylinders at a Fermi surface implies that one of the cylinders, when shifted over another, would be a close match in shape and size (see also Fig. 26

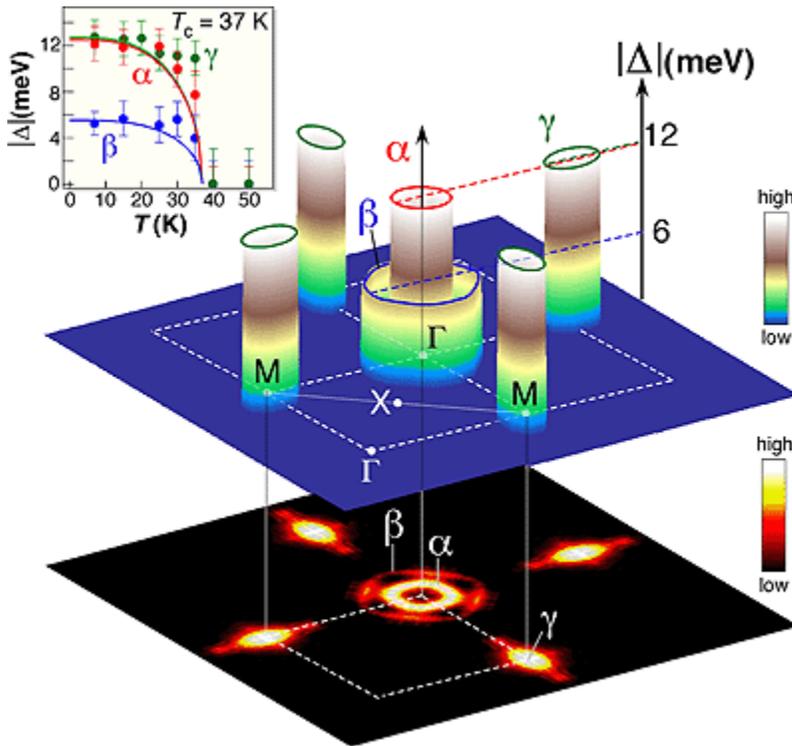

Fig. 27 (color online) Schematic picture of the Fermi surface in $Ba_{1-x}K_xFe_2As_2$ determined by ARPES measurements (H. Ding et al., 2008). The color bars denote the size of the energy gap, and the upper left inset displays the temperature dependence of the gaps on the three Fermi surface sheets (note the two different sized $\Delta$'s). The $\alpha$ hole-like pocket and $\beta$ hole-like sheet are both centered at the Brillouin zone center $\Gamma$ while the electron-like $\gamma$ Fermi sheet is centered at the M point.

where the hole and electron pockets in the idealized sketch show perfect nesting). The nesting between the cylinders in Fig. 27 and concomitant measured susceptibility peak at this wave vector are the motivation for the spin-fluctuation pairing mechanism in several theories. A large amount of nesting of states at the Fermi energy is not necessary for the applicability of



these theories (nesting changes with doping since the size of the cylinders changes with hole or electron addition to the respective pockets as discussed below when the ARPES data are reviewed). In fact, Platt, Thomale, and Hanke (2011) – using a theory that takes into account orbital dependent interactions – propose that LiFeAs, which as discussed in the experimental section just below has according to ARPES *no* nesting, also has an $s_{+-}$ order parameter caused by antiferromagnetic fluctuations. As discussed above in the INS section (IVA1) such fluctuations have now been experimentally found (Taylor et al., 2011).

Some theories have posited that p-wave (triplet) pairing is possible (X.-L. Qi et al., 2008; Lee and Wen, 2008; Brydon et al., 2011). Theories of the FePn/Ch superconductors are further split into subgroups depending on whether they involve strong or weak coupling of the magnetic excitations and whether the predicted pairing states are nodeless or have gaps. The predicted extended *s*-wave symmetry can be either nodeless or have nodes, depending on the interplay between intraband and interband interactions (Chubukov et al. 2009), which can be tuned by small changes in the electronic structure (Kemper et al., 2010), e. g., by moderate hole doping in $Ba_{1-x}K_xFe_2As_2$ (for a discussion, see Thomale et al., 2011) or by adjustment of the pnictogen height by substituting P for As (Kuroki et al., 2009)). Upon further hole doping in $Ba_{1-x}K_xFe_2As_2$ to $KFe_2As_2$ Thomale et al. (2011) argue that the modification of the Fermi surface by fulling gapping the electron pockets leads to nodal $d_{xy}$-wave behavior. Interestingly, at the other end of the doping spectrum, the 122* $A_xFe_{2-y}Se_2$ (which, according to ARPES data by L. Zhao et al., 2011 and references therein, have only electron pockets on the Fermi surface) are predicted (Maier et al., 2011; F. Wang et al.,



2011) to have nodeless $d_{x2-y2}$-wave pairing symmetry (although see Mazin, 2011 and Fang et al., 2011a for counterarguments).  Indeed, the richness of the Fermiology in the FePn/Ch involves more than just the large number of pockets (up to five) at the Fermi energy, their nesting, and their multi-orbital (see following experimental section for a discussion) character.  The variation of the gap structure and superconducting transition temperature across a particular phase diagram with doping adds another dimension to this richness.

**Experiment:**  Angle resolved photoemission spectroscopy (ARPES) on single crystals is a very powerful tool that resolves both the Fermi surface structure in momentum space and also the spectra of the electronic states near the Fermi energy.  For an early review of ARPES investigations of the FePn/Ch, see Liu et al. (2009a).

ARPES can show the size, shape, and position in momentum space of the predicted Fermi surface pockets, allowing the verification of the extent of Fermi surface nesting – important as discussed above in numerous theories for the role of spin fluctuations in the superconducting pairing mechanism.  As well, ARPES data can show the evolution of the Fermi surface pockets with doping, for example the hole pocket at the $\Gamma$ point in undoped $BaFe_2As_2/SrFe_2As_2$ expanding with K, i. e. hole, doping.  This evolution is, to a first approximation, describable by a rigid band model (C. Liu et al., 2008; Malaeb et al., 2009/Y. Zhang et al., 2009), although as discussed above in Section IIB2a the variation of $T_c$ with isoelectronic doping makes clear that such a rigid band picture is oversimplified.  Further, ARPES has been used to measure the *magnitudes* of the superconducting gap(s) in the FePn/Ch (see, for example, the inset in Fig. 27 for the two gaps found in K-doped $BaFe_2As_2$ by Ding et al. 2008).  Evtushinsky et al. (2009b)



list the magnitudes of the superconducting energy gaps determined via ARPES and other measurement techniques (for a discussion of some of these techniques, see section IVB below), showing good agreement between the methods.   For determining the *symmetry* of the gap in momentum space ARPES - due to the complexity and difficulty of the method as well as partially due to the inherent error bar (quoted to be ~ 20% in a work on 1111 material by Kondo et al., 2008) - is less used than other methods (section IVB).   It is interesting to note that one of the puzzles of the research to date in the FePn/Ch is that ARPES measurements – despite their success in the cuprates is finding nodes (Damascelli, Shen and Hussain, 2003) and despite there being (see Section IVB) a wealth of other experimental evidence for nodal behavior in the FePn/Ch – in general are interpreted as consistent with fully gapped behavior.

As with any measurement technique, ARPES measurements also have limitations, among them a resolution of at best several (sometimes as high as 15) meV, and a sensitivity to surface physics.   For a discussion of some of these experimental limitations, see Yi et al. (2009) and van Heumen et al. (2011), as well as the theoretical discussion of Kemper et al. (2010) on the sensitivity of the surface band structure in the FePn/Ch to small perturbations.   Van Heumen et al. show that the standard methods for preparing a clean surface for ARPES measurments (cleaving at low temperatures) in $BaFe_{2-x}Co_xAs_2$ create surface states which broaden the ARPES spectra and also cause a surface related band (which can be annealed away by warming to 150 K, following by recooling) not characteristic of the bulk.   This is similar to ARPES results for 1111's (Liu et al., 2010b).   It should be noted that the surface in LiFeAs – due to the surface chemistry - does not (Lankau et al., 2010) have such an influence on ARPES results.



There has been a large amount of ARPES work to characterize these new FePn/Ch superconductors. Work to date, because of the size and quality of the single crystals, has been focused in the 122 and 11 structures, which as an exception to the normal sequence in this review will be discussed first in this section, with some results in the 1111 (where of course for undoped LaFePO sizeable crystals exist but also including work on 200x200x50 μm crystals of $NdFeAsO_{0.9}F_{0.1}$, see Kondo et al., 2008), 111, 21311, and 122* materials discussed afterwards. As will be seen, and as follows a recurring theme in this review, there are important differences in the ARPES-determined Fermiology for the various structures, particularly for the nesting, which is important for the theories of spin-fluctuation-mediated superconductivity. ARPES data for the FePn/Ch, with their strong Fe conduction bands (width ~ 4 eV) which have significant densities of states at the Fermi energy, strongly contrast with those for the cuprates (for a review of ARPES in the cuprates, see Damascelli, Hussain, and Shen (2003).

**122:** In the early ARPES work of Ding et al. (2008) (Fig. 27), in K-doped $BaFe_2As_2$ the general topology of five Fermi surface sheets (vs one in the cuprates) matching the calculations was clearly revealed. The schematic nature of the pockets, i. e. the cylindrical shape, in the 122 compounds has been refined by more recent work of, e. g., Malaeb et al. (2009) in both $BaFe_2As_2$ and $BaFe_{1.86}Co_{0.14}$ to show significant variation of the size of the pocket in the $k_x$-$k_y$ plane along the z-axis – particularly around the Γ point, giving a 3D character. This 3D variation is seen even in the parent $BaFe_2As_2$ but is accentuated around both the Brillouin zone hole Γ center and electron M corner pockets in the doped compound. This 3D character in $BaFe_2As_2$ and its derivatives is consistent with ARPES work on the other 122's, see e. g. Hsieh et al. (2008) ($SrFe_2As_2$),



Kondo et al (2010) (CaFe$_2$As$_2$) and Zhou et al. (2010) (EuFe$_2$As$_2$) and with calculations, see e. g. Ma, Lu and Xiang et al. (2010) for DFT calculations on MFe$_2$As$_2$, M=Ba, Sr, Ca.

Another refinement of the Fermiology in K-doped BaFe$_2$As$_2$ was carried out by Zabolotnyy et al. (2009), using improved energy resolution. They found (in disagreement with calculations and the early ARPES work) – instead of the double walled electron pocket at the M point shown in Fig. 27 - a central circular pocket surrounded by four 'blade' shaped pockets, described as like the shape of a propeller. This result was refined by Evtushinsky et al. (2009a), see Fig. 28, who determined the superconducting gap in K-doped BaFe$_2$As$_2$ in all of these pockets, with the result that the gap on the inner barrel at Γ and in the inner circular pocket and outer blades at M was approximately the same at 9 meV, while the gap on the outer barrel at Γ was only ~ 4 meV.

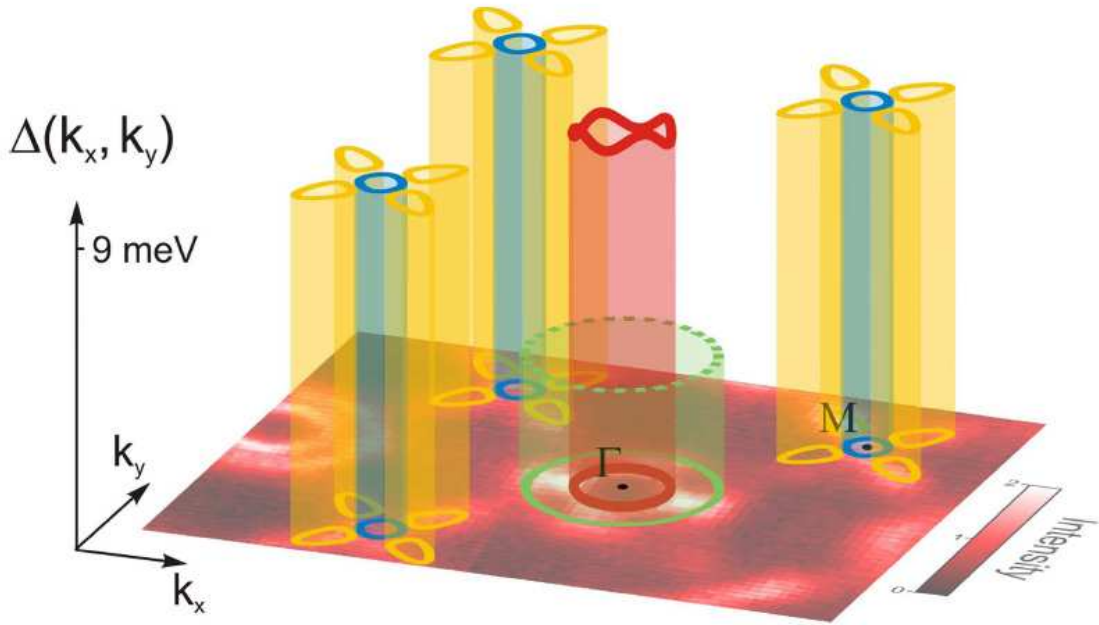



Fig. 28 (color online) ARPES determined Fermi surfaces in K-doped $BaFe_2As_2$ (Evtushinsky et al., 2009a). Note the propeller shaped five electron pockets at the M points.

The Fermiology in the parent compounds exhibits a temperature dependence due to the strong influence of the magnetic moment on the band structure below $T_{SDW}$. Yi et al. (2009) detail the Fermi surface reconstruction below ~135 K in $BaFe_2As_2$, with multiple new bands appearing. Below $T_{SDW}$ their ARPES data show, in addition to two hole pockets centered at the $\Gamma$ point, the appearance of four small surrounding 'petal shaped' electron pockets while at the M point four hole-like bands exist below $T_{SDW}$ that merge into one above. P. Richard et al. (2010) using ARPES find the creation of 'tiny Fermi surface pockets' below $T_{SDW}$ in $BaFe_2As_2$ due to a Dirac cone in the electronic structure below $T_{SDW}$. In a follow up ARPES work, Liu et al. (2010a) follow the evolution of Yi et al.'s magnetic-order-induced additional hole-like pockets at the M point in $BaFe_{2-x}Co_xAs_2$ as a function of Co-doping and find that they disappear at the point in the phase diagram where superconductivity appears. Liu et al. advance the plausible (but not conclusive) argument that the pairing interaction due to spin fluctuations is suppressed by the long range magnetic order, which is indicated by the additional Fermi surface features. They also show that there is no nesting between the $\Gamma$ and M point Fermi surface pockets at x=0.114 even though there is still superconductivity ($T_c$=12.8 K), yet another argument that nesting is not necessary for superconductivity (see the discussion of ARPES in the 111 and 122* materials below for a similar result.)

Recent ARPES work with improved (~ 10 meV) resolution (Yoshida et al., 2010a) on the strongly hole-doped end point of $Ba_{1-x}K_xFe_2As_2$, i. e. on pure $KFe_2As_2$, $T_c$



= 4 K, reveals three hole pockets (vs two in earlier work, Sato et al., 2009) at the zone center Γ point and, as expected from calculation, a small *hole* pocket (due to the strong hole doping) at the M points.  The Fermi surface pockets, in contrast to the other 122 results discussed above, are nearly 2D in character and, due to the strong hole doping, have no electron pockets (no nesting.)  An additional hole band near the hole center is seen in the ARPES data that is not in the calculation.  Yoshida et al. (2010a) speculate this may due to surface states, again illustrating the difficulties of this very surface-sensitive measurement.

As mentioned above in the introduction to this experimental ARPES section, ARPES data in general do not find nodal behavior in the FePn/Ch.  This is true, for example, in the prototypical (section IVB) nodal case, P-doped $BaFe_2As_2$, where Yoshida et al. (2010b), using synchrotron radiation with an energy resolution of 15 meV, find no evidence of nodes.

**11:**  ARPES studies of the 11 materials are to date more limited in number.  In the parent compounds, Xia et al. (2009) find in $Fe_{1+x}Te$ a hole pocket at the Γ point and four electron pockets at the corner M points - similar to calculations (Subedi et al., 2008) and to the experimental results for the 1111's and the 122's as sketched in Fig. 26.  Unlike other magnetically-ordered parent compounds, however, Xia et al. find no evidence for a SDW nesting-driven gap in the bands below $T_{SDW}$, ~ 70 K, in $Fe_{1+x}Te$.  This is consistent with the discussion above in section IIB (see Fig. 8) about the magnetic ordering wave vector in the 11's (1/2, 0) not being in the same direction (1/2, 1/2) as links the nested electron and hole pockets.   However, it is worth pointing out that – as discussed above in section IVA1 – the INS-determined spin resonance below $T_c$ in the doped 11 structure



superconductors shows spin fluctuations indeed in the electron-hole pocket nesting vector direction.

**1111:** As mentioned in the introduction to this section, ARPES data have been measured on $NdFeAsO_{0.9}F_{0.1}$, $T_c=53$ K, and LaFePO, $T_c=5.9$ K . In $NdFeAsO_{0.9}F_{0.1}$ Kondo et al. (2008) report the same Fermiology as reported for the 122's and concentrate on measuring the magnitude of the superconducting gap. The gap at the $\Gamma$ point is found to be 15 meV, with no measureable nodes or anisotropy within their error limits. Early ARPES work (Lu et al., 2008) on LaFePO found reasonable agreement with LDA calculations and the usual five Fermi sheets, with hole pockets centered at the $\Gamma$ point (based on $d_{xz}$ and $d_{yz}$ Fe orbitals for the inner pocket and based on Fe $d_{3z2-r2}$ states hybridized with P p orbitals and La orbitals for the outer pocket) and electron pockets at the M point.

**111:** Although the Fermi surface (Borisenko et al., 2010) of 111 LiFeAs has qualitative similarities to the 122, 11 and 1111 topologies just discussed (i. e. the requisite five Fermi surface pockets corresponding to the five Fe 3d bands, with three hole-like FS's around the $\Gamma$-point and two electron-like ones at the corner of the Brillouin zone, with 3D character somewhat reduced vs the 122 structure), there is one important difference. As Borisenko et al. (2010) point out, the disparate sizes of these five pockets at $\Gamma$ and M argues against any (1/2, 1/2) nesting at all. This could be used as an argument for nesting being important for magnetism (see the counter arguments of Johannes and Mazin, 2009, discussed in section IIB) since LiFeAs is not magnetic. Borisenko et al. (2010) further report an isotropic energy gap of ~ 3 meV in the double walled electron cylindrical pocket at the M point in LiFeAs.



**21311:** Single crystals of a few tenths of a mm on a side of $Sr_2VO_3FeAs$ have been measured using ARPES (Qian et al., 2011). The results show some nesting between the outer ($\beta$) of two circular hole pockets at the $\Gamma$ point and the outer ($\delta$) of two elliptical electron pockets at the M point, making the 21311's similar to the 122, 11, and 1111 structures in their nesting.

**122*:** In the early ARPES work on these superconductors there were sample quality issues. L. Zhao et al. (2011) report unifying results on single crystals of $K_{0.68}Fe_{1.79}Se_2$ and $(Tl_{0.45}K_{0.34})Fe_{1.84}Se_2$ (composition determined by EDX spectroscopy), $T_c$=32 and 28 K respectively. In both materials they find at the zone center $\Gamma$ two electron pockets, a small one they label $\alpha$ and a low intensity, larger pocket labeled $\beta$, and at the zone corner M an electron pocket labeled $\gamma$ similar in size to the $\beta$ pocket. The energy gaps for the $\gamma$ pocket in both materials are $\approx$8-9 meV and fairly isotropic. These features are similar to those found in ARPES measurements on $Tl_{0.58}Rb_{0.42}Fe_{1.72}Se_2$ (Mou et al., 2011, $\Gamma$ and M pockets gaps of 15 and 12 meV respectively) and in previous measurements of $Tl_{0.63}K_{0.37}Fe_{1.78}Se_2$ (X.-P. Wang et al., 2011, $\Gamma$ and M pockets gaps both $\approx$8 meV). Although all three of these ARPES works claim their results imply nodeless behavior in the 122*'s, due to sensitivity and energy resolution issues this is not conclusive. X.-P. Wang et al. report that there is a hole pocket approximately 50 meV below the Fermi energy at the $\Gamma$ point which F. Wang et al. (2011) note could have an important influence on the pairing interaction.

## B. Experimental Probes of the nodal structure

Understanding the pairing mechanism in the FePn/Ch superconductors is a central goal to the study of these materials. In a 'conventional' superconductor, the



superconducting gap – barring strong impurity effects – is nodeless, and the temperature dependence of a number of experimental probes is exponential, $\propto \exp(-\Delta/T)$. The nodal structure in the FePn/Ch superconductors is thus heavily studied deep in the superconducting state, $T<<T_c$, for clues about the pairing symmetry and thus the pairing mechanism although defect scattering can play an important role in the nodal structure, e. g. gapped behavior may arise through intraband defect scattering (Mishra et al., 2009a). In the discussion of the Fermiology above (Section IVA2), a number of theories and their predictions for the pairing symmetry were mentioned. The possible underlying pairing mechanisms are many and varied as discussed in the theory section above (in the introduction to Section IV and in Section IVA) and in the several reviews cited there. While exchange of spin fluctuations as the pairing mechanism has at present somewhat more experimental support (see Sections IVA and IVA1), there is certainly no consensus within sight at this time. Thus, the experimental study of the nodal structure is important to provide further clues to the pairing mechanism responsible for the rather high $T_c$ values found in the FePn/Ch.

The generally accepted fact that the FePn/Ch superconductors have multiple bands at the Fermi surface (see, e. g., the ARPES data in Figs. 28 and 29) creates a variety of possibilities for the gap structure. As has been pointed out by Kemper et al. (2010), this multiplicity of nearly compensated electron and hole Fermi surfaces (excluding of course the 122* structure and $KFe_2As_2$) and the concomitant sensitivity of various properties, including the nodal structure, to small changes in atomic and/or electronic structure makes the FePn/Ch 'quite special'. Further, Kemper et al. (2010) issue a warning that is important to remember during the remainder of this section: the sensitivity of the band



structure may cause surface probes of the nodal structure to return evidence for a nodeless, fully gapped superconductor while the bulk behavior might in fact be nodal. More generally, measurements that probe primarily the surface are sensitive to small changes that in the FePn/Ch can have important impact, see e. g. the discussion of ARPES above (Section IVA2) and the work by van Heumen et al. (2011) on surface reconstruction effects. Thus, in order to experimentally determine the nodal structure, it is important to consider more than just one experimental method, preferably including at least one bulk probe. Even when such multiple results exist, it is well to remember (P. Hirschfeld, 2011) that $\kappa$ and $\lambda$, unlike the specific heat, are weighted by the Fermi velocity, $v_F$, and may be dominated by nodal behavior from a small, high $v_F$ part of the Fermi surface - resulting in $\kappa$ and/or $\lambda$ measurements implying nodes in a system, while specific heat data imply a fully gapped superconductor. This may be more of an issue in the future as more specific heat data in field and as a function of angle become available.

Like ARPES just discussed, infrared optical spectroscopy - see e. g. Li et al., 2008, Dubroka et al., 2008, Cheng et al., 2011, Gorshunov et al., 2010, Tu et al., 2010, and the review by Dressel et al., 2010 - is more used to determine the *size* of the gap rather than its symmetry (although see Carbotte and Schachinger, 2010, for theoretical modeling of how optics could provide more information about the nodes in the FePn/Ch.) The experimental probes used in the study of the nodal structure in the FePn/Ch that will be discussed here are penetration depth ($\Delta\lambda(T)$), NMR spin lattice relaxation time ($1/T_1$), specific heat $C/T(T{\to}0)$ ($\gamma$), thermal conductivity ($\kappa/T$), Andreev spectroscopy, Josephson tunneling and Raman scattering. The results to date of these experimental probes are both numerous and often self-contradictory. Reasons for these contradictions



range from the trivial, including sample quality, to rather subtle.  As an example of the latter, the complicated Fermiology and multiple bands return different results to probes that measure differing parts of the Fermi surface.  Thus, measurement of the thermal conductivity, $\kappa$ (dominated by the light electron sheets on the Fermi surface) in P-doped $BaFe_2As_2$ up to 12 T ($H_{c2}(0)=52$ T) returns $\kappa \sim H^{1/2}$ which implies (Hashimoto et al., 2009b) a gap with nodes.  In contrast, the specific heat (dominated by the heavy hole sheets) on the same sample as a function of field up to 15 T appeared to result in $\gamma \sim H^1$ which implies (J. S. Kim et al., 2010a) fully gapped behavior.  Recent measurements in P-doped $BaFe_2As_2$ (Y. Wang et al., 2011) focused on the *low* field $\gamma$ (up to 4 T) does in fact reveal $\gamma \sim H^{1/2}$ and will be discussed below in the specific heat subsection, IVB3.

In order to provide a way to follow this involved discussion, it is useful to note that, despite all the disagreements, some compounds – as is thoroughly discussed below – show mostly concurring evidence for nodes, and for some there is fairly good agreement for fully gapped behavior.  As a short summary, a list of the **nodal** FePn/Ch superconductors and the supporting data would include **LaFePO** ($\Delta\lambda(T) \propto T$, analysis of $\kappa(T)$), **KFe_2As_2** ($\Delta\lambda(T) \propto T$, large value of $\kappa/T$ as $T \rightarrow 0$, $\kappa(H)/T \propto H^{1/2}$), **P-doped BaFe_2As_2** ($\Delta\lambda(T) \propto T$, $1/T_1 \propto T$, significant value of $\kappa/T$ as $T \rightarrow 0$, $\kappa(H)/T \propto H^{1/2}$, $\gamma \propto H^{1/2}$ for $H < 0.1 H_{c2}$), and **overdoped BaFe_{2-x}Co_xAs_2** ($\kappa(H)/T \propto H^{1/2}$, $\gamma \propto H^{0.7}$).  It is interesting to note that the first two of these are low $T_c$ materials, $T_c \approx 5$-6 and 3.4 K respectively, and that $KFe_2As_2$ has – due to K being monovalent -  a much different (Hashimoto et al., 2010a) Fermiology (including no nesting and 2D behavior, as discussed above in section IVA2) than the other 122 FePn/Ch superconductors.   In fact, as noted above in Section IIIB3 in the discussion of $\Delta C/T_c$, $KFe_2As_2$ may be more comparable to an electron-phonon



coupled superconductor. A list of the ***fully gapped*** materials would include ***Ba$_{1-x}$K$_x$Fe$_2$As$_2$*** (analysis of $\Delta\lambda(T)$, $\gamma\propto H^1$, $\kappa/T\approx 0$ as $T\rightarrow 0$) and ***underdoped BaFe$_{2-x}$Co$_x$As$_2$*** (analysis of $\Delta\lambda(T)$, $\kappa/T\approx 0$ as $T\rightarrow 0$).

Even within this short list, there are contradictions. For the supposed **nodal** systems, $\kappa/T\approx 0$ as $T\rightarrow 0$ (consistent with gapped behavior) for overdoped BaFe$_{2-x}$Co$_x$As$_2$. (However, note that nodes have been reported in c-axis thermal conductivity measurements for overdoped BaFe$_{2-x}$Co$_x$As$_2$, Reid et al., 2010; Mishra et al., 2011). For the putative ***fully gapped*** systems, some NMR $1/T_1$ data for Ba$_{1-x}$K$_x$Fe$_2$As$_2$ indicate nodal behavior and specific heat in field data for underdoped BaFe$_{2-x}$Co$_x$As$_2$ gives $\gamma \propto H^{0.7}$ over a broad field range just like in the overdoped, believed-to-be-nodal material.

It is notable that these conclusions about nodal structure are not consistent within a given structure, nor sometimes even within a given doping series, with underdoped BaFe$_{2-x}$Co$_x$As$_2$ different than overdoped (although not according to the $\gamma\propto H^{0.7}$ data).

Finally, before beginning the discussion of the experimental data, we list some caveats. In discussing systems where the experimental probes do not find exponential (fully gapped) temperature dependences, nodes caused by the underlying symmetry of the superconducting order parameter (of interest for understanding the superconducting pairing mechanism) should be distinguished from states in the superconducting gap caused by defects. In the case of realistic materials with unavoidable defects, states in the superconducting gap at the Fermi energy due to defects will of course cause a finite $\gamma_r$. Further, if these defect states are extended (offering a complete path in real space), then $\kappa/T$ will also be finite. Nodes in the s$_\pm$ scenario are accidental if they exist, and are not symmetry driven. Note that deep minima in the gap (see, e. g., Tanatar et al., 2010b) can



mimic nodal behavior in measurements done as a function of temperature unless measurements are done to very low (dilution refrigerator) temperature. On the other hand, measurements in fields of several Tesla in materials with deep minima in the gap will mimic nodal behavior at low ($\leq$ several Kelvin) temperature, since the field energy scale is much larger than the milliKelvin gap scale.

### 1.) Penetration Depth Measurements

The temperature dependence of the London magnetic field penetration depth below $T_c$ can give information about the superconducting gap structure. Various measurement techniques are employed, including rf tunnel diode cavity oscillators, $\mu$SR, scanning tunneling microscopy and small angle neutron scattering. For a fully gapped superconductor, $\Delta\lambda(T) \propto \exp(-\Delta/T)$. At sufficiently low temperatures ($T_c/T < 0.25$) the superfluid density of the superconducting electrons,

$\rho_{SF} = [1/(\lambda(T)/\lambda(0))]^2 = [1/(1 + (\lambda(T) - \lambda(0))/\lambda(0)]^2 = [1/(1 + \Delta\lambda(T)/\lambda(0)]^2$, can be approximated by just the leading correction term $(1 - 2\Delta\lambda(T)/\lambda(0))$ in the expansion:

$\rho_{SF} = (1 + \Delta\lambda(T)/\lambda(0))^{-2} \approx 1 - 2\Delta\lambda(T)/\lambda(0) + 3(\Delta\lambda(T)/\lambda(0))^2 - 4((\Delta\lambda(T)/\lambda(0))^3 + \ldots$  (1)

where $\Delta\lambda(T)$ is the temperature dependent penetration depth, $\lambda(T)$, minus the value of the penetration depth as $T \rightarrow 0$, $\lambda(0)$, i. e. $\Delta\lambda(T) = \lambda(T) - \lambda(0)$.

The temperature dependence of the superfluid density $\rho_{SF}$, which can be found by measurements of the penetration depth via eq. 1, indicates the nodal gap structure. For a gap function with nodes, $\lambda$ varies more rapidly with temperature, requiring higher order terms beyond the first correction term in eq. 1 or measurements to lower temperature.



**a. $\underline{\textbf{Δλ(T) ∝ T}}$** (or, equivalently, using this temperature dependence for Δλ(T) and just the first term in the expansion in (1) for the superfluid density, $\mathbf{ρ_{SF} ≈ 1\text{-const}*T}$), for temperatures much smaller than $T_c$ is clear indication of nodes (e. g. line nodes from d-wave pairing symmetry), with one proviso. Roddick and Stroud (1995) raised the possibility that Δλ(T) ∝ T could also be due to phase fluctuations, and estimated the magnitude of the effect on the coefficient, C, of the temperature in λ(T) - λ(0) = CT, as $C ≈ k_B[8πλ(0)^3]/ξ_0φ_0^2$, where $ξ_0$ is the coherence length and $φ_0 = 2.07 \, 10^{-7} \, Gcm^2$ is the flux quantum. For λ(0)=2000 Å and $ξ_0$=10 Å, Roddick and Stroud get C≈1 Å/K. Thus, any conclusions about nodal behavior in the FePn/Ch from Δλ(T) ∝ T (or $ρ_{SF}$ ≈ 1-const*T) should consider whether the slope, dλ/dT, of the measured variation of the penetration depth with temperature is comparable to the estimate for C from phase fluctuation effects. For the materials considered here C<1 Å/K (e. g. for LaFePO, λ(0) ≈ 2400 Å – Fletcher et al., 2009, $ξ_0$ ≈ 60 Å estimated from $H_{c2}$ – Yamashita et al., 2009, giving C≈0.3 Å/K) and dλ/dT is measured to be much larger. Thus, the conclusion that Δλ(T) ∝ T implies nodal behavior is valid in the FePn/Ch. The clean, linear decrease with increasing temperature of $ρ_{SF}$ for T<<$T_c$ can be smeared by slight disorder (Hashimoto et al., 2010b), see following discussion for Δλ(T)∝$T^2$.

**b. $\underline{\textbf{Δλ(T) ∝ T}^2}$** at low temperatures for both d-wave parity in the presence of strong scattering (Hirschfeld and Goldenfeld, 1993) as well as for a fully gapped $s_±$ state also with strong impurity scattering (Vorontsov, Vavilov, and Chubukov 2009). Thus, impurities/quality of sample can play an important role in being able to translate a 'simple' temperature dependence of Δλ(T) (or indeed any of the experimental probes of



nodal structure discussed below) into a firm conclusion as to the gap structure. As a further example of the difficulty in interpretation, $\Delta\lambda(T) \propto T^2$ has also been interpreted (Einzel, et al., 1986) as evidence for axial spin triplet, p-wave pairing in the heavy Fermion superconductor $UBe_{13}$.

Thus, as will be true of most of the experimental probes of the nodal structure discussed in this review, clear interpretation of a single probe may be difficult, particularly in the FePn/Ch superconductors with their complicated Fermiology whose implications for various measurements, including magnetic penetration depth, in the presence of scattering (see, e. g., Vorontzov, Vavilov, and Chubukov 2009) is still in the process of being understood theoretically. For a review of magnetic penetration depth in unconventional superconductors, see Prozorov and Giannetta (2006), while Gordon et al. (2010) provide an overview of such measurements in the FePn/Ch.

**a.) 1111 Structure:** Perhaps due to sample problems in the small (50 μm) single crystals available in the early investigation of the As-based 1111 FePn superconductors, or perhaps due to intrinsic differences between various rare earth 1111 compounds, there remains open discussion as to what to conclude about the gap structure in the 1111's from penetration depth measurements. There are reports of fully gapped behavior ($PrFeAsO_{1-x}$ - Hashimoto et al., 2009b and $SmFeAsO_{1-x}F_x$ - Malone et al., 2009) and a report of $\Delta\lambda(T) \propto T^2$ behavior interpreted as consistent with unconventional two gap superconductivity ($La/NdFeAsO_{0.9}F_{0.1}$ - Martin, et. al., 2009b).

In the $T_c \approx 6$ K 1111 superconductor LaFePO, there is agreement (Fletcher et al., 2009; Hicks et al., 2009a) that $\Delta\lambda(T) \propto T$, with analysis of this evidence for nodal structure leaving both d-wave and multi-band s-wave symmetry with nodes as possible



explanations.  Fletcher et al. find the slope of λ with temperature (with an exponent within 5% of $T^1$), proportional to the rate at which the gap grows away from the nodes, for their three samples to be 200 – 300 Å/K, while Hicks et al. (whose exponent, n, for $\Delta\lambda(T) \propto T^n$ data down to $0.06T_c$ varies between samples from 0.97 to 1.22) find dλ/dT to be 143 ± 15 Å/K.  Thus, since dλ/dT >> the Roddick and Stroud (1995) estimate for the contribution from phase fluctuations, the measured $\Delta\lambda(T) \propto T$ behavior in LaFePO is indicative of nodes in the gap.

**b.) 122 Structure:**  Although much larger crystals of 122 FePn superconductors were generally available than for the 1111 material (with the exception of LaFePO), there is a similar range of conflicting results on a priori similar samples.  Hashimoto et al. (2009a), for their cleanest K-doped $BaFe_2As_2$ crystal, find 2 band gaps, both fully gapped, consistent with ARPES data (section IVA2).  Khasanov et al. (2009a), using μSR, also find 2 gaps.  Martin et al. (2009a) for their samples of K-doped $BaFe_2As_2$ find $\Delta\lambda(T) \propto T^n$, with n≈2.

Work by the latter group on Co-doped $BaFe_2As_2$ (Gordon et al., 2009a,b) find n ranges from ≈ 2 for underdoped to about 2.5 in overdoped samples, which was interpreted to imply either gapless regions or point nodes in the superconducting gap.  Using magnetic force microscopy and scanning SQUID susceptometry, Luan et al. (2010) measure single crystal $BaFe_{1.90}Co_{0.10}As_2$ and describe their data ($\Delta\lambda(T) \propto T^{2.2}$) using a clean two-band fully gapped model, consistent with the $s_\pm$ model.

Work on $BaFe_{2-x}Ni_xAs_2$ found (Martin et al., 2010) in overdoped material, x=0.144, $T_c$≈7 K that λ in the c-axis direction behaved linearly with temperature (nodal), while $\lambda_{ab} \propto T^{1.6}$, i. e. anisotropy was present.  In the underdoped, x=0.066 and $T_c$=15 K,



and optimally doped regimes, x=0.092 and $T_c$=19.4 K, $\lambda$ was isotropic, with the temperature exponent being 2 or larger. This opened up the possibility of a three dimensional nodal structure (see the 3D spin fluctuation pairing calculations of Graser et al., 2010) in the (over) Ni-doped $BaFe_2As_2$, unlike what was seen in the Co-doped and unlike the underdoped-with-Ni case, i. e. indicating a true richness of behavior in these materials. Upon irradiation of a nearly optimally doped $BaFe_{2-x}Ni_xAs_2$ sample, $T_{c0}$=18.9 K, as $T_c$ decreases with irradiation (down to 15.9 K) the temperature exponent in $\lambda \propto T^n$ also decreases by about 15% (H. Kim et al., 2010a.) H. Kim et al. analyze these results – where disorder increases - as consistent with a nodeless $s_{+-}$ state in their optimally doped $BaFe_{2-x}Ni_xAs_2$ and in agreement with the result for a similar composition by Martin et al. (2010).

$\mu$SR determination of $\lambda$ in $SrFe_{1.75}Co_{0.25}As_2$, $T_c$ = 13 K, (Khasanov et al., 2009b) found 2 gaps. The size of the two gaps, when normalized as $2\Delta/k_BT_c$, agrees well with the general behavior of all the FePn/Ch (with the large/small $2\Delta/k_BT_c \approx 7/2.5$) based on all the measurement techniques as reviewed by Evtushinsky et al. (2009b).

Measurements of $\Delta\lambda(T)$ (Hashimoto et al., 2010a) in very clean (RRR≈1200) crystals of $KFe_2As_2$, the $T_c$=3.4 K endpoint of the $Ba_{1-x}K_xFe_2As_2$ phase diagram, result in linear with temperature dependence down to 0.1 $T_c$ with some admixture of $T^2$ due to impurity scattering below this temperature. They fit $\Delta\lambda(T)$ to $T^2/(T+T^*)$ with $T^* \approx 0.3$ K. The slope $d\lambda/dT \sim 550$ Å/K (i. e. much greater than the phase fluctuation contribution, almost a factor of four larger than in LaFePO), implying line nodes. Thus, the non-nested Fermiology at the K-endpoint in the $Ba_{1-x}K_xFe_2As_2$ phase diagram has perhaps surprisingly clear indication of nodal superconductivity. In a single crystal of



BaFe$_2$(As$_{0.7}$P$_{0.3}$)$_2$, T$_c$=30 K, Hashimoto et al. (2010b) find $\Delta\lambda$(T) $\propto$ T$^{1.1}$ (or $\propto$ T$^2$/(T+T*), with T*=1.3 K or 0.04T$_c$ – comparable to the value for KFe$_2$As$_2$) between 0.2 and 6 K with d$\lambda$/dT $\approx$ 25 Å/K. Using their NMR and thermal conductivity data, they conclude that there are lines nodes in the gap of a relatively clean superconductor (d-wave rather than impurity scattered s$_\pm$). (The Roddick and Stroud, 1995, phase fluctuation constant C is 0.4 Å/K - using $\lambda$(0)$\approx$2000 Å, typical of the FePn/Ch, and H$_{c2}$(0)=52 T from Hashimoto et al., 2010b, which implies $\xi_0$=25 Å - i. e. negligible compared to the d$\lambda$/dT of $\approx$ 25 Å/K from the penetration depth measurements of Hashimoto et al., 2010b.)

It is important to reiterate that $\Delta\lambda$(T) behaving approximately linearly with temperature (as discussed here for LaFePO, KFe$_2$As$_2$ and BaFe$_2$(As$_{0.7}$P$_{0.3}$)$_2$) is not only consistent with nodal behavior. It is – at least according to current theoretical understanding and as long as the phase fluctuation contribution is minimal – a proof thereof. However, the other power law behaviors for $\Delta\lambda$ (e. g. T$^2$) can either be interpreted as due to nodes or due to an s$_\pm$ scenario with strong impurity scattering (Vorontsov, Vavilov, and Chubukov 2009), as already mentioned above.

**c.) 111 Structure:** Measurements (Inosov et al., 2010b) of $\Delta\lambda$(T) determined from the magnetic field dependence of the form factor in small angle neutron scattering in a large single crystal of LiFeAs, T$_c$=17 K, imply a single isotropic superconducting gap. Imai et al. (2010), using microwave surface impedance, determined the in-plane penetration depth of single crystal LiFeAs, T$_c^{onset}$=19.0 K, and found their data to be consistent with two nodeless isotropic gaps. H. Kim et al. (2011), using single crystals of LiFeAs, T$_c$=17.5 K, found, via tunnel diode resonance, data in agreement with Imai et al., i. e. two nodeless isotropic gaps.



**d.) 11 Structure:** Measurements (Bendele et al., 2010 / Khasanov et al., 2008) of $\Delta\lambda(T)$ using $\mu$SR data on $Fe_{1.045}Se_{0.406}Te_{0.594}$ / $FeSe_{0.85}$ , $T_c$=14.6 / 8.3 K, were fit by a fully gapped two gap $s_\pm$ model. Measurements (H. Kim et al., 2010b) of $\Delta\lambda(T)$ using a tunnel diode oscillator on $Fe_{1.03}Se_{0.37}Te_{0.63}$ resulted in approximately $T^2$ behavior, which was interpreted as evidence for multi-gap superconductivity with scattering causing pair breaking and thus deviation from $exp(-\Delta/T)$ behavior.

### 2.) NMR/NQR Measurements

Measurements of the temperature dependence of $1/T_1T$, where $1/T_1$ is the nuclear-spin-lattice relaxation rate, in the superconducting state of the FePn/Ch compounds have been used to infer the presence or absence of a residual density of states, 'DOS', (gapless or nodal behavior.) Coupled with other experimental probes, such data contribute to a more complete understanding. Although the applied magnetic field used to carry out the NMR measurements can introduce normal regions, i. e. vortex cores (and thus evidence for a finite DOS), the upper critical fields in these materials are high enough that this is generally not a problem. Methods to avoid the field induced DOS include NMR data on $1/T_1$ taken as a function of field and extrapolated to H=0 and zero field NQR measurements of $1/T_1$. A peak in $1/T_1$ just below $T_c$, the Hebel-Schlichter coherence peak for a conventional isotropic gap open everywhere on the Fermi surface (simple s-wave symmetry), is in general not seen in the NMR/NQR measurements of all six structural families of the FePn/Ch superconductors. The lack of this coherence peak is discussed as theoretically consistent with the nodeless $s_\pm$ symmetry state by Parker et al. (2008). For spin singlet (s- or d-wave) pairing, the spin susceptibility part of the NMR Knight shift decreases to zero below $T_c$ in all crystalline directions - thus ruling out triplet



pairing. Thus, a strong decrease in the measured Knight shift below $T_c$, which as discussed below is sometimes seen in the FePn/Ch, can be used to argue for singlet pairing. However, the *lack* of such a strong decrease in the total Knight shift need not be due to triplet pairing, since there are often large contributions, e. g. van Vleck (interband) susceptibility, not affected by the superconductivity which mask the spin susceptibility. For a discussion of this, see Joynt and Taillefer (2002) and their review of $UPt_3$, which is an example of an unconventional superconductor whose very small Knight shift below $T_c$ has been interpreted as evidence for spin triplet pairing.

   **a.) 1111 Structure:** Grafe et al. (2008), Nakai et al. (2008) and Nakai et al. (2009) find $1/T_1 \sim T^3$ in $LaFeAsO_{0.9}F_{0.1}$, $T_c=26$ K, which they analyze as indicative of line nodes in the gap function. The lack of a significant residual density of states (no low temperature Korringa term in the NMR) was used by the latter authors to argue for s-wave pairing, since d-wave pairing in the presence of the scattering centers introduced by the F-doping would be expected to introduce a significant residual DOS. Similar data $(1/T_1 \sim T^3)$ and arguments have been put forward (Mukuda et al., 2008) for $LaFeAsO_{0.6}$, $T_c=28$ K. NMR $1/T_1$ data for $PrFeAsO_{0.89}F_{0.11}$ (Tc = 45 K) has been interpreted (Matano et al., 2008) as "$T^3$-like" just below $T_c$, with evidence for a second gap at lower temperatures, i. e. two gaps with nodes, while the strong decrease in the Knight shift below $T_c$ implied singlet pairing. NQR measurements (Kawasaki et al., 2008) on $LaFeAsO_{0.92}F_{0.08}$, $T_c=23$ K, were fit with a two gap model, where the gaps have either d-wave or $s_\pm$ symmetry.

   **b.) 122 Structure:** NMR data by Fukazawa et al., 2009b on $Ba_{1-x}K_xFe_2As_2$, $T_c=38$ K, gives $1/T_1 \sim T^{2.6}$ from 4-20 K, interpreted to mean that the sample's behavior is



similar to the NMR data for the 1111's, i. e. with possible nodal behavior.  In contrast, NMR data by Yashima et al. (2009) on $Ba_{0.6}K_{0.4}Fe_2As_2$, $T_c$=38 K, gives $1/T_1 \sim T^5$ from about 7-20 K, interpreted to imply a multiple fully gapped $s_{\pm}$ state.  Yashima et al. note that, based on the strong decrease of the Knight shift below $T_c$, their $Ba_{0.6}K_{0.4}Fe_2As_2$ is a spin singlet superconductor.   Both measurements were on polycrystalline samples. NMR data on a single crystal of $Ba_{0.72}K_{0.28}Fe_2As_2$, $T_c$=31.5 K found no simple power law behavior for $1/T_1$ and was interpreted (Matano et al., 2009) as coming from two gaps, of either d-wave or $s_{\pm}$ symmetry.  NQR of single crystal, RRR=67, $KFe_2As_2$ was analyzed (Fukazawa et al., 2009a) to indicate multiple gaps, but was unable to distinguish (see discussion of specific heat below) between nodal or fully gapped.  Nakai et al. (2010) used NMR to measure $1/T_1$ of single crystal $BaFe_2(As_{0.7}P_{0.3})_2$ and found a linear-in-T response between 0.1 and 4 K, clear evidence for a residual DOS at zero energy. Together with penetration depth and thermal conductivity measurements, Nakai et al., argue that their NMR data imply the existence of line nodes in the gap.  Unfortunately, Nakai et al. could not separate the spin susceptibility part of the Knight shift, leaving the question of singlet vs triplet pairing open from the NMR point of view.

**c.) 111 Structure:**  Measurements (Z. Li et al., 2010) of NMR and NQR on a polycrystalline sample of LiFeAs, $T_c$=17 K, are fit to a two gap, $s_{\pm}$ model.  Jeglic et at. (2010) find a Knight shift that $\rightarrow$0 as T$\rightarrow$0, consistent with spin singlet pairing.

**d.) 11 Structure:**  NMR measurements (Michioka et al., 2010) down to 2 K of $1/T_1$ on a single crystal of $Fe_{1.04}Se_{0.33}Te_{0.67}$, $T_c$=15 K, resulted in a roughly $T^3$ temperature dependence, and were interpreted as consistent with spin singlet superconductivity.



**e.) 122\* Structure:**   L. Ma et al. (2011) report an approximately 50% drop in the Knight shift below $T_c \approx 32$ K in single crystals of $K_{0.8}Fe_{2-y}Se_2$, consistent with singlet pairing.  In terms of the temperature dependence of $1/T_1$, they find an approximate $T^2$ dependence below $T_c/2$ which is unexplained.  Torchetti et al. (2011) find a 60% decrease in their Knight shift measured in both crystalline directions in single crystal $K_xFe_{2-y}Se_2$ below $T_c$, consistent with spin singlet pairing, while Kotegawa et al. (2011) find an 80% decrease in Knight shift in their $K_xFe_{2-y}Se_2$ below $T_c$.  Kotegawa et al. find that the best fit to their $1/T_1$ data below $T_c$ matches an $s_{+-}$ model.

### 3.) Specific Heat

Measurement of the specific heat, C, in the superconducting state can give information about the nodal structure in three ways.   One way to probe the superconducting gap using specific heat is to determine if the temperature dependence of $C \propto T^2$, which implies line nodes in the gap.  Although this is a well known theoretical result (Sigrist and Ueda, 1991) it is extremely difficult to verify experimentally due to the large contributions from other temperature dependences.  See the tour-de-force determination of $C \propto T^2$ in YBCO by Y. Wang, et al. (2001).

A second way to use specific heat as a probe of the superconducting gap structure is to measure the low temperature, $T \ll T_c$, $\gamma$ as a function of field - as long as the sample does not have a magnetic impurity phase (J. S. Kim, et al. 2009c) whose field response obscures that of $\gamma$.  For a fully gapped superconductor with only one gap, $\gamma$ will vary simply as H due to the localized Caroli-de Gennes-Matricon states in the vortex cores.  Moler et al. (1994) observed $\gamma \sim H^{1/2}$ up to 9 T while investigating the gap structure on YBCO, $H_{c2}(0) \sim 120$ T.   The theory of Volovik (1993) predicts $\gamma \sim H^{1/2}$ in a clean



superconductor with lines of nodes for H<<$H_{c2}$, while the theory of Kübert and Hirschfeld (1998) gives $\gamma \sim$ HlogH for a disordered superconductor with lines of nodes. The $H^{1/2}$ and HlogH laws arise from the Doppler shift of the low-energy nodal quasiparticles in the superflow field of the vortex lattice. Another physical explanation for a pure power law, $\gamma \sim H^{\alpha}$, $\alpha < 1$, in a superconductor is due to vortex-vortex interactions changing the size of the vortex cores, giving $\gamma \sim H^{0.66}$ in the T$\rightarrow$0 limit, as seen experimentally in the superconductor $NbSe_2$ up to about 0.3 $H_{c2}$ (Sonier et al., 1999).

However, studies of $\gamma$ vs H in superconductors are often more complicated than these simple, pure power law predictions. Although the Volovik theory is valid only in the low field limit, $\gamma \sim H^{1/2}$ has been found to higher field, e. g. up to $H_{c2}$ in both $LuNi_2B_2C$ (Nohara et al., 1997) and $YNi_2B_2C$ (Park et al., 2003). Another possible explanation for a sub-linear variation of $\gamma$ with H in the superconducting state is when the superconductor has two (or more) gaps (as found in all the FePn/Ch due to their Fermiology) - as reported experimentally, e.g., in the conventional superconductor $Na_{0.3}CoO_2$:$1.3H_2O$ (Oeschler et al., 2008) and discussed theoretically, e. g. by Bang (2010) where both gaps have conventional s-wave symmetry. Thus, two gaps of differing magnitudes can, depending on the ratio of $\Delta_{min}/\Delta_{max}$ (possibly but not necessarily including the nodal case where $\Delta_{min}$=0) mimic non-linear behavior of $\gamma$ with H due to nodes. As Nakai et al. (2004) point out, even in fully gapped superconductors the gap anisotropy (the ratio of $\Delta_{min}/\Delta_{max}$) can cause behavior similar to $\gamma \propto H^{1/2}$. Unfortunately, a rather large field range (to perhaps $H_{c2}$/2 or even higher) can be needed to distinguish between $\gamma \sim$ H, HlogH, $H^{1/2}$ and the non-linear field dependence $H^{\alpha}$, 0.5<$\alpha$<1, caused by have two



separate band gaps, as would come from, e. g., the $s_\pm$ model. Such high field work is in progress.

A third way to use specific heat as a probe of the superconducting gap structure is to measure $\gamma$ in field as a function of angle in the nodal plane, see Fig. 29, where the minima indicate the nodal directions. For field perpendicular to the nodal plane,

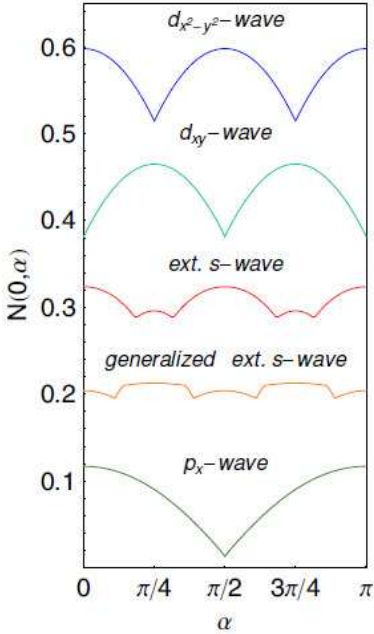

Fig. 29 (color online) Predictions for $\gamma(H)$ for field in the nodal plane of FePn/Ch superconductors of various pairing symmetries (Graser et al., 2008). The direction chosen in their coordinates is that the Fe-Fe direction determines [100], whereas some works choose the Fe-As direction as defining [100], causing a $\pi/4$ shift in nomenclature for the angle.

$\gamma$ varies as $H^{1/2}$. Although this technique has been used for other unconventional superconductors (for a review see Park and Salamon, 2004), due to its technical difficulty and the precision required (the experimental variation between maximum and minimum in $\gamma$ vs angle is typically only 2-4%) such measurements are just beginning for the FePn/Ch superconductors.

Unlike thermal conductivity, discussed in Section IVB4 just below, the residual gamma, $\gamma_r$, being finite is (as discussed above in Section IIIB3 when $\Delta C/T_c$ was discussed) generally *not* useful as a definitive sign of nodal behavior.

**a.) 1111 Structure:** Measurements (Mu et al., 2008b) of C/T down to 1.8 K and up to 9 T on polycrystalline LaFeO$_{1-x}$F$_x$ were found to vary at $H^{1/2}$, implying either nodal superconductivity due to the inherent gap symmetry or possibly (Bang, 2010) two full band gaps with scattering. The residual $\gamma$ in the superconducting state in this work was



0.7 mJ/moleK$^2$ which is possibly consistent with nodes broadened by defects, but may be due to extrinsic (sample quality) effects.

**b.) 122 Structure:** Measurements (Mu et al., 2009a) of C/T down to 1.8 K and up to 9 T on single crystal K-doped BaFe$_2$As$_2$, T$_c$=36.5 K showed a linear dependence on field, implying fully gapped behavior. However, the quality of the crystals may not have been optimal since the residual γ in the superconducting state was 7.7 mJ/moleK$^2$ and the magnetic field below 4 K induced anomalies in C. Work (Dong et al., 2008b) on polycrystalline Ba$_{0.5}$K$_{0.5}$Fe$_2$As$_2$, T$_c$=36 K, gave a residual γ of 9.1 mJ/moleK$^2$, which was described as possibly not intrinsic.

Unlike the status in K-doped BaFe$_2$As$_2$, where sample quality has hindered progress, the quality of samples in Ni- and Co-doped BaFe$_2$As$_2$ has been gradually improved such that a consistent picture of intrinsic behavior has emerged. Early work in measuring the specific heat down to 2 K of both unannealed single crystal Ni- and Co-doped BaFe$_2$As$_2$ gave (Bud'ko, Ni and Canfield 2009) a residual γ (T→0) of ~10 mJ/moleK$^2$. Specific heat (Gofryk et al., 2010a) down to 0.4 K on a range of compositions in self-flux grown unannealed single crystals of BaFe$_{2-x}$Co$_x$As$_2$ gave comparable γ(T→0) values ranging between 3.7 mJ/moleK$^2$ for optimally doped, x=0.16, up to 14.6 mJ/moleK$^2$ for overdoped, x=0.21. Gofryk et al. (2010a), based on their specific heat data as well as magnetic susceptibility shielding data, assigned the large residual γ values as being due to non-superconducting volume fractions in their unannealed samples. Later, these values were decreased markedly upon annealing: for optimally doped BaFe$_{2-x}$Co$_x$As$_2$, γ(T→0) =1.3 (0.25 on a second sample), and for



overdoped $\gamma(T\to 0)$ =3.8 mJ/moleK$^2$ for samples annealed at 800 $^o$C for 1 week (Gofryk et al., 2011a,b).

**Low field γ vs H**:  Gofryk et al.'s (2010a) measured (less than linear with) field dependence of γ up to 9 T in their unannealed BaFe$_{2-x}$Co$_x$As$_2$ samples was analyzed to be consistent with a two gap model, as discussed theoretically by Bang (2010) for the s$_\pm$ model with impurity scattering, with the ratios of the gap sizes in their analysis being independent of the doping.   Qualitatively, the amount of curvature in γ vs H up to 9 T was not markedly different in the annealed samples.  Jang et al. (2011) measured γ(H) up to 9 T on single crystals of unannealed overdoped BaFe$_{1.8}$Co$_{0.2}$As$_2$, T$_c$=19 K, and fit their data to a two gap model – an isotropic hole Fermi surface and an anisotropic nodal electron Fermi surface.  They also find that γ∝H$^{0.7}$ fit their data as well, cf. the high field γ vs H data from J. S. Kim et al. (2011b) discussed just below.  Mu et al. (2010) measured γ(H) up to 9 T on optimally doped BaFe$_{2-x}$Co$_x$As$_2$ and found non-linear behavior up to 1 T and essentially linear behavior above – too complicated a behavior to be analyzed by any of the simple existing models and in disagreement with the Gofryk et al (2011b) γ(H) results.  The possibility that the low field, ≤ 1T, behavior of Mu et al. (2010) was extrinsic was not discussed; Gofryk et al (2010a) only had one field point in that range.

**High field γ vs H**:  Measurement of underdoped and overdoped, annealed single crystals of BaFe$_{2-x}$Co$_x$As$_2$ in fields up to H$_{c2}$ ~ 18 and 25 T respectively show γ ~ H$^{0.7}$ over the whole field range of measurement (J. S. Kim et al., 2011b).  The same measurements in underdoped BaFe$_{2-x}$Ni$_x$As$_2$ (J. S. Kim et al., 2011d) also show γ ∝ H$^{0.5}$ up to at least 14 T.  The fact that γ vs H shows a relatively *pure* power law behavior all



the way up to $H_{c2}$ for several doping levels of BaFe$_{2-x}$(Co,Ni)$_x$As$_2$, like observed in YB$_2$Ni$_2$C (Park et al., 2003), in contrast to the Volovik effect, predicted to hold only for H<<$H_{c2}$, requires modeling with at least two gaps (cf. Jang et al., 2011, discussed just above) and variable anisotropy (ratio of $\Delta_{min}/\Delta_{max}$) therein as done by Bang (2010), Nakai et al. (2004), and Y. Wang et al. (2011).

In summary, clearly the $\gamma$(H) data for doped BaFe$_2$As$_2$ promise insights into the gap anisotropy. However, the sample quality is still being tuned with annealing, the data are still incomplete for the optimally doped composition and are still being analyzed with improved multi-gap models (see discussion of the P-doped BaFe$_2$As$_2$ $\gamma$ vs H work, Y. Wang et al., 2011, below), leaving implications for the gap structure of these materials at present still open.

A rather large residual $\gamma$ is found (Fukuzawa et al., 2009b) in a polycrystalline, RRR=67 sample of KFe$_2$As$_2$, $T_c$=3.4 K; analysis of the temperature dependence (two gap model) measured down to 0.4 K (not a large range of data below T/$T_c$<0.25) of the superconducting specific heat is somewhat hampered by the >50% ratio of the residual $\gamma_r$ vs the normal state extrapolation of C/T to T=0, $\gamma_n$. In addition, there is evidence (J. S. Kim et al., 2011c) that there is a magnetic transition in KFe$_2$As$_2$ at ~ 0.7 K, further complicating the two gap model analysis.

Specific heat in fields to 15 and down to 0.4 K of BaFe$_2$(As$_{0.7}$P$_{0.3}$)$_2$ gave (J. S. Kim et al., 2010a) $\gamma \sim H^1$, with a residual $\gamma$ of 1.8 mJ/moleK$^2$. Since this field result indicated a fully gapped material, the residual $\gamma$ was discussed as being not intrinsic. However, a follow up work (Y. Wang et al., 2011) showed that $\gamma \sim H^{1/2}$ in the low field,



H≤4 T (H<<$H_{c2}$=52 T) limit – consistent with nodal or at least deep gap minimum behavior, so that this $\gamma_r$ could be partially due to nodes with defect broadening.

**c.) 11 Structure:** The specific heat (Zeng et al., 2010a) as a function of angle, C(Θ), of self-flux grown single crystals of FeSe$_{0.4}$Te$_{0.6}$ $T_c$=14.5 K, was measured in the superconducting state (T ~ 2.6 K, or ~0.2 $T_c$) in 9 T. Four fold oscillations (minima at ±π/4) with an amplitude of 0.25 mJ/moleK$^2$ were observed and could be interpreted – see Fig. 29 - as either due to d$_{xy}$ pairing or due to a strongly anisotropic s-wave nodeless gap, with deep minima at or near particular high-symmetry "hot spots". Based on thermal conductivity data (discussed below in the following section, IVB4), the authors conclude that the second explanation is correct. However, theoretical work by Vorontsov and Vekhter (2006 and 2010), as well as experimental work on C(Θ) in field in the nodal superconductor CeCoIn$_5$ (An et al., 2010) points out that the maxima and minima in C(Θ) *invert* upon going into the low temperature (< 0.1 $T_c$) limit at low magnetic fields and only then show the correct nodal direction. Thus, the identification of the nodal directions from the C(Θ) data measured by Zeng et al. (2010) in a 9 T field at 0.2 $T_c$ was questioned by Vorontsov and Vekhter (2010), who argue that the nodes will occur at π/4 away from the direction assigned by Zeng et al., and are therefore consistent with d$_{x2-y2}$ pairing. Zeng et al. (2010b), using improved data (sharper, more distinct minima) and correcting an error in their identification of the angular minima/maxima with respect to the crystallographic axes, have reiterated their conclusion that an extended s-wave state (s$_\pm$ state) best fits their data. This work is the first report of C(H,Θ) in the FePn/Ch, is a tour-de-force of measurement technique, and highlights the dynamic interaction of theory and experiment in this field. The measurement (Hu et al., 2011) of the specific heat γ up



to 9 T ($H/H_{c2}(0)=0.2$)) on the high quality single crystal $FeSe_{0.43}Te_{0.57}$, $T_c^{mid}=14.2$ K, shows $\gamma \propto H^1$, also consistent with nodeless behavior.

**d.) 122\* Structure:** Zeng et al. (2011) report $\gamma$ vs H up to 9 T in single crystals of $K_xFe_{2-y}Se_2$, $T_c=32$ K and $H_{c2}(0)=48$ T, all fields in the c-axis direction. Their data show a large change in slope at 3 T, with $\gamma \propto H^1$ both above and below this point. Wang, Lei, and Petrovic (2011b) report $\gamma$ vs H up to 9 T in single crystals of sulfur doped $K_{0.8}Fe_{2-y}Se_{1.68}S_{0.32}$ ($T_c=31.4$ K and $H_{c2}(0)=45$ T) and $K_{0.8}Fe_{2-y}Se_{1.01}S_{0.99}$ ($T_c=21.4$ K and $H_{c2}(0)=13$ T. Both sets of data show $\gamma \propto H^1$ over the whole field range from 0 to 9 T which, at least in the lower critical field, higher S-doped sample seems conclusive evidence for lack of nodes.

**4.) Thermal Conductivity:**

Thermal conductivity, $\kappa$, is similar to specific heat in its probing of nodal structure. A zero $\kappa/T$ as T→0 indicates a fully gapped superconductor, while a finite value can indicate either nodal structure due to the pairing symmetry, gapless behavior due to scattering, or non-intrinsic contributions connected throughout the sample. In the nodal case, the field dependence of $\kappa/T$ (~HlogH) is also similar in cause to that of the specific heat ($H^{1/2}$). Although the specific heat residual $\gamma$ in the FePn/Ch superconductors has not yet been reported to be smaller than 0.7 mJ/moleK$^2$ in a 1111 material ($LaFeAsO_{1-x}F_x$, Gang Mu et al., 2008b) or 0.25/1.78 mJ/moleK$^2$ in the 122's (in annealed optimally doped $BaFe_{1.84}Co_{0.16}As_2$, Gofryk et al., 2011b/in unannealed P-doped $BaFe_2As_2$, J. S. Kim et al., 2010a) and is typically 4-10 mJ/moleK$^2$, several reports of $\kappa/T$ ~ 0 within the error bar of the measurement (typically ≈1 $\mu$W/K$^2$cm in the c-axis



direction and ≈10-20 μW/K$^2$cm in the a-axis direction, Reid et al., 2010) are discussed

below, taken as clear evidence for fully gapped behavior.

   **a.) 1111 Structure:**  Thermal conductivity in Sn-flux grown single crystal

LaFePO, T$_c$=7.4 K, RRR=28, was measured in the ab-plane down to 0.46 K, with

κ(T→0)/T = 3000 μW/K$^2$cm, possibly at least partially due to extrinsic contributions

(Yamashita et al., 2009).  The rather complex field dependence of the low temperature

thermal conductivity was analyzed in a multi-band model, with at least one band with

nodal behavior.

   **b.) 122 Structure:**  Measurements (Luo et al., 2009) of the thermal conductivity,

κ, in zero magnetic field result in a negligible residual linear term in κ/T as T → 0 in self

flux grown crystals of Ba$_{1-x}$K$_x$Fe$_2$As$_2$, x=0.25 and 0.28, T$_c$=26 and 30 K.  This was

interpreted as showing that there are no zero-energy quasiparticles and hence the

superconducting gap has no nodes in the ab-plane anywhere in this composition range.

However, the authors find that a small magnetic field can induce a large κ/T, interpreted

to imply that there is a deep minimum in the size of the gap somewhere on the Fermi

surface.   For a theoretical discussion of this scenario, see Mishra et al. (2009b).  In

BaFe$_{2-x}$Co$_x$As$_2$, 0.048 ≤ x ≤ 0.114, measurements (Tanatar et al., 2010b) of the thermal

conductivity  in zero magnetic field result in a negligible residual linear term in κ/T as T

→ 0 at all x.  This was interpreted just as in the results for K-doped BaFe$_2$As$_2$:  no zero-

energy quasiparticles and hence the thermal currents in the ab-plane are not carried by

nodal quasiparticles.  Also, a small magnetic field can induce a large κ/T, again implying

that there is a deep minimum in the size of the gap somewhere on the Fermi surface.



Follow up measurements in Co-doped $BaFe_2As_2$, with $0.038 \le x \le 0.127$, by the same group (Reid et al., 2010) found a finite residual $\kappa/T$ as $T \to 0$ (implying states in the gap, or nodal behavior) with the thermal current along the c-axis *away* from optimal doping, while in the ab-plane $\kappa/T$, within the error bar, vanishes as $T \to 0$ for the whole composition range. A field of $H_{c2}/4$ induces a finite $\kappa/T$ as $T \to 0$ along the a-axis as well and brings the c- and a-axis data back into agreement. The field behavior of $\kappa/T$ in the overdoped $BaFe_{2-x}Co_xAs_2$, x=0.127, where the sample has a residual $\kappa/T$ (evidence for nodes) along the c-axis, shows the same sub-linear rise with H in both the c-axis and ab-plane directions as does the d-wave superconductor $Tl_2Ba_2CuO_{6-x}$. However, $\kappa(H)/T$ for the nearly optimally doped $BaFe_{2-x}Co_xAs_2$, x=0.074, where there was no residual $\kappa/T$, shows $\kappa/T \sim H$ in both directions. The appearance of nodal quasiparticles carrying c-axis thermal currents as composition is moved away from optimal doping is used (Reid et al., 2010) to imply that the gapless behavior is 'accidental', i. e. not imposed by symmetry but instead by scattering, and therefore consistent with, for example, $s_\pm$ symmetry. For a discussion of the theory, see Mishra et al. (2011). Thermal conductivity data (Dong et al., 2010a) for overdoped $BaFe_{2-x}Co_xAs_2$, x=0.27, in the ab-plane also show $\kappa/T$ (T→0) equal to zero within their error bar, and $\kappa(H)/T$ behavior like d-wave $Tl_2Ba_2CuO_{6-x}$.

Thermal conductivity of single crystal $BaFe_{1.9}Ni_{0.1}As_2$, $T_c$=20.3 K, was measured (Ding et al., 2009) down to 0.07 K. The results that the residual $\kappa/T$ (T→0) was negligible, and $\kappa(H)/T \sim H^\alpha$, $\alpha$>1, were interpreted as consistent with nodeless multiple gaps.

Thermal conductivity of single crystal $BaFe_2(As_{0.7}P_{0.3})_2$ was measured (Hashimoto et al., 2010b) in zero and applied fields down to 0.1 K. A significant residual



κ/T (T→0) of 250 μW/K$^2$cm and κ(H)/T ~ H$^{1/2}$ up to 13 T are found, analyzed to be consistent with nodal behavior.  Thermal conductivity in the same material as a function of angle and field has been measured and found consistent with s-wave symmetry, with nodal structure on the electron pockets (Yamashita et al., 2011).

Thermal conductivity of self-flux grown single crystal KFe$_2$As$_2$, RRR=86, down to 0.07 K and up to H$_{c2}$ was measured (Dong et al., 2010b), resulting in a large residual κ(T→0)/T = 2270 μW/K$^2$cm and a field dependence comparable to that of d-wave Tl$_2$Ba$_2$CuO$_{6-x}$.

**c.) 111 Structure:**  Thermal conductivity of single crystal LiFeAs, T$_c$≈18 K, was measured (Tanatar et al., 2011) down to 0.05 K in both ⊥c-axis and ∥c-axis directions.  The residual κ(T→0)/T ≈0 and the field dependence were interpreted to mean that LiFeAs has a 3D isotropic gap without nodes or deep minima.

**d.) 11 Structure:**  Thermal conductivity of vapor self transport grown single crystal FeSe$_{≈1}$, T$_c$=8.8 K, was measured (Dong et al., 2009) in plane down to 0.12 K and up to 14.5 T (~ 0.75 H$_{c2}$).  The residual κ(T→0)/T found was 16 μW/K$^2$cm, only 4% of the normal state value.  Together with a dependence on field similar to that of NbSe$_2$, these thermal conductivity data were interpreted as evidence for nodeless multi-gap s-wave superconductivity.

### 5.)  Andreev Spectroscopy, Tunneling, Raman Scattering

Point contact Andreev reflection spectroscopy applied to polycrystalline samples of the 1111 structure finds evidence for a conventional, single gap (T. Y. Chen et al., 2008) or multiple gaps (Y. L. Wang et al., 2009; Gonnelli et al., 2009; Samuely et al., 2009b; Yates et al., 2008) with possible unconventional behavior in one of the gaps.



Work on Andreev spectroscopy on the 122 structure has found a single gap in single crystal K-doped $BaFe_2As_2$, but the authors suggest that their c-axis tunneling direction could be missing bands mostly in the ab-plane (Lu et al., 2009). Andreev spectroscopy by Szabo et al. (2009), also on single crystal $Ba_{0.55}K_{0.45}Fe_2As_2$, found two gaps in the a-b plane. Work on Co-doped $BaFe_2As_2$ finds (Samuely et al., 2009a) a single gap. Andreev spectroscopy on thin film Co-doped $BaFe_2As_2$ (Sheet et al., 2010) finds evidence for unconventional pairing with fluctuations up to 1.3 $T_c$. For an early review on Andreev spectroscopy in the 122 superconductors, see Samuely et al. (2009a).

C.-T. Chen et al. (2010) study Josephson tunneling in a novel composite Nb–$NdFeAsO_{0.88}F_{0.12}$ superconducting loop and find evidence (1/2 integer quantum flux transitions) for a sign change in the superconducting order parameter on the Fermi surface. C.-T. Chen et al. (2010) then put forward arguments that this implies $s_{\pm}$ pairing. In a similar hallmark experiment, Hanaguri et al. (2010) in $FeSe_{1-x}Te_x$, $T_c \sim 14$ K, used scanning tunneling microscopy in 10 T to conclude $s_{\pm}$ pairing. Josephson tunneling has been used to infer s-wave pairing in K-doped $BaFe_2As_2$ (X. Zhang et al., 2009).

Scanning SQUID microscopy on polycrystalline $NdFeAsO_{0.94}F_{0.06}$, $T_c = 48$K, detected (Hicks et al., 2009b) no paramagnetic Meissner effect (Wohlleben effect). This was analyzed as consistent with s-wave (including $s_{\pm}$) pairing or s-wave with a slight admixture of d-wave. Scanning tunneling microscopy (STM) on a similar composition, $NdFeAsO_{0.86}F_{0.14}$, with the same $T_c$ by Jin et al. (2010) showed only a single gap, with $2\Delta(0)/k_BT_c \sim 4.3$. In general (see in addition, e. g., the work by Massee et al., 2009 on optimally doped $BaFe_{1.86}Co_{0.14}As_2$ and the review by Evtushinsky et al., 2009b), STM and scanning tunneling spectroscopy measurements of the FePn/Ch only reveal one gap –



in most cases the large, $2\Delta/k_B T_c \sim 7$, one. For a review of scanning tunneling microscopy and spectroscopy in the cuprates, see Fischer et al. (2007).

Muschler et al. (2009) measured $BaFe_{2-x}Co_xAs_2$ at two compositions around optimal doping using Raman spectroscopy, which is in principle sensitive to different Fermi surface sheets, and found evidence for nodes on the electron pockets. A follow up theoretical paper (Boyd, Hirschfeld, and Devereaux, 2010) analyzed the results of Muschler et al. and found that Co functions primarily as an intraband scatterer.

In contrast to the results of Muschler et al., Sugai et al. (2010) investigated the pairing symmetry of $BaFe_{2-x}Co_xAs_2$ using Raman scattering and argued that their similar data rather indicate nodes on the hole pockets. In the introduction to this section (IVB), it was stated that the experimental probes often give contradictory answers for the nodal structure and these Raman data provide a last example thereof.

A. M. Zhang et al. (2011a) have performed Raman spectroscopy measurements on single crystals of $K_{0.8}Fe_{1.6}Se_2$, $T_c$=32 K, and find a large number (14) of phonon modes which they analyze as consistent with the Fe-vacancy ordering proposed by Bao et al. (2011a,b). Interestingly, one of the observed phonon modes (with $A_g$ symmetry) shows a change in frequency at $T=T_c$, indicating a connection between the superconductivity and a very limited subset of the phonon modes. A follow up work by A. M. Zhang et al. (2011b) also reported Raman data for $Tl_{0.5}K_{0.3}Fe_{1.6}Se_2$ ($T_c$=29 K) and $Tl_{0.5}Rb_{0.3}Fe_{1.6}Se_2$ ($T_c$=31 K) as well as for the insulating compound $KFe_{1.5}Se_2$. Consistent with the similar $T_c$ values, they find that the alkali metal substitution does not cause distortion (change the phonon frequencies) in the Fe-Se layers (where presumably the superconductivity occurs).



### V.  Sample Preparation:

The cornerstone on which the study of the FePn/Ch rests is well prepared and well characterized samples.  The discovery of superconductivity at 26 K by Kamihara et al. (2008) in $LaFeAsO_{1-x}F_x$ excited the imagination of the materials physics community, and led to concerted efforts by researchers worldwide to understand the new phenomena.  However, it is not just the initial discovery of superconductivity in a given structure or at a particular composition that rewards insight and creativity in sample preparation, but also very importantly the ensuing characterization drives the sample growers.  Any hope of understanding the basic physics of these new materials depends strongly on the sample quality. The preceding sections discussed case after case where sample quality was key in deciding on the intrinsic behavior -   the role of defects and disorder in discovering the true nodal behavior is just one example.  Here in the final section before the conclusions we discuss a representative subset of the efforts in sample preparation, and the wide panoply of techniques being brought to bear, including budding efforts at producing materials for applications – certainly years ahead compared to the time frame required for application of the previous high $T_c$ discovery in the cuprates.  See Putti et al., 2010 for an overview of the FePn/Ch properties relevant for application.

Progress in the sample preparation of the FePn/Ch superconductors has been impressive.  After the original discovery (Kamihara et al., 2008) that F-doped LaFeAsO was superconducting at 26 K, it was only several months until Ren et al. (2008b) succeeded in prepared electron-doped LaFeAsO without F via oxygen deficiency using high pressure synthesis.   Single crystals of 122 $Ba_{1-x}K_xFe_2As_2$ were produced and characterized (Ni et al., 2008a) using Sn-flux within two weeks of the original discovery



(Rotter, Tegel and Johrendt, 2008) of $T_c$=38 K in polycrystalline $Ba_{1-x}K_xFe_2As_2$. Faced with sample difficulties due to inclusions from the Sn flux, the community responded with creative flux alternatives that have led to bigger and cleaner single crystals. Further work found systems where Sn-flux did not degrade the properties. Below is a small synopsis of these ongoing efforts in sample preparation – which is resulting in not only discovery of new systems but also improvement in quality to reveal the intrinsic physics in known systems.

## A. Polycrystalline

The discovery work in the six structures discussed in this review was in each case using polycrystalline samples: Kamihara et al. (2008) in $LaFeAsO_{1-x}F_x$; Rotter, Tegel and Johrendt (2008) in $Ba_{1-x}K_xFe_2As_2$; X. C. Wang et al. (2008) in LiFeAs; Hsu et al. (2008) in FeSe; Ogino et al. (2009) in $Sr_2ScO_3FeP$; Gou et al. (2010) in $K_{0.8}Fe_{2-y}Se_2$. The powder preparation techniques used are fairly standard, as an example consider the Kamihara et al. (2008) preparation of the discovery compound, $LaFeAsO_{1-x}F_x$. Polycrystalline samples were prepared by first mixing the appropriate stoichiometric amounts of lanthanum arsenide, iron arsenide, and dehydrated $La_2O_3$ powders, with $LaF_3$ and La added to achieve the proper fluorine content. Pressed pellets of the starting materials were then heated in a quartz tube under partial pressure of Ar gas at 1250 °C for 40 hours. Certain polycrystalline preparation involves high pressures to keep in a volatile component during the sintering process, e. g. X. C. Wang et al. (2008) sintered their LiFeAs samples under 1 to 1.8 GPa for 1 hour at 800 °C, with the starting material already containing prereacted (at 800 °C for 10 hours) FeAs, so-called "precursor" material. High pressure polycrystalline synthesis is also used to achieve more



homogeneous non-equilibrium concentrations, for example in oxygen deficient $LnFeAsO_{1-x}$ by Ren et al. (2008a). Pre-sintered LnAs powder, As, Fe and $Fe_2O_3$ powders were mixed in the appropriate stoichiometric amounts, ground thoroughly, and pressed into small pellets. These were sealed in boron nitride crucibles and sintered under 6 GPa pressure at 1250 $^o$C for two hours.

Disadvantages of polycrystalline sintered material include, e. g.: the contribution of grain boundary resistance to the determination of $\rho$ (perhaps increasing the absolute value of $\rho$ by a factor of two in some cases); the inability to determine direction dependence of properties (including, e. g., critical fields, resistivity, thermal conductivity); the inability to do elastic neutron scattering determinations which are useful – when sufficient single crystal mass is available – for example to determine small magnetic moments; lack of homogeneity – important for determining the microscopic coexistence of superconductivity and magnetism; and potential increased reactivity of surfaces due to increased surface areas. For a recent study, and discussion of sample difficulties, of the intergranular current density of polycyrstalline sintered and hot isostatically pressed ("HIPped) $SmFeAsO_{1-x}F_x$, see Yamamoto et al. (2011).

On the other hand, polycrystalline sample preparation is often easier, and - turning the small grain size into an advantage - can make samples where the diffusion of some component is the limiting factor so that powder winds up being *more* homogeneous than a large single crystal. Also, stoichiometry is often easier to control in a polycrystalline sample, as shown in the definitive work of Williams, McQueen and Cava (2009) where the correct stoichiometry of superconducting FeSe (not deficient, but instead essentially 1:1 in stoichiometry) was determined in polycrystalline samples.



**B. Superconducting Thin Films/Wire and Their Possible Application**

Since these new superconductors are metals, since some of them are quite malleable ($CaFe_2As_2$ has a small bending radius, Canfield, 2009), and since modern thermoelectric coolers can reach 10 K quite efficiently, preparation of superconducting thin films or wires of the FePn/Ch holds out the possibility of achieving applications of these materials. There has been a continuing effort in the superconducting thin film/application area almost since Kamihara et al.'s initial discovery in the 1111 structure.

Considering first thin films of FePn/Ch compounds which are known to be bulk superconductors, there is sufficient work to data to merit considering the results for the 1111, 122, and 11 materials in separate sub-sections.

**1111:** Backen et al. (2008) used pulsed laser deposition (PLD) onto room temperature $LaAlO_3$ and MgO substrates to prepare 600 nm thick films of $LaFeAsO_{1-x}F_x$. After a post-anneal of four hours at 1030 $^o$C the films shows $T_c^{onset}$=11.1 K, but – possibly due to non-superconducting islands in the film - $\rho$ did not fall entirely to zero. PLD work on epitaxial films of LaFeAsO using a target of $LaFeAsO_{0.9}F_{0.1}$ reported two weeks earlier by Hiramatsu et al., 2008b, - despite post-annealing – saw no superconductivity. Thus, it was clear in the beginning of this effort that conditions for producing superconducting films were not easy to achieve. More than a year later, the current state of the art of thin film preparation of 1111 superconductors has shown significant progress. Haindl et al. (2010), using PLD and post-annealing, prepared homogeneous (pore free) polycrystalline films of $LaFeAsO_{1-x}F_x$ with $T_c^{onset}$=28 K, $\rho(0)$ ~



0.6 mΩcm, RRR~4, and a 2 K critical current density around $2x10^3$ A/cm$^2$.  Kidszun et al. (2010), also using PLD and post-annealing, have succeeded in preparing 200 nm thick epitaxial films of LaFeAsO$_{1-x}$F$_x$ with T$_c$=25 K and RRR=6.8.  T. Kawaguchi et al. (2010), using molecular beam epitaxy (MBE) on GaAs substrates at 650 $^o$C, have now achieved T$_c^{onset}$ = 48 K in NdFeAsO$_{1-x}$F$_x$ films, with ρ=0 by 42 K, i. e. a complete transition, - without, it should be stressed, the ex-situ second annealing step necessary in the PLD works.   The resistivity of their best films is ~1000 μΩ-cm at room temperature.

**122:**  Excellent progress has also been made in preparing thin films of doped 122 FePn/Ch superconductors, essentially getting to the point where applications are possible. Just as in the thin film work in the 1111's, much initial work was needed to improve the thin film quality.  Hiramatsu et al. (2008a) succeeded early on using PLD in growing epitaxial, superconducting films of SrFe$_{2-x}$Co$_x$As$_2$ with no post-annealing with T$_c$~20 K, RRR~1.5, and ρ(0)~300 μΩcm. This resistivity is comparable to that of polycrystalline material at the same temperature (270 μΩ-cm, Leithe-Jasper et al., 2008).  This work – concurrent in time with the early, non-superconducting 1111 films reported by the same group (Hiramatsu et al., 2008b) - illustrates the relative ease with which 122 films can be grown vs 1111 films.   Attacking the grain boundary/weak link problem (see Lee et al., 2009, for a discussion of this in Co-doped BaFe$_2$As$_2$) to increase the critical current density, a number of groups including Maiorov et al. (2009) and Choi et al. (2009) continued using PLD to make thin (450-750 nm) SrFe$_{1.8}$Co$_{0.2}$As$_2$ films, T$_c$=18.9 K, with one film of Maiorov et al. showing a critical current density of 0.5 10$^6$ A/cm$^2$.



Lee et al. (2010b), using PLD of K-doped $BaFe_2As_2$ onto single crystal $Al_2O_3$ substrates and post-annealing at 700 $^o$C for six hours, have achieved $T_c^{onset}$=40 K (a new record for 122 $T_c$'s) with $\rho$=0 at 37 K, $\rho$(300 K)=2500 $\mu\Omega$-cm, and RRR>25 in 1 $\mu$m films of $Ba_{0.6}K_{0.4}Fe_2As_2$. The higher $T_c^{onset}$ in the film vs bulk material is discussed as possibly due to strain in the a-axis direction. Strain as a way to increase $T_c$ in Co-doped $BaFe_2As_2$ thin films has been also investigated by Iida, et al. (2009). Baily et al. (2009), in a study of upper critical magnetic field, reported the preparation of 180 nm thick $SrFe_{1.8}Co_{0.2}As_2$ epitaxial films on mixed perovskite (La,Sr)(Al,Ta)$O_3$ ("LSAT") substrates at 670 $^o$C, with $T_c^{mid}$=17.1 K and $\rho_n$(30 K)=330 $\mu\Omega$-cm. These $SrFe_{1.8}Co_{0.2}As_2$ films were reported to have rough surfaces, granular morphology and be unstable against reaction with the water vapor in the air. To improve on this, for increased critical current density and possible application, Katase et al. (2009) prepared, using PLD, 500 nm thick films of $BaFe_{2-x}Co_xAs_2$ deposited at 700 $^o$C. These films, with $T_c^{onset}$ = 20 K, were optically flat, of better crystallinity, and much more resistant to reaction with water vapor than Co-doped $SrFe_2As_2$ films. The room temperature resistivity, $\rho$(300 K) was 1300 $\mu\Omega$-cm, or about four times larger than that of a single crystal. The report did not address critical current questions for applications. However, in follow up works Lee et al. (2010c) and Katase et al. (2010a) were able to break through the $10^6$ A/cm$^2$ barrier considered necessary for Josephson junctions by continuing the work with $BaFe_{2-x}Co_xAs_2$. Lee et al. (2010c) report critical current densities of 4.5 $10^6$ A/cm$^2$ (~ 10 times that reported for single crystals, Yamamoto et al., 2009) in epitaxial thin films of Co-doped $BaFe_2As_2$, $T_c$ ($\rho\rightarrow$0)=21.5 K, grown using PLD on single crystal intermediate layers of $SrTiO_3$ or $BaTiO_3$ between the single crystal perovskite substrate and the



superconducting film. The residual resistivity in these ~350 nm films is $\rho(0) \approx 75$ $\mu\Omega$cm, and the films are fully strain relaxed. Katase et al. (2010a) achieved critical currents of 4 $10^6$ A/cm$^2$ in thin films of BaFe$_{2-x}$Co$_x$As$_2$ using PLD, again on single crystal perovskite substrates but without the buffer layer of Lee et al. (2010c).

Based on these PLD BaFe$_{2-x}$Co$_x$As$_2$ thin films, Katase et al. (2010b) succeeded in making initial thin film Josephson junctions across bicrystal grain boundaries, a critical step for potential application. (See section IVB5 above for Josephson tunneling work on bulk specimens.) Katase et al. (2010c) have also succeeded in fabricating the first Superconducting QUantum Interference Devices (SQUIDs) using this thin film technology, although the devices are still in the development stage with flux noise levels ~40 higher than in typical dc-SQUIDs using epitaxial YBCO films.

In summary, the thin film work in the 122 FePn superconductors has now been brought, in under three years, to the application stage, with clear ideas on how to proceed and improve the process parameters to optimize performance.

**11:** FeSe thin films have been grown on semiconducting substrates for spintronic applications for over a decade (Takemura et al., 1997 evaporation/MBE on GaAs; Hamdadou, Bernede and Khelil, 2002), without measurements below room temperature and without superconductivity being discovered. After the discovery of superconductivity in FeSe (Hsu et al., 2008) M. J. Wang et al. (2009) reported the preparation of thin films of FeSe using PLD. Films of ~100 nm thickness grown on an MgO substrate at 500 $^o$C exhibited superconducting resistive transitions starting around 9 K. According to Nie et al. (2009), FeSe films under tensile strain have their



superconductivity suppressed. Jung et al. (2010) have succeeded in growing high quality films of $FeSe_{0.9}$ using PLD with $T_c$ onset above 11 K, RRR ~ 4, and $H_{c2}(0)$ ~ 50 T. Huang et al. (2010), using PLD, prepared 400 nm films of $FeSe_{0.5}Te_{0.5}$, with the optimal $T_c^{onset}$=15 K and $\rho$=0 at 11 K achieved on 310 $^o$C MgO substrates. Huang et al., varied the substrate temperature to vary the stress applied to their epitaxial films and thus to vary the lattice structure. They conclude that the chalcogenide height is the controlling parameter for $T_c$ in their films. Bellingeri et al. (2009), using PLD, prepared ~ 50 nm films of $FeSe_{0.5}Te_{0.5}$ and also found that they could control $T_c$ on their $SrTiO_3$ substrates using substrate temperature, with their best $T_c$ (17 K) occurring on a 450 $^o$C substrate.

Now superconducting thin films of non-bulk superconducting material are summarized. As discussed in section II, FeTe in the 11 structure has coincident $T_S$ and $T_{SDW}$ transitions at 72 K and is non-superconducting. Han et al. (2010), using PLD, prepared ~ 100 nm thick FeTe films under tensile stress on a variety of substrates at ~540 $^o$C and achieved $T_c^{onset}$ of 13 K.. In order to compensate for Te losses, the targets used had the stoichiometry $FeTe_{1.4}$. The tetrahedral bond angles were changed from the non-superconducting bulk sample values, and the c-axis lattice parameter was uniformly decreased. Resistive, susceptibility, and Hall effect anomalies associated with the structural/magnetic transitions in the films were all broadened and occurred at slightly higher temperatures than in the bulk, indicating coexistence of magnetism and superconductivity but not necessarily on a microscopic scale. It was not clear from the description if the 20 % superconducting fraction was a shielding or a Meissner expulsion fraction, but phase separation of the magnetic and superconducting domains is in any case a possibility. A second thin film work that achieved superconductivity in a material



otherwise normal was by Hiramatsu et al. (2009). In that work (see also the discussion of the Co-doped $SrFe_2As_2$ films above) they discovered that 200 nm films of $SrFe_2As_2$ grown using PLD on 700 $^o$C LSAT single crystal substrates, displayed a full resistive superconducting transition at $T_c^{onset}$=25 K, $\rho$=0 at 21 K, after exposure to water vapor for six hours. A more recent work in pressed pellets of $FeTe_{0.8}Se_{0.2}$ powder by Mizuguchi et al. (2010a) found an improvement in the temperature where $\rho \rightarrow 0$, the resistive transition width as well as an increase in the diamagnetic shielding, upon exposure to water vapor. The exact mechanism of the water exposure causing superconductivity is not yet clarified. However, the surface of the $SrFe_2As_2$ film (see also Katase et al., 2009) after exposure to water has a $Fe_2As$ impurity phase present after the reaction with the water vapor.

**Wires:** Gao et al. (2008) prepared $SmFeAsO_{0.65}F_{0.35}$ wires by filling 0.008 m diameter Ta tube, 0.001 m wall thickness, with stoichiometric amounts of the constituent reactant powders (powder-in-tube, or PIT method). The tube was then swaged down to 0.00225 m diameter and reacted at ~1170 $^o$C for 45 hours. The resultant wire had $T_c^{onset}$=52 K, a global critical current density of 3.9 $10^3$ A/cm$^2$ at 5 K, and $H_{c2}(T \rightarrow 0) \approx 100$ T using the WHH formula. The rather low critical current in this early attempt at a practical FePn superconducting wire is affected by impurity phases and weak links between grains. Using the PIT method, Ozaki et al. (2011) prepared single and seven core $FeTe_xSe_{1-x}$ wires, $T_c^{onset} \approx 11$ K, with critical currents at 4 K of order 200 A/cm$^2$. As a comparison, although single crystals are not a practical form for a conductor, Kashiwaya et al. (2010) find a critical current density, $j_c$, in single crystal $PrFeAsO_{0.7}$, $T_c = 35$ K, in the c-axis direction of 2.9 $10^5$ A/cm$^2$. Prommapan et al. (2011) found $j_c$(2 K) in single



crystals of LiFeAs of $\approx 2 \ 10^6$ A/cm$^2$. Ma et al. (2009) also discuss the PIT process, with

Nb or Fe tubes in addition to Ta. L. Wang et al. (2010) prepared $Sr_{0.6}K_{0.4}Fe_2As_2$, $T_c$=34

K, in tape form with Ag sheathing with a critical current of $1.2 \ 10^3$ A/cm$^2$ at 4.2 K.

## C. Single Crystals

Although single crystals of the 122's could be grown larger than those for the

1111's for a few months, the surge of effort in making larger single crystals has now also

extended to the 1111 structure, with a flux developed by Yan et al. (2009) achieving

crystals of several mm in size, vs the old 50-100 μm size in the beginning. At present,

five of the six discovered structures (1111, 122, 111, 11, and 122*) of the FePn/Ch

superconductors can be grown in mm-sized single crystal form, and the 21311 structure

has been prepared in 0.2x0.2 mm$^2$ crystals (Qian et al., 2011). Some measurement

techniques always can benefit from ever larger crystal mass: Goko et al. (2009)

measured μSR of a collection of over 100 single crystals (each with a mass of ~10 mg) of

$CaFe_2As_2$ prepared in Sn flux. Pratt et al. (2009b) measured inelastic neutron scattering

under pressure of a collection of 300 single crystals (each with a mass of ~ 5 mg), again

of Sn-flux grown $CaFe_2As_2$. However, it is important to understand that a "single

crystal" is not a guarantee of a lack of impurities, perfect lattice order, lack of twinning

(see Tanatar et al., 2010a for strain detwinning of $CaFe_2As_2$ and $BaFe_2As_2$ below the

tetragonal-orthorhombic structural phase transitions), or indeed of representative intrinsic

behavior in the particular measurement of interest to a researcher. As discussed above in

the specific heat section (IVB3), annealing of single crystals of Co-doped $BaFe_2As_2$ at

800 $^o$C for one week has led to significant changes in their measured properties, including



both an increase in $T_c$ at a given composition and also to changes in the measured specific heat $\gamma$. Rotundu et al. (2010) found that the residual resistivity ratio in a single crystal of $BaFe_2As_2$ increased from 5 to 36 with 30 days of annealing at 700 $^o$C. Starting with a short overview of flux growth, a summary of some of the various methods used to prepare single crystal FePn/Ch superconductors is given here, along with comparisons of sample quality.

**1.) Flux growth:** In general, if the thermodynamics and stabilities of the various possible compounds involved are heeded, growing crystals via the flux method is straightforward, see reviews by Fisk and Remeika (1989) and Canfield and Fisk (1992) on the use of molten *metal* fluxes. (As will be seen below, fluxes for the FePn/Ch need not be metallic.) The flux method consists of loading stoichiometric amounts of the elements desired in the final crystals into a ceramic crucible (perhaps alumina or MgO) with an excess of the material serving as the flux, with, for example, a molar ratio of 20-40 Sn-flux: 1 $Ba_{0.6}K_{0.4}Fe_2As_2$. The crucible, sealed in quartz, or the more expensive welded Nb or Ta vessels to more securely contain the hazardous arsenic or volatile phosphorous or lithium, is then heated to some high temperature (typically 850-1150 $^o$C) where all the constituent elements are dissolved in the molten flux. The solubility of each of the constituents with the flux can be checked via compendia of binary phase diagrams if the flux is an element. The crucible is then slowly cooled (~5 $^o$C/hr) and at some point the constituent elements form a supersaturated solution and crystals begin to nucleate out of the molten flux. Depending on the flux and the crystals, separation of the crystals from the flux is accomplished via dissolving of the flux (e. g. NaAs flux dissolves in water), decanting/centrifuging of the flux above the flux's melting point ($T_M$ for Sn is



232 $^{\circ}$C), harvesting of the crystals from the crucible on a hot plate ($T_M$ for In is only 157 $^{\circ}$C), mechanical separation, and others. For the FePn/Ch's, all of the activities performed when the material is not sealed away from the atmosphere in quartz or Nb/Ta are best done in an inert atmosphere glove box until the sensitivity to air (high, e. g., in LiFeAs) is determined.

2.) **Development of Fluxes/Progress in Crystal Growing:** The first discovered FePn/Ch superconductor was in the 1111 structure (Kamihara et al., 2008), and the search for higher sample quality and the ability to measure directionally dependent intrinsic properties such as resistivity, critical field, and penetration depth led to early efforts to produce single crystals. Zhigadlo et al. (2008) succeeded in growing single crystals of SmFeAsO$_{1-x}$F$_x$ in the 100 μm size regime using a NaCl/KCl flux technique at high (3 GPa) pressure. At about the same time, the first single crystals (~3x3x0.2 mm$^3$) of the 122 superconducting compound Ba$_{1-x}$K$_x$Fe$_2$As$_2$ were grown using Sn flux (Ni et al., 2008a), with an incorporation of ~1% Sn (see Su, et al., 2009 for a report of up to 5% Sn) into the crystals, not just as inclusions but at least partly into the lattice as an impurity. It was clear in the Ni et al. work that Sn from the metal flux had an important influence on the properties of crystals of the parent compound, BaFe$_2$As$_2$, depressing T$_S$/T$_{SDW}$ from the known polycrystalline value of 140 K to 85 K. The Sn incorporated in Ba$_{1-x}$K$_x$Fe$_2$As$_2$ also affects the low energy spin fluctuations in the NMR measurements (Baek et al., 2008; Sun et al., 2009) and causes a large upturn in the low temperature specific heat divided by temperature, C/T (J. S. Kim et al., 2009a). Rb-doped BaFe$_2$As$_2$ crystals grown in Sn-flux have as much as 9% Sn included (Bukowski et al., 2009). Contrary to this experience of Sn inclusion in the BaFe$_2$As$_2$ crystals, it became clear later that Sn-flux



crystal growth was not in general detrimental to most FePn/Ch sample's intrinsic

properties, and has been used quite successfully in the crystal growth of various other

$MFe_2As_2$, 1111, and 111 compounds.  In fact, a recent report (Urbano et al., 2010) using

a revised Sn-flux growth procedure finds little or no suppression of $T_S/T_{SDW}$ in

underdoped $Ba_{0.86}K_{0.14}Fe_2As_2$ from values in self-flux-grown samples.  However, due to

the initial experience with Sn a number of other fluxes were quickly tried.

One of these, somewhat unique to iron arsenide materials, is the so-called FeAs

"self flux."  X. F. Wang et al. (2009a) grew $BaFe_2As_2$ crystals using pre-reacted FeAs

powder as the flux, thus avoiding contamination from an extraneous element.  An excess

(factor of two) of the FeAs "precursor" material is used with Ba, placed in an alumina

crucible sealed in quartz, then heated to 700 $^o$C to "soak" for three hours, then to 1100 $^o$C

to react for ~30 hours, then slowly cooled to 900 $^o$C, then relatively rapidly cooled to

room temperature.  The 2x2x0.1 mm$^3$ crystals were mechanically removed, since the

compound FeAs melts at 1030 $^o$C, and a $T_S/T_{SDW}$ of 136 K is reported.  Using FeAs self

flux is not without negative consequences, since FeAs, which is magnetic, can be

contained in the crystals as an inclusion.  In terms of magnetic properties, Sn flux grown

crystals, on the other hand, can have elemental Sn inclusions (not just in the lattice

atomically but as small regions) which superconduct at 3.7 K (Colombier et al., 2009).  A

1 cm crystal of Ni-doped $SrFe_2As_2$ grown in FeAs self flux is shown in Fig. 30.



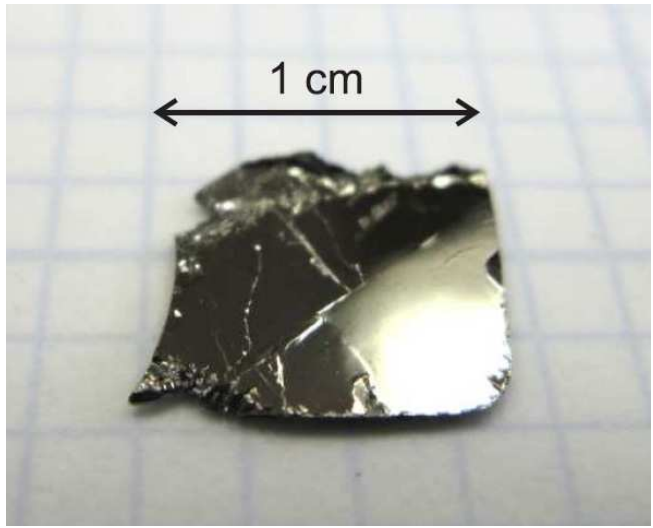

Fig. 30 (color online) As-grown single crystal of Ni-doped $SrFe_2As_2$ harvested from FeAs flux. Note the optically flat surfaces. The plane of the crystal is in the ab-plane, while the c-axis is perpendicular to the plane of the crystal. This is the typical growth habit for flux grown tetragonal 122 crystals. Size is limited by the size of the crucible (Saha et al., 2009a).

Other fluxes that have been used to grow the 122's include In (J. S. Kim et al., 2009a), where ~0.4 at% In is included in $BaFe_2As_2$ crystals and $T_S/T_{SDW}$ = 137 K (J. S. Kim, 2009) and Pb for growing $BaNi_2As_2$ (Ronning et al., 2009) and $BaRh_2As_2$ (Singh et al., 2008). Before ending the discussion of single crystal growth of the 122's, it is instructive to compare $T_{SDW}$ in various samples of undoped $SrFe_2As_2$ to gain an idea of how the properties can vary with differing preparation techniques. $T_{SDW}$ was reported to be 201.5 ± 0.25/198/200 K in single crystals from self flux - FeAs (H. Li, et al., 2009/Saha et al., 2010a/Matsubayashi et al., 2009), 203/205/220 K in polycrystalline material, (Schnelle et al., 2009/Kaneko et al., 2008/Shi et al., 2009), 198/200/220 K in single crystals from Sn flux (Yan et al., 2008/G. F. Chen et al., 2008/Zhao et al., 2008c). Self flux appears to give the most consistency in the result for $T_{SDW}$, while the values for



Sn flux single crystals and polycrystalline samples vary by 10%. In any case, Sn flux does not suppress $T_{SDW}$ in $SrFe_2As_2$ like it does in $BaFe_2As_2$.

With all this effort in developing flux growth of single crystals in the 122 structure, workers had not lost focus on the more difficult, but higher $T_c$, 1111 FePn/Ch superconductors. Crystal size had grown from the initial 100 μm size to ~600 μm (CeFeAsO from Sn flux, Jesche et al., 2009) when Yan et al. (2009) reported a breakthrough in crystal growth using NaAs flux for growing mm-sized crystals of LaFeAsO, $LaFeAsO_{1-x}F_x$, and $LaFe_{1-x}Co_xAsO$. Just as the case for the FeAs flux material, Na is prereacted with As, but in a sealed Ta tube at 600 $^o$C for 12 hours. For preparing LaFeAsO, the appropriate stoichiometric amounts of prefired LaFeAsO, LaAs, $Fe_2O_3$ and Fe are mixed in the molar ratio of 20 NaAs:1 LaFeAsO (similar to the large molar ratio using Sn as a flux) and then sealed in a Ta tube. For the crystals containing F NaAs is partially replaced by NaF, for crystals containing Co the Co partially replaces the iron. The material is then reacted at 1150 $^o$C for 24 hours, and cooled at 3 $^o$C down to 600 $^o$C to allow the crystals to form out of the NaAs flux. Harvesting of the crystals, of typical size 3x4x0.05-0.3 mm$^3$, from the flux is done by dissolving the NaAs flux in water.

Growth of single crystals in the more recently discovered 111 and 122* structures benefitted from the efforts in growing crystals of 1111 and 122 samples. $Na_{1-\delta}FeAs$ crystals have been grown from self flux (G. F. Chen et al., 2009) while LiFeAs crystals have been grown by a Bridgman technique (Song et al., 2010) and from both self flux and Sn flux (Borisenko et al., 2009). Both Bridgman and self flux techniques were used to



grow the 122*'s beginning already in the discovery works (see e. g. Krzton-Maziopa et al., 2011 and Fang et al., 2011b) after Guo et al.'s (2010) initial discovery work in polycrystalline $K_{0.8}Fe_2Se_2$. $FeSe_{1-x}$ crystals have been grown using a vapor self-transport method, as iodine vapor transport was found to be ineffective (Patel et al., 2009). $FeSe_{1-x}Te_x$ crystals have been grown by optical zone melting techniques (Yeh et al., 2009) and a modified Bridgman technique (Sales et al. 2009). For a review of single crystal growth in the 11 structure, see Wen et al. (2011). The 111 structure crystals can exceed 5 mm in lateral dimension, while the 11 structure crystals can exceed 10 mm.

## D. Outlook

Much work remains to be done from a materials point of view. The thin film and wire application-oriented work is still just beginning. Superconducting transition widths are sometimes several Kelvin wide (in the case of $Na_{1-\delta}FeAs$, as much as 15 Kelvin wide), and residual resistivity ratios of undoped superconducting compounds are seldom over 10. Upon doping, the residual resistivity ratios, due to the scattering centers introduced by the doping, fall even further. Certainly greater homogeneity, possibly by long term annealing, may affect much that has been discussed herein, not least of all the temperature dependences of various measures of nodal behavior. A study to reduce defects in certain systems, e. g. in $Na_{1-\delta}FeAs$, - as was carefully done in FeSe by Williams, McQueen and Cava (2009), would be useful. On the other hand, the controlled introduction of defects (e. g. see H. Kim et al., 2010a for $\Delta\lambda(T)$ measurements on superconducting doped $BaFe_2As_2$ irradiated with heavy ions) also is useful for understanding the influence of defects. After the initial rush to dope everything possible into the 122's, now is a good time to gain a perspective on what all these data mean for



the fundamental physics and the mechanism of superconductivity. "Isoelectronic" doping, e. g. P for As or Ru for Fe, has revealed interesting behavior (not found in the cuprates), and should be further pursued in more systems. Systems near a magnetic instability that show non-Fermi liquid behavior are perhaps of critical importance to further understand FePn/Ch superconductivity. In the end, superconducting samples of new Fe-containing structures would also greatly help the search for commonality and therefore deeper understanding of the entire class of materials.



## VI. Summary and Conclusions

The discovery of superconductivity in systems not just containing iron, but in systems where the magnetic behavior of iron appears to play a dominant role in the superconducting properties, has caused an "iron rush" of research. Up until this discovery of Kamihara et al. (2008) of $T_c$=26 K in F-doped LaFeAsO, the preponderance of superconductors seemed conventional, phonon-mediated-pairing types with a few unconventional, low $T_c$ heavy Fermion superconductors and the cuprates as exceptions. Now, this new class of materials, with frequent examples of phase diagrams with clearly coexistent magnetism and superconductivity, makes the previously known unconventional superconductors seem to be less like exceptions and more like harbingers of what superconductivity is really like.

Much of this review has been spent presenting evidence for magnetism/magnetic fluctuations being linked with the superconducting pairing mechanism in the FePn/Ch materials. See sections IIC and IVA for partial overviews of the results pertaining to this central issue. Interesting goals/questions/observations raised by this review for further understanding the superconductivity, the magnetism, and their possible "linkage" include the following.

1. As discussed in section IIIA, G. M. Zhang et al., 2009 initially proposed that strong fluctuations in these materials cause $\chi \sim T$ based on data up to ~300 K for the LaFeAsO$_{1-x}$F$_x$ and MFe$_2$As$_2$, M=Ba, Sr and Ca. Susceptibility data varying linearly with temperature above $T_c$ have also been measured in additional FePn/Ch's (SrFeAsF, Co-doped BaFe$_2$As$_2$, Na$_{1+\delta}$FeAs, FeSe$_{0.5}$Te$_{0.5}$) up to temperatures as high as 700 K. It would be useful if the lack of $\chi$ vs T data above 50 K in the three superconducting 21311 and in

the reported 43822 FePn/Ch compounds, as well as the lack of $\chi$ vs T data above $T_N \approx 540$ K in the 122*, could be corrected. Presumably such $\chi$ data could serve as another metric for measuring the strength of the magnetic fluctuations in these materials, as well as to function as a potential differentiator in their fundamental behavior.

2. The fact that this $\chi \sim T$ behavior persists in $LaFeAsO_{1-x}F_x$ even after $T_{SDW}$ is suppressed with increasing F-doping (Fig. 20) while $\chi \sim T$ behavior disappears upon the suppression of $T_{SDW}$ for $BaFe_{2-x}Co_xAs_2$ (Fig. 21) is intriguing. Does this indicate that the 1111's have stronger magnetic fluctuations than the 122's? This would be consistent with their higher $T_c$'s if indeed this linkage between superconductivity and magnetism is correct, and seems straightforward to further investigate by a more microscopic measure (e. g. INS) of the fluctuation strength.

3. The idea of Jesche et al. (2009) discussed in section IIB1b that $T_S$ will coalesce with $T_{SDW}$ with increasing sample quality in the 1111's is certainly worth pursuing to see if the 1111's in their undoped states are indeed intrinsically different from the undoped 122's.

4. The idea that quantum criticality can play a role in the FePn/Ch superconductivity has support from the resistivity data for several materials, section IIIA. A typical scenario for a quantum critical point is that a second order magnetic transition (such as antiferromagnetism) has been suppressed to T=0 at that point in a phase diagram. This is certainly a fertile field of investigation in these materials where there are so many examples of magnetism being suppressed by doping. Better quality samples, with attention to reducing magnetic impurities, need to be made so that possible non-Fermi



liquid behavior in the low temperature magnetic susceptibility – a mainstay of determining quantum criticality – can be investigated.

5.  In addition to aiding the investigation of intrinsic $\chi$ behavior, there are other areas where sample quality is central to understanding the FePn/Ch's. Knowledge of the nodal structure, as discussed in section IV, is key to understanding the superconducting pairing mechanism. Presently, the consensus of the data indicates that several nodal FePn/Ch superconductors exist, while several fully gapped compounds also exist – with a larger number of disputed systems. Reduction of defects in the samples, e. g. to clarify the temperature dependences in penetration depth measurements, will advance this investigation markedly. Cleaner samples will help determine what the low temperature limiting values are for the specific heat $\gamma$ and the thermal conductivity divided by temperature, $\kappa/T$, as well as allowing correct determination of the field dependences of $\gamma$ (often made difficult by magnetic-impurity-phase-caused anomalies at $\approx 2$ K) and $\kappa/T$ at low temperature. Whether $T_S$ remains equal to $T_{SDW}$ in doping on the M-site in 122 $MFe_2As_2$ – unlike for most doping on the Fe and As sites - needs to be checked in homogeneous samples, which K-doped $BaFe_2As_2$ is not.

6.  Specific heat was discussed in sections IIIB and IVB3. Angle resolved specific heat in field to help determine the pairing symmetry, specific heat $\gamma$ to fields greater than 9 T so that $H_{c2}(0)/2$ can be reached to look into two (or more) band anisotropy questions, as well as more high-precision low field data to try to distinguish $H^{1/2}$ from HlogH (clean vs defects) Volovik effect would be interesting. Measuring $\Delta C$ in higher $T_c$ 1111 compounds now that crystals of sufficient mass for such measurements are beginning to



be available, as well as $\Delta C$ data for higher quality 122* samples would extend the check on the correlation $\Delta C \sim T_c^3$, section IIIB3.

7.  Although clearly difficult, it would be nice to settle the question of whether the isotope effect (section IVA) is positive or negative in some model FePn/Ch system.

8.  Pressure is an ideal method in these materials to scan the phase diagram, but only a few of the extant measurements have been able to track the $T_S/T_{SDW}$ anomalies due to sample quality issues and perhaps strain broadening from non-ideal pressure media.

9.  Crystals of LiFeAs are reportedly easily grown, and doping larger atoms on the Li-site to expand the lattice and try to increase $T_c$, based on the monotonic suppression of $T_c$ with pressure discussed in section IID, might provide interesting insights.

10.  Several routes to achieve higher $T_c$ seem to offer promise.  Introducing additional layers, or layers with different structure and/or chemistry, between the $Fe_2As_2$ layers (Ogino et al., 2010a) and trying new compounds using theoretical insight are two such.

In summary, the central question of the relationship between magnetism and superconductivity in this new class of superconductor remains open, although the INS data on the spin fluctuations below $T_c$ in particular are intriguing.  There have been interesting suggestions for the key organizing parameter to link the known FePn/Ch materials and their $T_c$'s, such as pnictide height or tetrahedral angle.   As discussed herein it appears that a single parameter will prove insufficient.  Certainly understanding the FePn/Ch puzzle and how these structures interrelate could benefit from discovering more examples of this unusual form of superconductivity intertwined with magnetism.   Faced with the large number of possible 1111, 122, and 21311/43822/? compounds containing magnetic ions and pnictides or chalcogenides as a starting point for such a search for new



superconducting FePn/Ch, more theoretical input from band structure calculations, e. g. similar to Zhang and Singh's (2009) prescient DFT work on $TlFe_2Se_2$ as a possible parent compound for superconductivity, would certainly be welcome. For example, Yan and Lu (2010) have proposed that CaClFeP might exhibit high temperature superconductivity under doping or high pressure. The work underway to increase $T_c$ by expanding the c-axis, going from the 21311 to the 43822 structure and beyond, is another promising route.

In summary, hopefully researchers in the field can benefit from this review to help their future work. There seems much more to be done. For those not directly involved in the FePn/Ch, the goal was to introduce a rather complex set of results in an approachable fashion, with sufficient references to guide further study.


Acknowledgements: I would like to thank B. Andraka, S. Bud'ko, N. Butch, P. Canfield, X. H. Chen, K. Gofryk, P. Hirschfeld, J. Joyce, S. Kasahara, J. S. Kim (Univ. Florida), J. S. Kim (Postech), K. H. Kim, P. Kumar, M. B. Maple, Y. Matsuda, D. Morr, J. P. Paglione, F. Ronning, T. Shibauchi, I. Vekhter, T. Vojta and H. H. Wen, for useful discussions and sometimes collaborations as well. Special thanks to K. Samwer, who hosted me during the month when this review was started, and to K. H. Kim, who hosted me towards the end of writing this review. This work is dedicated to the memory of Sung-Ik Lee. Work at Florida performed under the auspices of the U. S. Department of Energy, Contract No. DE-FG02-86ER45268.

142203.